\documentclass[paper]{JHEP3}
\pdfoutput=1
\usepackage{amsmath,amssymb,amsthm,amscd,graphicx,epsfig}
\input epsf.sty

\newtheorem{theorem}{Theorem}[section]

\theoremstyle{definition}

\newtheorem{example}[theorem]{Example}

\newtheorem{remark}[theorem]{Remark}

\newcommand{\CA}{{\cal A}}

\newcommand{\CC}{{\cal C}}
\newcommand{\CE}{{\cal E}}
\newcommand{\CF}{{\cal F}}
\newcommand{\CG}{{\cal G}}

\newcommand{\CI}{{\cal I}}

\newcommand{\CN}{{\cal N}}
\newcommand{\CO}{{\cal O}}

\def\IZ{{\mathbb Z}}
\def\IR{{\mathbb R}}
\def\IC{{\mathbb C}}
\def\IP{{\mathbb P}}

\def\IS{{\mathbb S}}

\def\IN{{\mathbb N}}

\newcommand{\tr}{{\rm Tr}}
\newcommand{\re}{{\rm e}}
\newcommand{\ri}{{\rm i}}
\newcommand{\rd}{{\rm d}}

\newcommand{\be}{\begin{equation}}
\newcommand{\ee}{\end{equation}}
\newcommand{\ba}{\begin{aligned}}
\newcommand{\ea}{\end{aligned}}
\newcommand{\ben}{\begin{eqnarray}\displaystyle}
\newcommand{\een}{\end{eqnarray}}

\newcommand{\sectiono}[1]{\section{#1}\setcounter{equation}{0}}

\newdimen\tableauside\tableauside=1.0ex
\newdimen\tableaurule\tableaurule=0.4pt
\newdimen\tableaustep
\def\phantomhrule#1{\hbox{\vbox to0pt{\hrule height\tableaurule width#1\vss}}}
\def\phantomvrule#1{\vbox{\hbox to0pt{\vrule width\tableaurule height#1\hss}}}
\def\sqr{\vbox{%
  \phantomhrule\tableaustep
  \hbox{\phantomvrule\tableaustep\kern\tableaustep\phantomvrule\tableaustep}%
  \hbox{\vbox{\phantomhrule\tableauside}\kern-\tableaurule}}}
\def\squares#1{\hbox{\count0=#1\noindent\loop\sqr
  \advance\count0 by-1 \ifnum\count0>0\repeat}}
\def\tableau#1{\vcenter{\offinterlineskip
  \tableaustep=\tableauside\advance\tableaustep by-\tableaurule
  \kern\normallineskip\hbox
    {\kern\normallineskip\vbox
      {\gettableau#1 0 }%
     \kern\normallineskip\kern\tableaurule}%
  \kern\normallineskip\kern\tableaurule}}
\def\gettableau#1{\ifnum#1=0\let\next=\null\else
\squares{#1}\let\next=\gettableau\fi\next}

\newcommand{\twoVgraph}{\raisebox{0pt}{
                 \begin{picture}(18,18)(-9,-5)
                 \put(0,0){\circle{16}} \put(-8,0){\line(1,0){16}}
                 \end{picture}}}

\newcommand{\fourVgraph}{\raisebox{0pt}{
                 \begin{picture}(18,26)(-9,-9)
                 \put(0,0){\oval(16,24)} \put(-8,4){\line(1,0){16}}
                 \put(-8,-4){\line(1,0){16}}
                 \end{picture}}}

\newcommand{\sixVgraph}{\raisebox{0pt}{
                 \begin{picture}(26,24)(-13,-8)
                 \put(-9,8){\circle{6}} \put(9,8){\circle{6}}
                 \put(-6,8){\line(1,0){12}} \put(0,-8){\circle{6}} 
                 \put(-9,5){\line(2,-3){7}} \put(9,5){\line(-2,-3){7}}
                 \end{picture}}}

\newcommand{\eightVgraphI}{\raisebox{0pt}{
                 \begin{picture}(26,26)(-13,-9)
                 \put(-9,9){\circle{6}} \put(9,9){\circle{6}}
                 \put(-9,-9){\circle{6}} \put(9,-9){\circle{6}} 
	      	 \put(-6,9){\line(1,0){12}}
                 \put(-9,6){\line(0,-1){12}}
                 \put(-6,-9){\line(1,0){12}}
                 \put(9,6){\line(0,-1){12}}
                 \end{picture}}}

\newcommand{\eightVgraphII}{\raisebox{0pt}{
                 \begin{picture}(28,28)(-14,-10)
                 \put(-13,13){\line(1,0){26}}
                 \put(-13,-13){\line(1,0){26}}
                 \put(-13,-13){\line(0,1){26}}
                 \put(13,-13){\line(0,1){26}}
                 \put(-3,3){\line(1,0){6}}
                 \put(-3,-3){\line(1,0){6}}
                 \put(-3,-3){\line(0,1){6}}
                 \put(3,-3){\line(0,1){6}}
                 \put(-13,13){\line(1,-1){10}}
                 \put(-13,-13){\line(1,1){10}}
                 \put(13,13){\line(-1,-1){10}}
                 \put(13,-13){\line(-1,1){10}}
                 \end{picture}}}

\newcommand{\tenVgraphI}{\raisebox{0pt}{
                 \begin{picture}(26,29)(-13,-9)
                 \put(-9,9){\circle{6}} \put(9,9){\circle{6}}
                 \put(-9,-9){\circle{6}} \put(9,-9){\circle{6}}
                 \put(0,9){\circle{6}}
                 \put(-6,9){\line(1,0){3}}
                 \put(6,9){\line(-1,0){3}}
                 \put(-9,6){\line(0,-1){12}}
                 \put(-6,-9){\line(1,0){12}}
                 \put(9,6){\line(0,-1){12}}
                 \end{picture}}}

\newcommand{\tenVgraphII}{\raisebox{0pt}{
                 \begin{picture}(28,31)(-14,-10)
                 \put(-13,13){\line(1,0){10}}
                 \put(0,13){\circle{6}}
                 \put(13,13){\line(-1,0){10}}
                 \put(-13,-13){\line(1,0){26}}
                 \put(-13,-13){\line(0,1){26}}
                 \put(13,-13){\line(0,1){26}}
                 \put(-3,3){\line(1,0){6}}
                 \put(-3,-3){\line(1,0){6}}
                 \put(-3,-3){\line(0,1){6}}
                 \put(3,-3){\line(0,1){6}}
                 \put(-13,13){\line(1,-1){10}}
                 \put(-13,-13){\line(1,1){10}}
                 \put(13,13){\line(-1,-1){10}}
                 \put(13,-13){\line(-1,1){10}}
                 \end{picture}}}

\newcommand{\figref}[1]{Fig.~\protect\ref{#1}}

\title{Lectures on non-perturbative effects in large $N$ gauge theories, matrix models and strings}

\author{
Marcos Mari\~no
\\
D\'epartement de Physique Th\'eorique et Section de Math\'ematiques,\\
Universit\'e de Gen\`eve, Gen\`eve, CH-1211 Switzerland\\
\\
\email{marcos.marino@unige.ch}}

\abstract{In these lectures I present a review of non-perturbative instanton effects in quantum theories, with a focus on large $N$ gauge theories and matrix models. I first consider the structure of these effects in the case of ordinary differential equations, which provide a model for more complicated theories, and I introduce in a 
pedagogical way some technology from resurgent analysis, like trans-series and the resurgent version of the Stokes phenomenon. After reviewing instanton effects in quantum mechanics and quantum field theory, I address general aspects of large $N$ instantons, and 
then present a detailed review of non-perturbative effects in matrix models. Finally, I consider two applications of these techniques in string theory.}

\begin{document}

\sectiono{Introduction}

Many series appearing in Physics and in Mathematics are not convergent. For example, most of the series 
obtained by perturbative methods in Quantum Field Theory (QFT) turn out to be asymptotic, rather than convergent. The asymptotic 
character of these series is typically an indication that non-perturbative effects should be ``added" in some way to the original perturbative series.

The purpose of these lectures is to provide an introduction to asymptotic series and non-perturbative effects. Since this is already quite a 
wide topic, we will restrict our discussion in various ways. First or all, we will consider non-perturbative effects of the 
instanton type, i.e. effects due to extra saddle points in the path integral (in particular, we will not consider effects of the renormalon type). 
Secondly, we will be particularly interested in the interaction between non-perturbative effects 
and large $N$ expansions. Finally, in our discussion we will rely heavily on ``toy models" of non-perturbative effects, and in particular on matrix models. This is not as restrictive as one could think, since matrix models underlie many interesting quantities in string theory and supersymmetric QFTs. 

A typical asymptotic series in a coupling constant $g_s$, with instanton corrections, has the following form, 
\be
\label{simpleseries}
\sum_n a_n g_s^n + \re^{-A/g_s}\sum_n a_n^{(1)} g_s^n+\CO(\re^{-2A/g_s}). 
\ee
Here, the first sum is the original divergent series. The second term is an one-instanton contribution. It has an overall, non-pertubative exponential in $g_s$, multiplying another series (which in general is also asymptotic). The third term indicates higher instanton contributions. 
In order to understand this type of quantities, we will develop a three-step approach: 

\begin{enumerate}

\item {\it Formal}: we want to be able to compute the terms in the original, perturbative series, as well as the ``non-perturbative quantities" characterizing the instanton 
corrections. These include the instanton action $A$, as well as the series multiplying the exponentially small terms. Notice that the resulting object is a series with two 
small parameters, $g_s$ and $\re^{-A/g_s}$, which should be regarded as independent. Such series are called {\it trans-series}. Therefore, the first step in the 
understanding of non-perturbative effects is the formal 
calculation of trans-series. 

\item {\it Classical asymptotics}: the series (\ref{simpleseries}) above is a purely formal expression, since already the first series (the perturbative one) is divergent. One way of making sense of this perturbative series is by regarding it as an asymptotic expansion of 
a well-defined function. Finding such a representation, once the original function is given, is the problem addressed by what 
I will call ``classical asymptotics."  In classical asymptotics, exponential corrections are ill-defined and are never written down explicitly, but they are 
lurking in the background: indeed, it might happen that, as we move in the complex plane of the coupling constant, an instanton correction which used to 
be exponentially small becomes of order one. This is the basis of the {\it Stokes phenomenon}. 

\item {\it Beyond classical asymptotics}: once the classical asymptotics is understood, one can try to go beyond it and use the full information in the formal trans-series 
to reconstruct the original function {\it exactly}. A general technique to do this is Borel resummation. The combination of the general theory of 
trans-series with Borel resummation gives the {\it theory of resurgence} of Jean \'Ecalle, which is the most powerful method to address this whole circle of questions. 

\end{enumerate}
 
 In these lectures we will follow this three-step program in problems with increasing order of complexity, from simple ones, like ordinary differential equations (ODEs),  to more difficult ones, like matrix models and string theory. Let us quickly review these different problems in relation to the approach sketched above:

\begin{enumerate}

\item In the theory of {\it ODEs} with irregular singular points, formal solutions are typically divergent series. In this case, 
the three steps are well understood. 
The first step is almost automatic, since the terms in the trans-series can be 
computed recursively, including the instanton corrections. The understanding of the classical asymptotics of these solutions is a venerable subject, 
going back to Stokes. The treatment beyond classical asymptotics is 
more recent, but has been well established in the work of \'Ecalle, Kruskal, Costin, and others, and we will review some of this work here. A closely related example is the calculation of integrals by the method of steepest descent. Here, the different trans-series are given by the 
asymptotic expansion of the integrals around the different critical points. The classical asymptotics of these series is also a well studied subject, 
and its resurgent analysis has been also considered by Berry and others. 

\item In {\it Quantum Mechanics} (QM) and {\it Quantum Field Theory}, the formal procedure to generate the leading asymptotics series is simply standard perturbation theory. Instanton corrections are obtained by identifying saddle-points of the classical action, and by doing perturbation theory around this instanton background, one obtains exponentially small corrections and 
their associated series. The study of classical asymptotics and beyond is more difficult, although many results have been obtained in QM. In realistic QFTs, perturbation theory is so wild that it is not feasible to pursue the program, but in some special QFTs --namely, those without renormalons, 
like Chern--Simons (CS) theory in 3d or $\CN=4$ Yang--Mills theory-- there are some partial results. In these lectures I will consider some aspects of the problem for CS theory. 

\item An interesting, increasing level of complexity appears when we consider 
{\it large $N$ gauge theories}. Here, the computation of formal trans-series becomes more complicated, since 
in the $1/N$ expansion we typically have to resum an infinite number of diagrams at each order. For example, in standard QFT, the instanton action is obtained by finding 
a classical solution of the equations of motion (EOM) with finite action, while the action of a large $N$ instanton is given by the sum of an infinite number of diagrams. 
Another way of understanding this new level of 
complexity is simply that in large $N$ theories we have two parameters in the game, $N$ and the coupling constant $g_s$, or 
equivalently $g_s$ and the 't Hooft parameter $t=N g_s$. Due to this, the Stokes phenomenon of classical asymptotics becomes more complicated 
and leads to {\it large $N$ phase transitions}\footnote{In fact, although it is not widely appreciated, standard phase transitions are examples of the Stokes phenomenon; see \cite{ps} for a nice discussion of this.}. However, in the simple toy case of large $N$ matrix models, the $1/N$ expansion is still very close to the ordinary asymptotic 
evaluation of integrals, and we will be able to offer a rather detailed picture of non-perturbative effects.

\item Finally, in {\it string theory} things are even more complicated. Even at the formal level we 
cannot go very far: there are rules to compute the genus expansion, but the rules to do (spacetime) instanton calculations are rather {\it ad hoc}. We know that in general instanton corrections involve D-branes or $p$-branes, 
and we know how to obtain some qualitative features of their behavior, but a precise framework is still missing due to the lack of a non-perturbative definition. 
An alternative avenue is to use large $N$ dualities, relating 
string theories to gauge theories, in order to deduce some of these effects from the large $N$ duals. This makes possible to compute formal trans-series for non-critical strings and some simple string models. 
\end{enumerate}

The plan of these lectures is the following: in section \ref{ODEs}, I will review some aspects of  asymptotics series and in particular of the series appearing in the context of ODEs. This includes formal trans-series, Borel resummation, the Stokes phenomenon, and the connection between large order behavior and trans-series. I have tried to provide as well some very elementary ideas of the theory of resurgence. In section 3, I review non-perturbative effects in QM and in QFT. The results in QM are well known to the expert; they have been analyzed in detail and made rigorous in for example \cite{delabaerepham}. The study of non-perturbative effects in QFT from the point of view advocated in these lectures is still in its infancy, and I have contented myself with explaining some of the results for CS theory, which are probably not so well known. In section 3 I also introduce some general aspects of instanton effects in large $N$ theories which are probably well known to experts, but difficult to find in the literature, and I illustrate them in the simple case of matrix quantum mechanics. Section 4 is devoted to the study of non-perturbative effects in matrix models, starting from the pioneering works of F. David in \cite{david} and explaining as well more recent results where I have been involved. In section 5, I present two applications of the techniques of section 4 to string theory, by using large $N$ dualities. Finally, in section 6 I make some concluding remarks.

\sectiono{Asymptotics, non-perturbative effects, and differential equations}
\label{ODEs}

\subsection{Asymptotic series and exponentially small corrections}
\label{smallcorr}
A series of the form 
\be
\label{aseries}
S(w)=\sum_{n=0}^{\infty} a_n w^n
\ee
is asymptotic to the function $f(w)$, in the sense of Poincar\'e, if, for every $N$, the remainder after $N+1$ terms of the series is much smaller than the last 
retained term as $w\rightarrow 0$. More precisely, 
\be
\lim_{w\to 0} w^{-N}\left( f(w)-\sum_{n=0}^{N} a_n w^n\right)=0, 
\ee
for all $N>0$. In an asymptotic series, the remainder does not necessarily go to zero as $N \rightarrow \infty$ for a fixed $w$, in contrast to what happens in convergent series. 
Analytic functions might have asymptotic expansions. For example, the Stirling series for the 
Gamma function 
\be
\label{stirling}
\Bigl( {z \over 2\pi}\Bigr)^{1/2} \Bigl( {z\over e}\Bigr)^{-z}\Gamma(z) = 1+ {1\over 12 z} +{1\over 288 z^2} +\cdots
\ee
is an asymptotic series for $|z| \rightarrow \infty$. Notice that different functions may have the same asymptotic expansion, since 
\be
\label{expcorr}
f(w) + C\re^{-A/w} 
\ee
has the same expansion around $w=0$ than $f(w)$, for any $C$, $A$. 

In practice, asymptotic expansions are characterized by the fact that, as we vary $N$, the partial sums 
\be
\label{partialsum}
S_N(w)=\sum_{n=1}^N a_n w^n
\ee
will first approach the true value $f(w)$, and then, for $N$ sufficiently big, they will diverge. A natural question is then to obtain the partial 
sum which gives the best possible estimate of $f(w)$. To do this, one has to find the $N$ that truncates the asymptotic expansion 
in an optimal way. This procedure is called {\it optimal truncation}. Usually, the way to find the optimal value of $N$ is to retain terms up to the smallest 
term in the series, discarding all terms of higher degree. Let us assume (as it is the case in all interesting examples) that the coefficients $a_n$ in (\ref{aseries}) grow factorially at large $n$, 
\be
\label{aas}
a_n \sim A^{-n} n! \, .
\ee
The smallest term in the series, for a fixed $|w|$, is obtained by minimizing $N$ in 
\be
\left|a_N w^N \right| =c N! \left|{w \over A}\right|^N.  
\ee
By using the Stirling approximation, we rewrite this as
\be
c \exp \left\{ N  \left( \log \, N  -1 -\log \,  \left|{A \over w}\right| \right) \right\}.
\ee
The above function has a saddle at large $N$ given by
\be
\label{optimalN}
N_*= \left|{A \over w}\right|. 
\ee
If $|w|$ is small, the optimal truncation can be performed at large values of $N$, but as $|w|$ increases, less and less terms of the series can be used. 
We can now estimate the error made in the optimal truncation by evaluating the next term in the asymptotics, 
\be
\label{nonpertamb}
\epsilon(w) =C_{N_*+1} |w|^{N_*+1} \sim \re^{-|A/w|}.
\ee
Therefore, the maximal ``resolution" we can expect when we reconstruct a function $f(w)$ from an asymptotic expansion is of order $\epsilon(w)$. This ambiguity 
is sometimes called a {\it non-perturbative ambiguity}. The reason for this name is that perturbative series are often asymptotic, therefore they do not 
determine by themselves the function $f(w)$, and some additional, non-perturbative information is required. Notice that the absolute value of $A$ gives the 
``strenght" of this ambiguity. 

\begin{figure}[!ht]
\leavevmode
\begin{center}
\includegraphics[height=4cm]{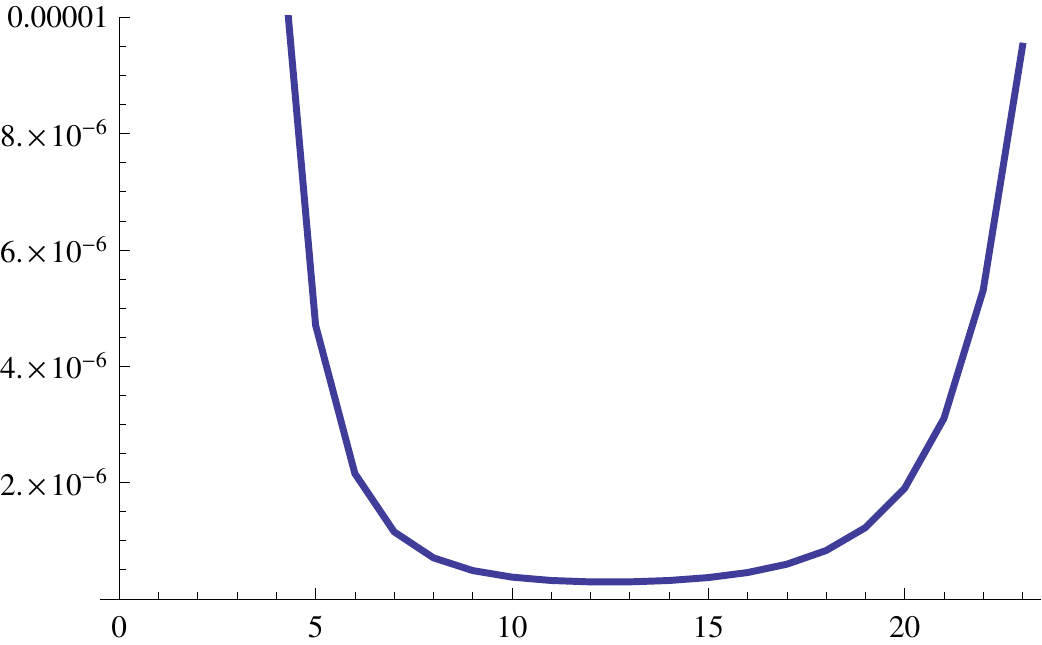}\qquad \includegraphics[height=4cm]{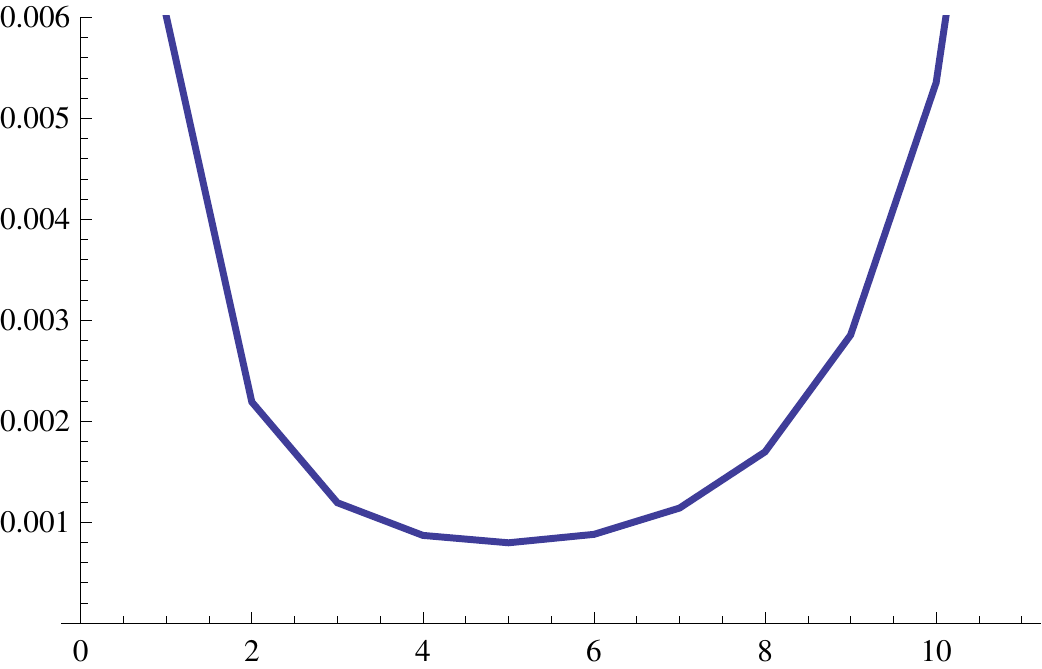}
\end{center}
\caption{We illustrate the method of optimal truncation for the quartic integral (\ref{quarticint}) by plotting the difference (\ref{diffexact}) between the integral and the partial sum of order $N$ of its asymptotic expansion, as a function of $N$, for $g=0.02$ (left) and $g=0.05$ (right).}
\label{optimal}
\end{figure}
It is instructive to see optimal truncation at work in a simple example. Let us consider the quartic integral
\be
\label{quarticint}
I(g) = {1\over {\sqrt {2\pi}}} \int_{-\infty}^{\infty} \rd z \, \re^{-z^2/2 -g z^4/4},
\ee
which is well-defined as long as ${\rm Re}(g)>0$. The asymptotic expansion of this integral for small $g$ can be obtained simply by first expanding 
the exponential of the quartic perturbation and then integrate the resulting series term by term. One obtains, 
\be
\label{quarticseries}
I(g) =\sum_{k=0}^{\infty} a_k g^k, 
\ee
where
\be
a_k  = {(-4)^{-k} \over  {\sqrt {2\pi}}} \int_{-\infty} ^{\infty} \rd z {z^{4k} \over  k!} \re^{-z^2/2} =(-4)^{-k} {(4k-1)!! \over k!}. 
\ee
This series has a zero radius of convergence and provides an asymptotic expansion of the integral $I(g)$. The asymptotic behavior 
of the coefficients at large $k$ is obtained immediately from Stirling's formula 
\be
\label{zkasym}
a_k \sim (-4)^k k!,
\ee
therefore $|A|=1/4$ in this case. In \figref{optimal} we plot the difference 
\be
\label{diffexact}
\left| I(g) -S_N(g)\right|
\ee
as a function of $N$, for two values of $g$. The optimal values are seen to be $N_*=12$ and $N_*=5$, in agreement with the estimate (\ref{optimalN}). 

Perhaps the most important problem in asymptotic analysis is how to go beyond optimal truncation, incorporating in a systematic way the small exponential effects (\ref{expcorr}). In this first section we will address this problem in the context of asymptotic series appearing in ODEs. 

\subsection{Formal power series and trans-series in ODEs}  

Very often, the solution to a physical or mathematical problem is given by an asymptotic series, and one has a systematic procedure to calculate this series at any desired order (like for example in perturbation theory in QM). The first question we want to ask is the following: can we calculate, at lest {\it formally}, non-perturbative effects of exponential type (i.e. like the one in (\ref{expcorr})) which must be added to an asymptotic series? 
Since these terms are typically not taken into account in classical asymptotics, in order to include them we have to consider generalizations of asymptotic series. 
The resulting objects are called {\it trans-series} and they were first considered in a systematic way by Jean \'Ecalle in his work on ``resurgent analysis" \cite{ecalle}.  

 An important class of asymptotic series and trans-series are the formal solutions to ODEs with irregular singular points. The simplest example is probably Euler's equation
 \be
 \label{eulereq}
  {\rd \varphi \over \rd z}+A \varphi(z) ={A\over z}
 \ee
 which has an irregular singular point at $z=\infty$. 
 There is a formal power-series solution to this equation of the form 
 \be
 \label{solEuler}
 \varphi_0(z)=\sum_{n=0}^{\infty} {a_n \over z^{n+1}}, \qquad a_n=A^{-n} n!.
 \ee
This solution is an asymptotic series, with zero radius of convergence, since the coefficients grow factorially with $n$. 
It is easy to see that one can construct a family of formal solutions to the Euler's ODE based on $\varphi_0(z)$,
 \be
 \label{eulerts}
 \varphi(z) =\varphi_0(z) +C\re^{-Az}
 \ee
 where $C$ is an arbitrary constant parametrizing the family of solutions. This is our first example of a {\it trans-series}, and it has three important properties:
 
   \begin{enumerate}
 
 \item The term added in (\ref{eulerts}) is non-analytic at $z=\infty$, and it goes beyond the standard solution in the form of an asymptotic series given by (\ref{solEuler}). 
 It is invisible in the ``perturbative" expansion around $z=\infty$, and it is therefore a ``non-perturbative" correction. 
 
 \item The resulting formal expression has {\it two} small parameters, $1/z$ and $\re^{-A z}$. 
 
 \item There is a relation, already pointed out above in the context of optimal truncation, between the ``strength" of 
 the non-perturbative effect, given by $A$, and the divergence of the asymptotic series. 
 Namely, $A$ encodes the next-to-leading behavior of $a_n$ at large $n$. 
\end{enumerate} These properties are typical of formal trans-series and will reappear in many examples. 
 
A more complicated example of an ODE with a trans-series solution is the well-known {\it Airy equation}, 
\be
\varphi''=x \varphi.  
\ee
The solutions to this equation, called Airy functions, are ubiquitous in physics. We are now interested in formal power series solutions to this equation around $x=\infty$. 
It is easy to see that one such a solution is given by 
\be
\label{fullai}
Z_{\rm Ai}(x)={1\over 2x^{1/4} {\sqrt{\pi}}} \re^{-2 x^{3/2}/3}
\sum_{n=0}^{\infty} a_n x^{-3n/2},
\ee
where
\be\label{anairy}
a_n = {1\over 2\pi} \Bigl( -{3\over 4} \Bigr)^n {\Gamma(n+{5\over 6}) \Gamma(n+{1\over 6} ) \over n!}. 
\ee
Here, the coefficients grow as
\be
\label{aigrow}
a_n \sim A^{-n} n!, \qquad A =-{4\over 3}.
\ee
What is the trans-series solution to this equation? As it is well-known, there is another, independent formal power series solution to the Airy equation, given by
\be
\label{fullbi}
Z_{\rm Bi}(x)={1\over 2x^{1/4} {\sqrt{\pi}}} \re^{2 x^{3/2}/3}
\sum_{n=0}^{\infty} (-1)^n a_n x^{-3n/2}.
\ee
The general ``trans-series solution" to the Airy differential equation is just a linear combination of these two formal asymptotic series, 
 \be
C_1  Z_{\rm Ai}(x)+C_2 Z_{\rm Bi}(x).
\ee
Notice that $Z_{\rm Ai}(x)$ and $Z_{\rm Bi}(x)$ have different leading exponential behavior. Moreover, the relation between these behaviors is in accord with the third property noted above in the context of Euler's equation: the growth (\ref{aigrow}) of the coefficients suggests that the trans-series added to the asymptotic solution (\ref{fullai}) should have a relative exponential weight of  
\be
\re^{4 x^{3/2}/3} 
\ee
as compared to (\ref{fullai}), which is indeed the case. 

In the case of non-linear ODEs the structure of trans-series solutions is much richer: linear ODEs have trans-series with a finite number of terms, while in nonlinear ODEs they have an infinite number of terms. A class of important examples which are relevant in many physical applications are the 
{\it Painlev\'e transcendants}. We will focus on the cases of Painlev\'e I (PI) and Painlev\'e II (PII).  

\begin{example} {\it Painlev\'e I}. 
 The PI equation is
 \be
 \label{p1}
u(\kappa)^2-{1\over 6}u''(\kappa)=\kappa.
\ee
 This equation appears in many contexts. In particular, it gives the all-genus solution to two-dimensional quantum gravity (see \cite{dfgzj} for a review). There is a formal solution to this equation which goes like $u(\kappa) \sim {\sqrt{\kappa}}$ as $\kappa \rightarrow \infty$:
 \be
 \label{asymp1}
u^{(0)}(\kappa)=\kappa^{1/2 } \sum_{n=0}^\infty u_{0,n} \kappa^{-5n/2}=\kappa^{1/2} \biggl( 1-{1\over 48} \kappa^{-{5 \over 2}} -{49 \over 4608} \kappa^{-5} -{1225 \over 55 296} \kappa^{-{15\over 2}}+\cdots\biggr). 
\ee
The {\it trans-series} solution to Painlev\'e I  is a one-parameter family of solutions to 
(\ref{p1}) which includes exponentially suppressed terms as $\kappa\rightarrow \infty$:
\be
\label{transseries}
u(\kappa) = \sum_{\ell=0}^{\infty}C^{\ell} u^{(\ell)}(\kappa)=
 {\sqrt {\kappa}}  \sum_{\ell=0}^{\infty} C^{\ell} \kappa^{-{5 \ell \over 8}} \re^{-\ell A \kappa^{5/4}} \epsilon^{(\ell)}(\kappa), 
\ee
where $C$ is a parameter, the constant $A$ has the value
\be
\label{aaction}
A={8 {\sqrt 3}\over 5}
\ee
and 
\be
\label{el}
\epsilon^{(\ell)}(\kappa)=\sum_{n=0}^{\infty} u_{\ell,n} \kappa^{-5n/4}
\ee
are asymptotic series. Since we have introduced an arbitrary constant $C$ in (\ref{transseries}), we can 
normalize the solution such that $u_{1,0}=1$. We will refer to the above series $u^{(\ell)}(\kappa)$ with $\ell \ge 1$ as the $\ell$-th instanton 
solution of PI, while the solution $u^{(0)}(\kappa)$ will be referred to as the ``perturbative" solution. 

It is easy to see that the $\ell$-th instanton solutions $u^{(\ell)}$ in the trans-series satisfy linear ODEs. For example, one has for $\ell=1$, 
\be
\left(u^{(1)}\right)''(\kappa)=12 u^{(1)}(\kappa) u^{(0)}(\kappa), 
\ee
and from this one finds, 
\be
\epsilon^{(1)}(\kappa)= 1 - {5 \over 64 {\sqrt {3}}} z^{-{5\over 4}} 
+ {75 \over 8192} z^{-{5\over 2}} - {341329 \over 23592960 {\sqrt{3}}} z^{-{15\over 4}} +\cdots. 
\ee
Recursion relationships for the coefficients $u_{\ell,n}$ in (\ref{el}) can be found in \cite{gikm}. 
\end{example}

\begin{example} \label{exp2} {\it Painlev\'e II}. The Painlev\'e II equation is
\be
\label{p2}
 u''(\kappa)-2 u^3(\kappa)+ 2\kappa u(\kappa)=0
\ee
This equation is of fundamental importance in, for example, random matrix theory and non-critical string theory. It appears in the celebrated Tracy--Widom law governing the statistics of the largest eigenvalue in a Gaussian ensemble of random matrices \cite{tw}, in the double-scaling limit of unitary matrix models \cite{ps} and of two-dimensional Yang--Mills theory \cite{gm}, and it also governs the all-genus free energy of two-dimensional supergravity \cite{kms}. As in the case of PI, there is a formal solution to PII which goes like $u(\kappa) \sim {\sqrt{\kappa}}$ as $\kappa \rightarrow \infty$:
\be
\label{pertp2}
u^{(0)}(\kappa)= {\sqrt{\kappa}} - 
  \frac{1}{16\,\kappa^{\frac{5}{2}}} - \frac{73}{512\,\kappa^{\frac{11}{2}}}- 
  \frac{10657}{8192\,\kappa^{\frac{17}{2}}}  - \frac{13912277}{542888\,\kappa^{\frac{23}{2}}} 
+\cdots, \qquad \kappa \rightarrow \infty.
\ee
One can consider as well exponentially suppressed corrections to this ``perturbative" behavior and construct a formal trans-series solution with the structure,
\be
\label{p2trans}
u(\kappa) =\sum_{\ell=0}^{\infty} C^{\ell} u^{(\ell)}(\kappa)={\sqrt {\kappa}}  
\sum_{\ell=0}^{\infty} C^{\ell} \kappa^{-{3 \ell \over 4}} \re^{-\ell A \kappa^{3/2}} \epsilon^{(\ell)}(\kappa), \quad \kappa \rightarrow \infty,
\ee
where
\be
A={4\over 3}
\ee
and
\be
\label{eltwo}
\epsilon^{(\ell)}(\kappa)=\sum_{n=0}^{\infty} u_{\ell,n} \kappa^{-3n/2}.
\ee
As before, we normalize the solution with $u_{1,0}=1$. 
The perturbative part $u^{(0)}(\kappa)$ is given by (\ref{pertp2}). The instanton expansions can be easily found by plugging the trans-series 
ansatz in the Painlev\'e II equation. One finds, for example, for the one-instanton solution, 
\be
\epsilon^{(1)}(\kappa)=1-{17\over 96} \kappa^{-3/2} + {1513 \over 18432} \kappa^{-3} -\cdots \, . 
\ee
\end{example}

After these examples, we can now give some more formal definitions (see \cite{costin}). Let 
\be
{\boldsymbol \varphi}'={\bf f}(z,{\boldsymbol \varphi}),
\ee
be a rank $n$ system of non-linear differential equations. We assume that ${\bf f}(z,{\boldsymbol \varphi})$ is analytic at $(\infty,{\bf 0})$. 
Let $\lambda_i$, $i=1,\cdots, n$, be the eigenvalues of the linearization
\be
\hat \Lambda= -\Bigl( {\partial f_i \over \partial \varphi_j} (\infty,{\bf 0})\Bigr)_{i,j=1, \cdots, n}. 
\ee
By using various changes of variables, one can always bring the system to the so-called {\it normal} or {\it prepared} form
\be
\label{matrices}
{\boldsymbol \varphi}'=-\Lambda {\boldsymbol \varphi} -{1\over z} B {\boldsymbol \varphi}+ {\bf g}(z, {\boldsymbol \varphi}), 
\ee
where 
\be
\Lambda={\rm diag} \, (\lambda_1, \cdots, \lambda_n), \qquad B={\rm diag}(\beta_1, \cdots, \beta_n), 
\ee
and by construction ${\bf g}(z, {\boldsymbol \varphi})= \CO(|{\boldsymbol \varphi}|^2, z^{-2}{\boldsymbol \varphi})$. We also choose variables in such a way that $\lambda_1>0$. 

\begin{example} {\it Airy equation in prepared form}. 
Define
\be
\label{xairy}
z=x^{3/2}, 
\ee
so that the Airy equation reads
\be
\varphi''(z)={4\over 9} \varphi(z) -{1\over 3z} \varphi'(z). 
\ee
Let us consider the matrix
\be
S(z)=\begin{pmatrix} -3/2 & 3/2 \\ 1-{1\over 4 z} & 1 +{1\over 4 z} \end{pmatrix}.
\ee
If we write
\be
{\bf u}(z)=S^{-1}(z) {\boldsymbol \varphi} (z), \qquad {\boldsymbol \varphi}(z)=\begin{pmatrix} \varphi(z) \\ \varphi'(z) \end{pmatrix},
\ee
we find 
\be
{\bf u}'(z) =-\Lambda {\bf u} -{1\over z} B {\bf u} +{\bf g}(z,{\bf u}), 
\ee
with 
\be
\Lambda=\left(
\begin{array}{ll}
 2/3 & 0 \\
 0 & -2/3
\end{array}
\right), \qquad B=\left(
\begin{array}{ll}
 \frac{1}{6 } & 0 \\
 0 & \frac{1}{6 }
\end{array}
\right), \qquad {\bf g}(z,{\bf u})={5\over 48 z^2}\begin{pmatrix} 1 & -1 \\ 1& -1 \end{pmatrix} {\bf u}.
\ee
\end{example}

The {\it formal trans-series solution} to (\ref{matrices}) is of the form
\be
\label{trans}
 {\boldsymbol \varphi}= {\boldsymbol \varphi}_0 + \sum_{ {\bf  k} \in \IN^n\backslash \{ 0\} } {\bf C}^{\bf k} \re^{-{\bf k} \cdot {\boldsymbol \lambda} z} z^{-{\bf k} \cdot {\boldsymbol \beta}}  {\boldsymbol \varphi}_{\bf k},
\ee
where
\be
{\bf C} =(C_1, \cdots, C_n)
\ee
are free parameters and
\be
{\bf C}^{\bf k}=C_1^{k_1}\cdots C_n^{k_n}.
\ee
 The functions $ {\boldsymbol \varphi}_0$ and ${\boldsymbol \varphi}_{\bf k}$ are formal power series, of the form
\be
{\boldsymbol \varphi}_{\bf k}=\sum_{n\ge 0} {\boldsymbol \varphi}_{{\bf k};n} z^{-n}
\ee
We will also denote
\be
|{\bf k}|=\sum_i k_i. 
\ee
\begin{remark} For linear systems, like the Airy equation, all the trans-series ${\boldsymbol \varphi}_{\bf k}$ vanish when $|{\bf k}|\ge 2$, so the general trans-series solution is of the form 
\be
{\boldsymbol \varphi} = {\boldsymbol \varphi}_0 + \sum_{i=1}^n C_i \re^{- \lambda_i z} z^{-\beta_i}  {\boldsymbol \varphi}_{i}.
 \ee
\end{remark}

\subsection{Classical asymptotics and the Stokes phenomenon}

\begin{quotation}
Stokes, by mathematical supersubtlety, transformed Airy's integrals... 
\begin{flushright}
Lord Kelvin, ``The scientific work of Georges Stokes."
\end{flushright}
\end{quotation}

\begin{figure}[!ht]
\leavevmode
\begin{center}
\includegraphics[height=6cm]{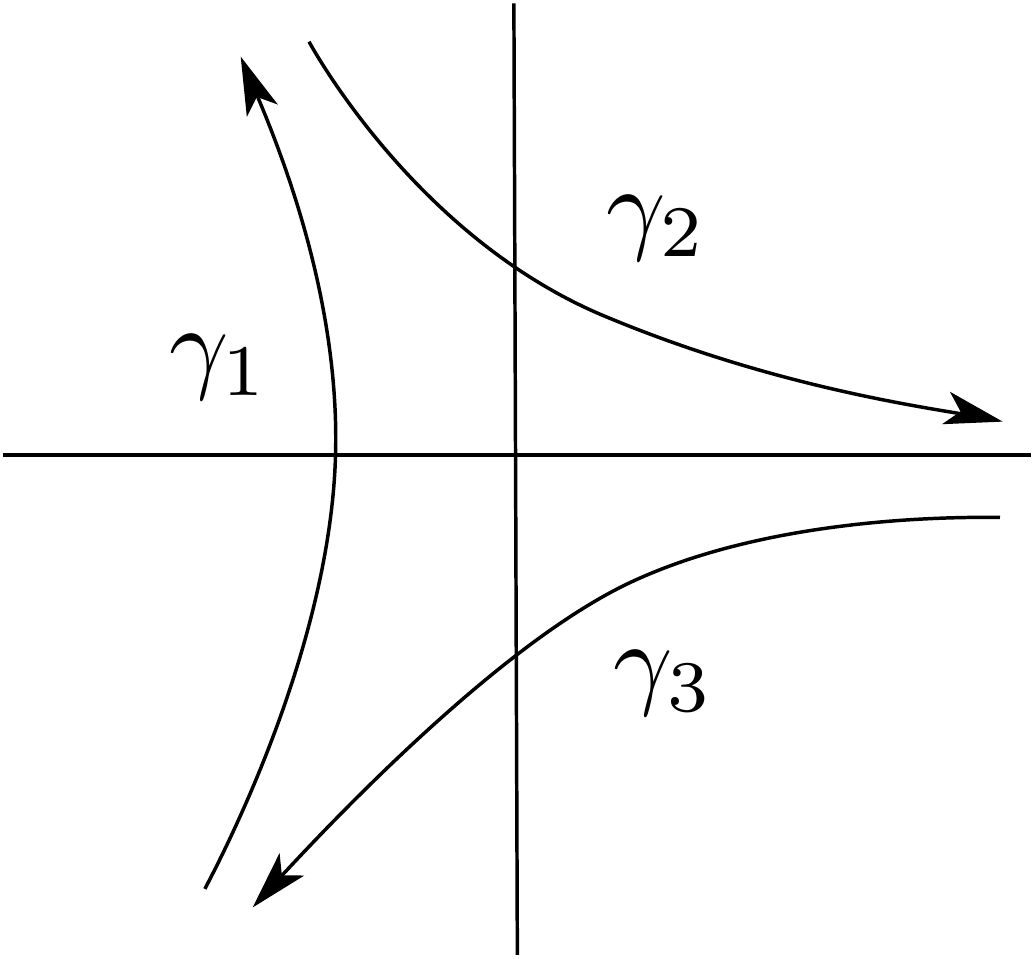}
\end{center}
\caption{Three paths which lead to solutions of the Airy equation.}
\label{airycontours}
\end{figure} 

\begin{figure}[!ht]
\leavevmode
\begin{center}
\includegraphics[height=8cm]{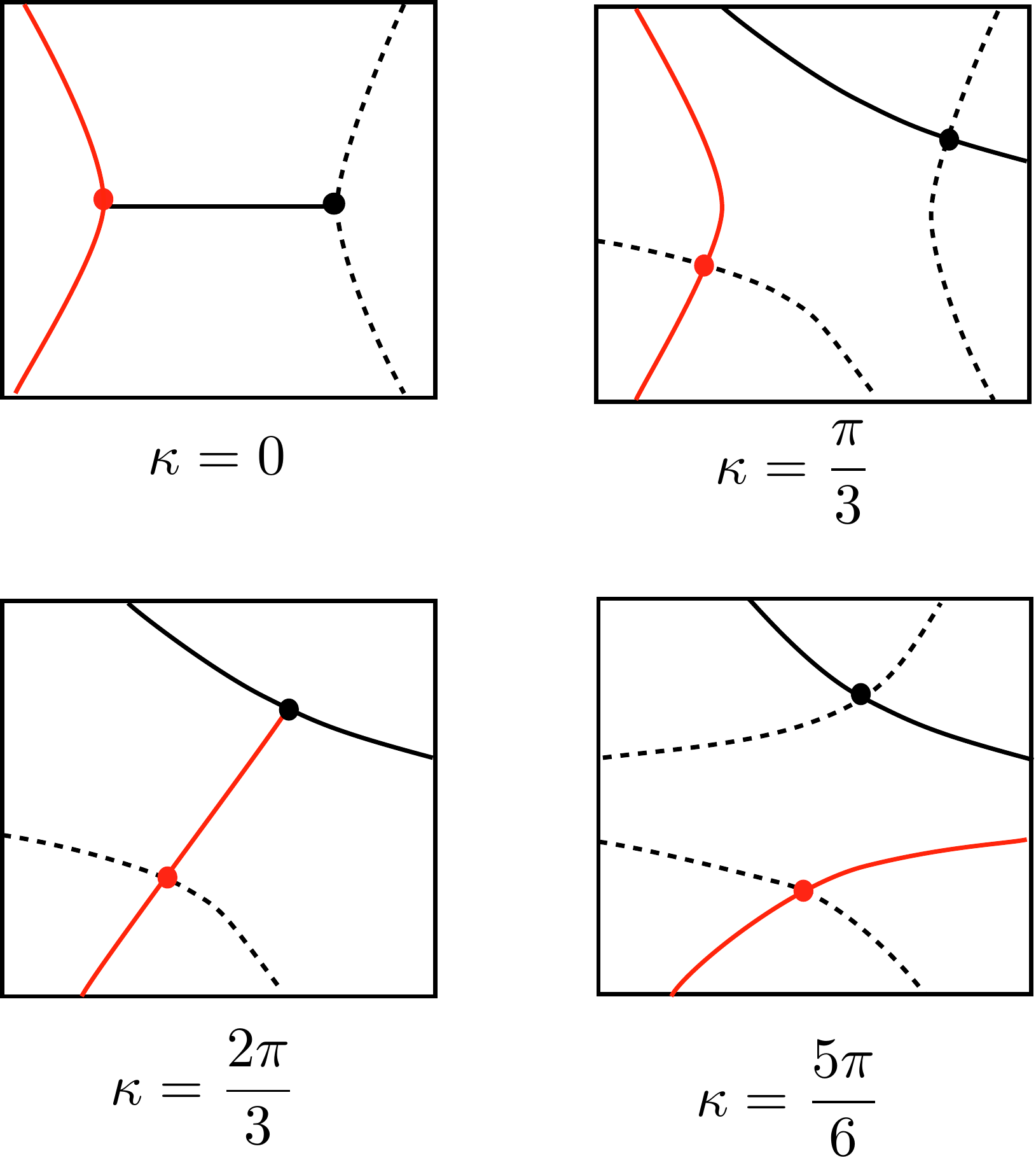}
\end{center}
\caption{Saddle-point analysis of the integral (\ref{lamint}) for different values of $\kappa$. The red dot on the left is the critical point $u^{\rm L}$, while the 
black point on the right is the critical point $u^{\rm R}$. The continuous lines represent paths of steepest descent, while the dashed lines are paths of steepest ascent.}
\label{thimbles}
\end{figure}

So far we have worked at a formal level and we have obtained formal solutions, in terms of asymptotic series, to ordinary differential equations. 
A natural questions is: what is the meaning of these formal power series? In order to answer this question, it is useful to consider in detail the case of the Airy equation and the Airy function, and to use an integral representation (we follow here the discussion in chapter 4 of \cite{miller}). Let us consider the integral
\be
\label{zoneg}
I_\gamma={1\over 2\pi \ri} \int_{\gamma} \rd z \, \re^{S(z)}, \qquad S(z)=x z-{z^3\over 3},
 \ee
where $\gamma$ is a path which makes the integral convergent. Three such paths are shown in \figref{airycontours}, but not all of them are independent, since
 \be
 \gamma_1+\gamma_2 + \gamma_3=0. 
 \ee
It is easy to see that the above integral (\ref{zoneg}) gives a solution to the Airy differential equation. We will now focus on the function defined by the path $\gamma_1$:
\be
{\rm Ai}(x)={1\over 2\pi \ri} \int_{\gamma_1} \rd z \, \re^{x z-{z^3\over 3}},
\ee
which defines the Airy function ${\rm Ai}(x)$. This function is analytic in the complex plane. We set 
\be
x=r\re^{\ri \kappa}
\ee
and rescale the integrand 
\be
z  = u r^{1/2}. 
\ee
We find in this way
\be
{\rm Ai}(x)={r^{1/2} \over 2\pi \ri} \int_{\gamma_1} \re^{r^{3/2} (\re^{\ri \kappa} u -u^3/3)} \rd u. 
\ee
We now want to study the behavior of this function for $|x| \gg 1$, by using the saddle-point method. We then focus on the integral 
\be
\label{lamint}
\int_{\gamma_1} \re^{\lambda S_{\kappa}(u)} \rd u
\ee
where
\be
S_{\kappa}(u) = \re^{\ri \kappa} u -{u^3\over 3}. 
\ee
There are two saddle points
\be
u^{\rm R}=\re^{\ri \kappa/2}, \qquad u^{\rm L}=-\re^{\ri \kappa/2}
\ee
with ``actions"
\be
S_{\kappa}(u^{{\rm R},{\rm L}})=\pm {2\over 3} \exp\left( {3 \ri \over 2} \kappa\right).
\ee
It is easy to see that the formal asymptotic power series around the saddle point $u^{\rm L}$ is precisely (\ref{fullai}), while the one around $u^{\rm R}$ is (\ref{fullbi}). 
We introduce the notation 
\be
R_{\kappa}^{ {\rm R}, {\rm L}} ={\rm Re}\left(S_{\kappa}(u^{{\rm R},{\rm L}})\right), \qquad 
I_{\kappa}^{ {\rm R}, {\rm L}} ={\rm Im}\left(S_{\kappa}(u^{{\rm R},{\rm L}})\right).
\ee
The paths of steepest descent (ascent) passing through these points are those where the function ${\rm Re}(S_{\kappa}(u))$ decreases (respectively, increases) most rapidly, as we move away from the critical point. We will also denote by $\gamma^{R,L}_\kappa$ the steepest descent paths through $u^{{\rm R},{\rm L}}$, respectively. We show some paths of steepest descent and ascent in \figref{thimbles}. 

Let us now study what happens as we change the angle $\kappa$. For 
\be
 |\kappa| < {2\pi \over 3}
\ee
the path $\gamma_1$ can be deformed into a path of steepest descent through the saddle point at $u^{\rm L}$ (for $\kappa=0, \pi/3$, the steepest descent paths are shown in 
\figref{thimbles}.) We therefore have 
\be
\label{oneint}
{\rm Ai}(x)=I_{\gamma_L^\kappa}, \qquad |\kappa| < {2\pi \over 3}, 
\ee
which leads to the asymptotics, 
\be
\label{haione}
{\rm Ai}(x) \sim Z_{\rm Ai}(x), \qquad  |\kappa| < {2\pi \over 3}.
\ee
When 
\be
|\kappa|=2\pi/3 
\ee
the steepest descent path coming from the saddle at $u^{\rm L}$ runs right into the other saddle point. At this angle we have 
\be
\label{eqim}
{\rm Im}(S_{\kappa}(u^L))={\rm Im}(S_{\kappa}(u^R)).
\ee
Values of $\kappa$ for which this happens are called {\it Stokes lines}. This is the place where the second saddle might start contributing to the integral. In fact, for 
\be
{2\pi \over 3}<| \kappa|<\pi
\ee
the contour $\gamma_1$ gets deformed into a steepest descent path passing through $u^L$ {\it together} with a steepest descent path 
passing through $u^R$, and 
\be
\label{stokesints}
{\rm Ai}(x)=I_{\gamma_L^\kappa}+ I_{\gamma_R^\kappa}. 
\ee
However, in this range of $\kappa$ the saddle $u^R$ gives an exponentially suppressed contribution to the asymptotics. In classical 
asymptotic analysis, subleading exponentials are not taken into account, and the asymptotics is still given by the contribution from $u^L$:  
\be
{\rm Ai}(x) \sim Z_{\rm Ai}(x), \qquad  |\kappa| < \pi.
\ee
However, when $x<0$, both saddles have the same real part
\be
{\rm Re}(S_{\kappa}(u^L))={\rm Re}(S_{\kappa}(u^R)).
\ee
A line where this occurs is called an {\it anti-Stokes line}. Therefore, both saddles contribute to the asymptotics, which is then given by a linear combination of the two trans-series 
$Z_{\rm Ai}$ and $Z_{\rm Bi}$ (the precise combination can be obtained by a more detailed analysis, see \cite{miller}). One finally obtains an {\it oscillatory} asymptotics:
\be
\label{oscila}
{\rm Ai}(x)\sim {|x|^{-1/4} \over {\sqrt{\pi}}} \cos \Bigl( {2\over 3} |x|^{3/2} -{\pi \over 4}\Bigr), \qquad x<0, \, \, |x| \rightarrow \infty.
\ee
The fact that different asymptotic formulae hold on different directions for the same analytic function is called the {\it Stokes phenomenon}. 
From the point of view of saddle-point analysis, 
what is happening is that the saddle point which appeared on the Stokes line, at $\kappa=2\pi/3$, is no longer subdominant at $\kappa=\pi$, and it has to be included in the 
asymptotics.

\begin{figure}[!ht]
\leavevmode
\begin{center}
\includegraphics[height=6.5cm]{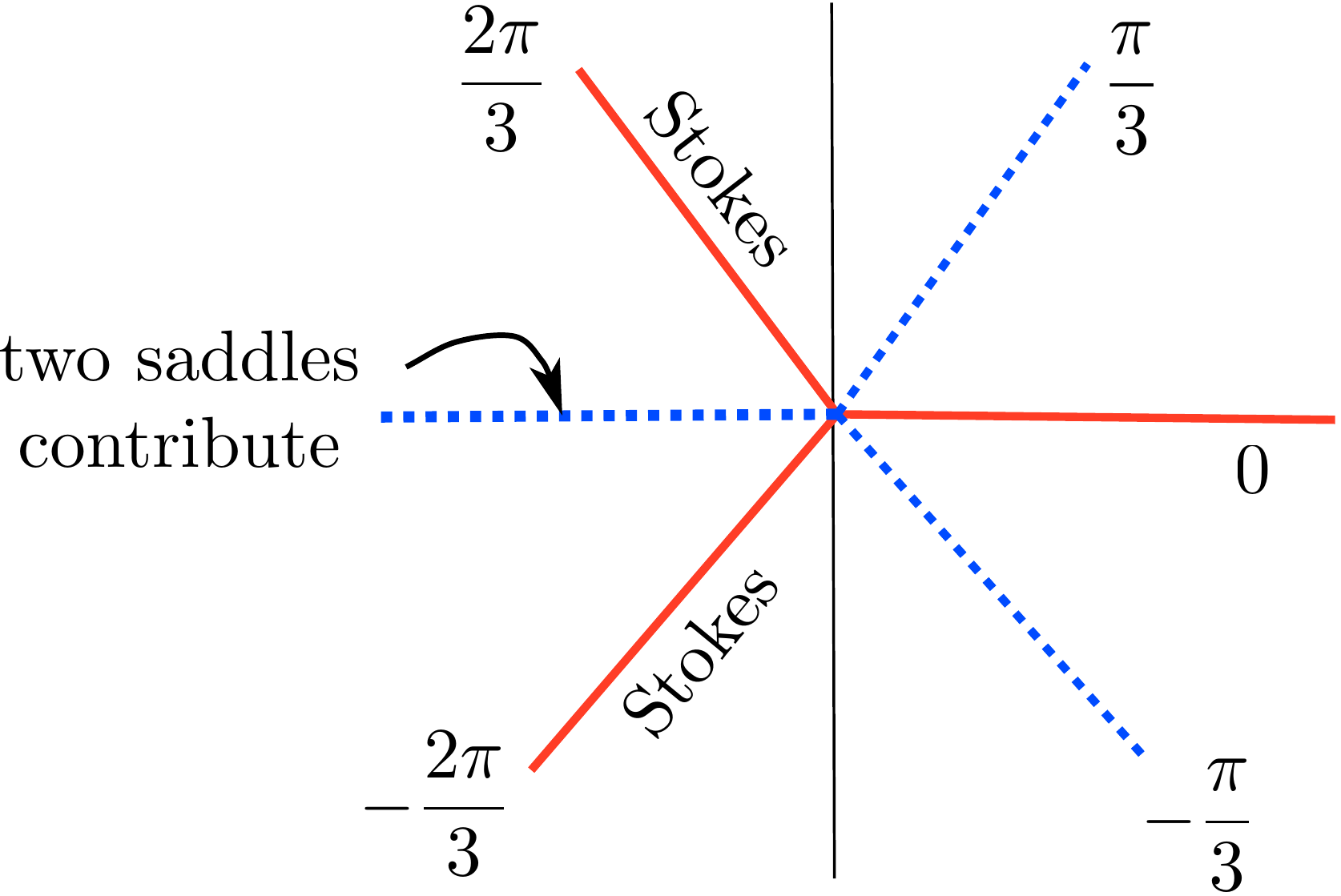}
\end{center}
\caption{Saddle-point analysis of the Airy function ${\rm Ai}(x)$. Full lines (in red) are Stokes lines, while 
dashed lines (in blue) are anti-Stokes lines. On the Stokes lines $\kappa =\pm 2\pi/3$, a second saddle appears in the 
integration contour. This saddle is subdominant when $2\pi/3\le |\kappa|<\pi$ and does not contribute to classical asymptotics. However, at $\kappa=\pi$, the saddle 
is not subdominant anymore and leads to an oscillatory asymptotics.}
\label{airystokes}
\end{figure}

The saddle-point analysis of the Airy integral is summarized in \figref{airystokes}. There are Stokes lines at
\be
\text{Stokes}: \qquad {\rm arg}(x)=0, \qquad {\rm arg}(x)=\pm {2 \pi \over 3},
\ee
and anti--Stokes lines at 
\be
\text{anti--Stokes}: \qquad {\rm arg}(x)=\pi, \quad \pm {\pi \over 3}. 
\ee

In light of the example of the Airy function, we can now understand the meaning of the formal trans-series solutions to ODEs. ODEs have ``true" solutions which are, generically, meromorphic functions on the complex plane. The classical asymptotics of these solutions is given by particular trans-series solutions, i.e. by linear 
combinations of formal solutions to the ODE. Due to the Stokes phenomenon, this combination changes as we change the angular sector where we study the asymptotics. Therefore, 
formal trans-series solutions are the ``building blocks" for the asymptotics of actual solutions.

In the context of systems of ODEs, Stokes and anti--Stokes lines are defined as follows. 
Let us consider our system in prepared form (\ref{matrices}). The directions where an exponential is purely oscillatory, i.e. 
\be
{\rm Re}(\lambda_i z)=0 
\ee
are called {\it anti-Stokes lines}. Along these directions, terms which used to be exponentially suppressed become of the same order than the leading term. This leads to an oscillatory asymptotic behavior. The directions where 
\be
\label{antistokes}
{\rm Im}(\lambda_i z)=0, 
\ee
are called {\it Stokes lines}. These are the directions where subleading exponentials start contributing to the asymptotics. If we consider the Airy equation in prepared form, the eigenvalues are $\pm 2/3$. In terms of the variable $z$ defined in (\ref{xairy}), the Stokes lines are at ${\rm arg}(z)=0, \pi$, while the anti--Stokes lines occur at 
${\rm arg}(z)=\pm \pi/2$, which, when translated to the variable $x$, give the structure found above. 

\begin{figure}[!ht]
\leavevmode
\begin{center}
\includegraphics[height=3.5cm]{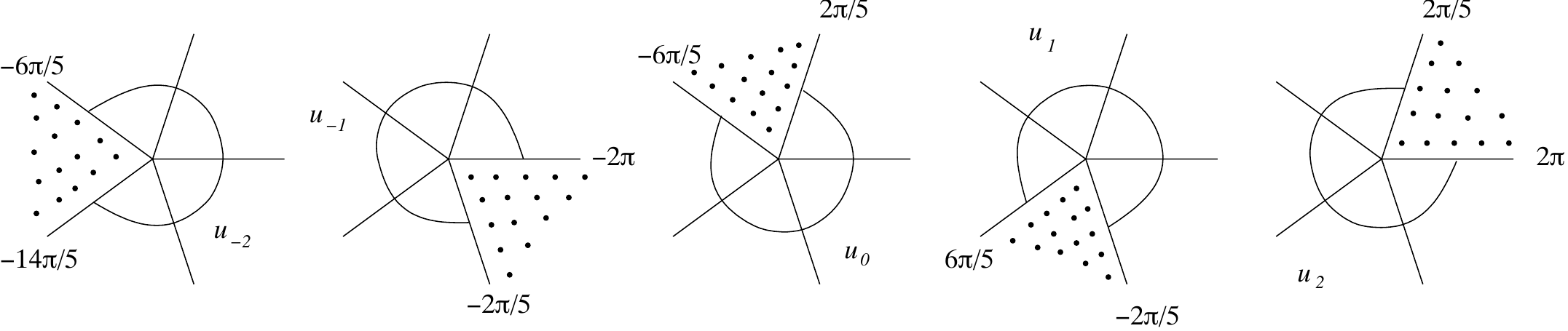}
\end{center}
\caption{The sectors of the $\kappa$-complex plane of opening $8\pi/5$ where the tritronqu\'ee 
solutions $u_{-2}$, $u_{-1}$, $u_0$, $u_1$, $u_2$ are 
represented by the formal series $u^{(0)}(\kappa)$. The dots in the remaining sector of opening $2\pi/5$ represent the infinite number of poles that the tritronqu\'ee 
solutions have there.}
\label{tritronquee}
\end{figure}

\begin{example} {\it Painlev\'e I and its tritronqu\'ee solutions}. 
The meaning of the formal power series solution to PI (\ref{asymp1}) can be clarified by looking at actual solutions of the Painlev\'e I ODE. 
It can be shown  (see for example \cite{painlevebook} and \cite{kapaev}) that there 
exist five different genuine meromorphic solutions of (\ref{p1}) with the 
asymptotic power expansion (\ref{asymp1}) as $z\to\infty$ in one of the sectors of the 
$z$-complex plane of opening $8\pi/5$, see \figref{tritronquee}:
\be
\label{u0_gen}
\ba
u_0(z) &\sim  u^{(0)}(z), \quad 
\arg z\in\bigl(-\tfrac{6\pi}{5},\tfrac{2\pi}{5}\bigr),
\\
u_k(z) &= \re^{-\ri\frac{8\pi}{5}k}u_0\bigl(\re^{-\ri\frac{4\pi}{5}k}z\bigr)
\sim u^{(0)}(z), \quad
\arg z\in\bigl(-\tfrac{6\pi}{5}+\tfrac{4\pi}{5}k,\tfrac{2\pi}{5}
+\tfrac{4\pi}{5}k\bigr), \quad k=1, \cdots, 5. 
\ea
\ee
Along the remaining sector of opening $2\pi/5$, the asymptotics involves elliptic functions and the solutions have an infinite number of poles. 
\end{example}

\subsection{Beyond classical asymptotics: Borel resummation}

So far we have discussed formal power series and trans-series. These formal power series give asymptotic approximations of well-defined functions 
which in the case of ODEs are their ``true" solutions. Our next question is: to which extent can we recover the original, ``non-perturbative" solution, from its 
asymptotic representation? Since asymptotic series are divergent, the answer is not obvious.

\begin{figure}[!ht]
\leavevmode
\begin{center}
\includegraphics[height=7cm]{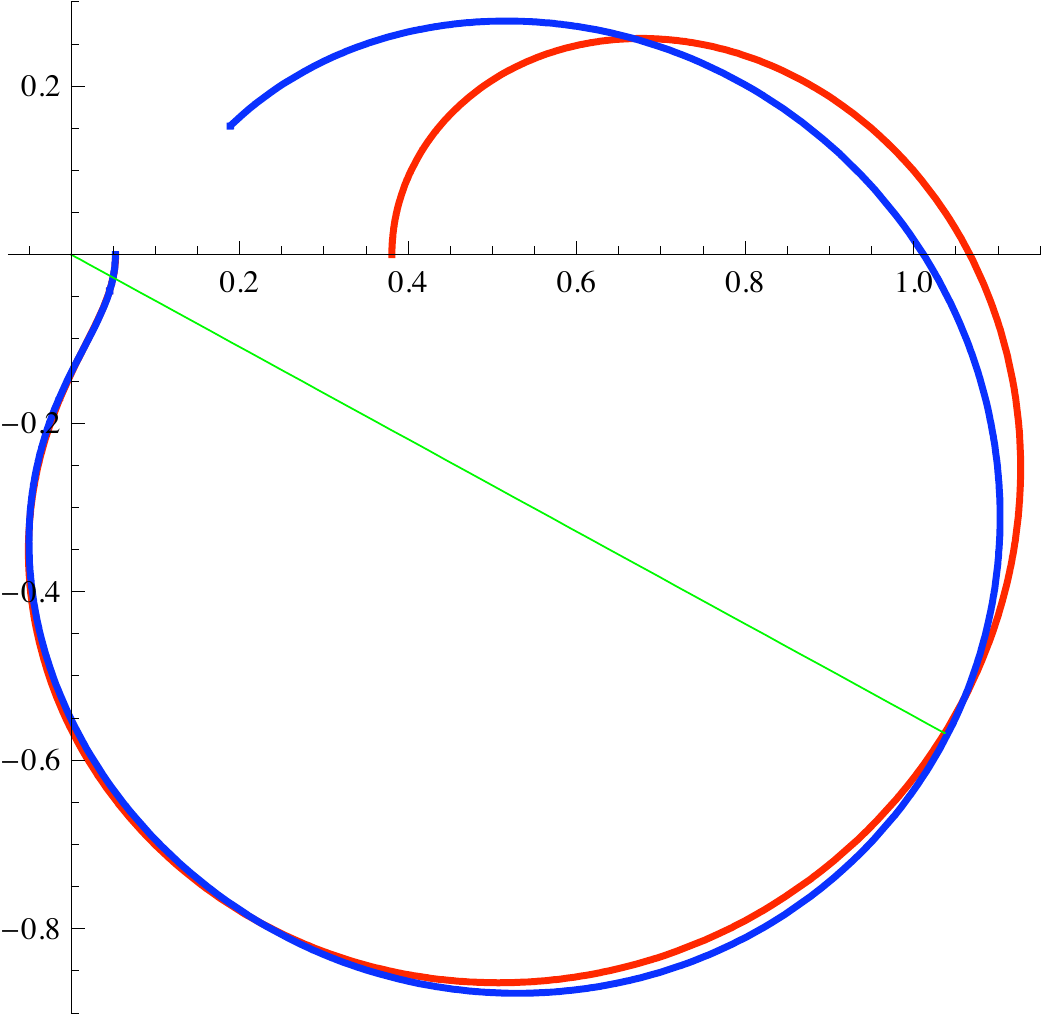}
\end{center}
\caption{Parametrized plot of the real and imaginary parts of the exact Airy function (red) compared to the optimal truncation of the asymptotic approximation (blue). The thin green line signals the angle $\kappa =2 \pi/3$.}
\label{seca}
\end{figure}

In fact, various answers have been proposed to this question. The more traditional answer is to use the optimal truncation procedure discussed in section \ref{smallcorr}. This gives a reasonable approximation to the original function in some regions of the complex plane, but it typically becomes a bad one in other regions. A nice illustration is provided again by the Airy function. Let us 
run the following numerical experiment, proposed by Berry in \cite{berry}. In optimal truncation, we approximate
\be
\label{optimalapp}
{\rm Ai}(x) \approx {1\over 2 {\sqrt {\pi}} x^{1/4}} \re^{ -2/3 x^{3\over 2}} \sum_{n=0}^{N^*} 
a_n x^{-3n/2},
\ee
where
\be
N^* = \Bigl[ 4/3 |x|^{3\over 2}\Bigr], 
\ee
and $[ \, ]$ denotes the integer part. Let us now plot the parametrized curve
\be
({\rm Re}({\rm Ai}(|x|\re^{\ri \kappa})), {\rm Im}({\rm Ai}(|x| \re^{\ri \kappa}))), \qquad 0< \kappa < \pi,
\ee
in the complex plane, for fixed $|x|=1.7171$, and let us compare it with the corresponding curve computed by using 
the r.h.s. of (\ref{optimalapp}). The result is shown in \figref{seca}. 
The approximation is quite good in the region 
\be
0<{\rm arg}(x) < {2\pi \over 3}
\ee
and it becomes worst and worst as we approach ${\rm arg}(x)=\pi$. The reason is simple: we are missing an exponentially small correction! This correction is born on the 
Stokes line ${\rm arg}(x) = {2\pi \over 3}$ and becomes more and more important as we approach the anti-Stokes line, where it is of the same order than the term we are keeping. It is clear that, in order to reproduce the full function, we must find a way to incorporate these exponentially small corrections. Notice as well that, in optimal truncation, only a finite number of terms in the asymptotic expansion are actually used, and the remaining terms cannot be used to improve the estimate. 

The most powerful way to go beyond optimal truncation and solve the above problems is probably the technique of {\it Borel resummation}. Let 
\be
\label{seriesone}
\varphi(z)=\sum_{n\ge 0} {a_n \over z^{n}}, 
\ee
be a factorially divergent series with $a_n \sim n!$. Its {\it Borel transfom} is defined by 
\be
\label{boreltrans}
\widehat \varphi (\zeta)=\sum_{n\ge 1} a_{n} {\zeta ^{n-1} \over (n-1)!}.
\ee
This series defines typically a function which is analytic in a neighboorhood of the origin. If the resulting function can be analytically continued to a neigbourhood 
of the positive real axis, in such a way that the Laplace transform 
\be
\label{borelr}
\int_0^{\infty} \re^{-z \zeta} \widehat \varphi (\zeta) \, \rd \zeta 
\ee
converges in some region of the $z$-plane, then the series $\varphi(z)$ is said to be {\it Borel summable} in that 
region. In that case,
\be
s(\varphi)(z) =a_0 +\int_0^{\infty} \re^{-z \zeta} \widehat \varphi (\zeta) \, \rd \zeta. 
\ee
defines a function whose asymptotics 
coincides with the original, divergent series $\varphi(z)$, and $s(\varphi)(z)$ is called the {\it Borel sum} of $\varphi(z)$. 

\begin{remark} There are other, equivalent definitions of Borel transform in the literature. In most of the physics literature (like for example \cite{caliceti-report}) 
the Borel transform of the series (\ref{seriesone}) is defined as 
\be
B_\varphi(\zeta)= \sum_{n \ge 0} {a_n \over n!} \zeta^n
\ee
and we have the relationship 
\be
\widehat \varphi (\zeta)= {\rd B_\varphi(\zeta) \over \rd \zeta}.
\ee
\end{remark}

\begin{remark} Sometimes we want to perform the Borel resummation along an arbitrary direction in the complex plane, specified by an angle $\theta$. 
It is then useful to introduce the generalized Borel 
resummation 
\be
\label{thetaresiduum}
s_\theta (\varphi)(z) =a_0 +\int_0^{\re^{\ri \theta} \infty} \re^{-z \zeta} \widehat \varphi (\zeta) \, \rd \zeta. 
\ee
\end{remark}

\begin{figure}[!ht]
\leavevmode
\begin{center}
\includegraphics[height=4cm]{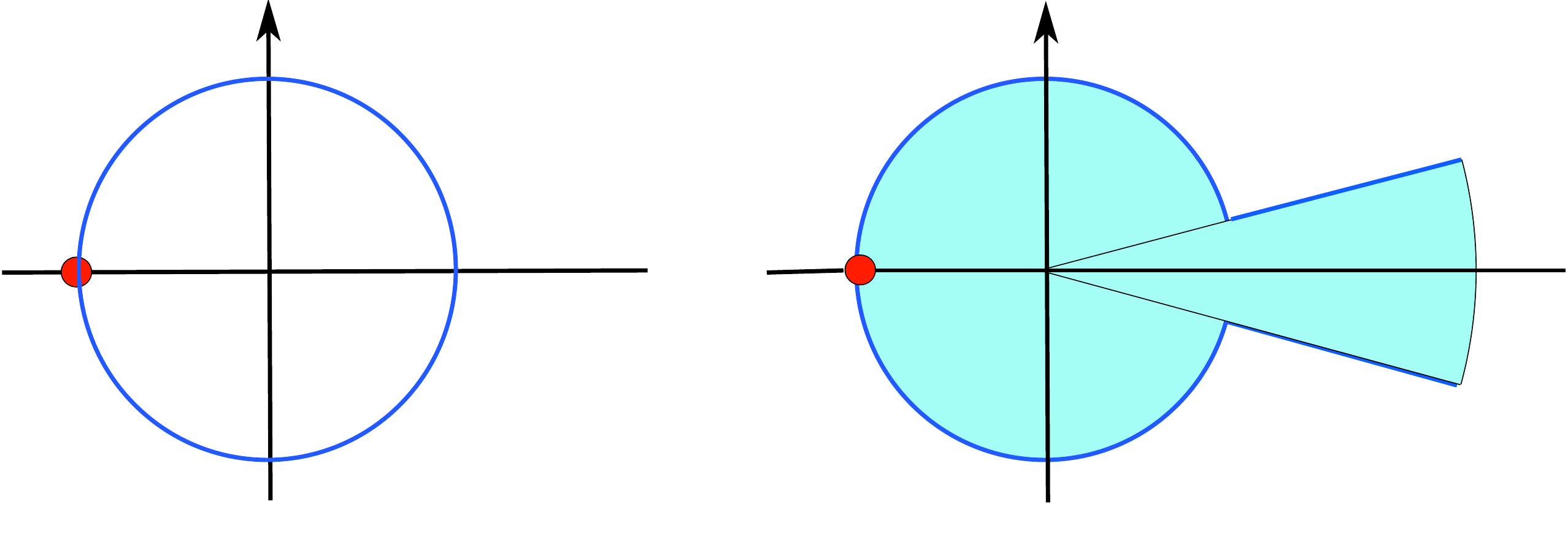}
\end{center}
\caption{The Borel transform defines an analytic function in a neighbourhood of $\zeta=0$, of radius $\rho=A$. There is a singularity on the circle $|\zeta|=A$, shown here as a red point on the negative real axis. If we can analytically 
continue this function to a neighbourhood of the positive real axis, and its Laplace transform exists, we say that the series is 
Borel summable.}
\label{borelan}
\end{figure}

A crucial issue in the analysis of Borel resummation is the location of the singularities of $\widehat \varphi(\zeta)$. It is easy to see that, if 
\be
a_n \sim A^{-n} n! ,
\ee
then $\widehat \varphi (\zeta)$ is analytic in an open neighborhood of radius $A$ around $\zeta=0$. There is a singularity at $z=A$, which can be a pole or a branch point. If the singularity is not on the positive real axis, the integral (\ref{borelr}) defining the Borel resummation is typically well defined and reconstructs the original function, see \figref{borelan}. 

\begin{example} {\it Borel resummation and Euler's equation}. Let us consider Euler's equation (\ref{eulereq}) with $A=-1$. Then, the formal solution is the asymptotic series
\be
\varphi (z) = \sum_{n\ge0} {(-1)^n n!\over z^{n+1}}.
\ee
Its Borel transform is
\be
\label{boreuler}
\widehat \varphi(\zeta)= \sum_{n=1}^{\infty}  (-1)^{n-1} \zeta^{n-1} ={1\over 1+\zeta}. 
\ee
Since $\widehat \varphi(\zeta)$ has no singularities on the positive real axis, we can define the Borel resummation
\be
s \left(\varphi\right)(z)=\int_0^{\infty} \re^{-z \zeta} {\rd \zeta \over 1+ \zeta}, 
\ee
which defines an analytic function in the region
\be
{\rm Re}\, z>0
\ee
and reconstructs a true solution to the original differential equation. 
\end{example}

\begin{example} {\it Borel resummation and the $c=1$ string}. Let us consider the following asymptotic series,
\be
\label{B-ser}
\varphi(z)=\sum_{n\ge 1} {B_{n+2} \over n (n+2)} z^{-n}, 
\ee
where $B_n$ are the Bernoulli numbers. Since $B_{2k}=0$ for $k\ge 1$, only even powers of $z$ appear. This series appears often in string theory. It gives the 
genus expansion of the $c=1$ string at the self-dual radius (see for example \cite{klebanov}), and it also appears in the asymptotic $1/N$ expansion of the 
partition function of the Gaussian matrix model (see for example \cite{mmleshouches}). To see that the series is asymptotic, notice that 
\be
\label{bernoulli-as}
B_n =- 2 (-1)^{n/2} (2 \pi)^{-n} n! \zeta(n), \qquad n\ge 2.
\ee
The zeta function behaves at infinity as, 
\be
\zeta(n)\approx 1, \qquad n\rightarrow \infty, 
\ee
up to exponentially small corrections in $n$, so the series (\ref{B-ser}) is alternating and factorially divergent. Its Borel transform can be computed explicitly, 
\be
\widehat \varphi(\zeta)=\sum_{n\ge 1} {B_{n+2} \over  (n+2) n!} \zeta^{n}={1\over \zeta^2} -{1\over 12} -{1\over 4} {\rm csch}^2\left({\zeta\over 2}\right). 
\ee
It has singularities along the imaginary axis, at the points $\zeta=2 \pi m \ri$, $m=\pm 1, \pm 2, \cdots$. It has no singularities along the positive real axis, 
and the integral of the Borel transform 
\be
s(\varphi)(z)=\int_0^\infty \rd \zeta \, \re^{-z \zeta} \left[ {1\over \zeta^2} -{1\over 12} -{1\over 4} {\rm csch}^2\left({\zeta\over 2}\right)\right], 
\ee
gives the Borel resummation of the original series for $z>0$ (see \cite{ps} for more details on this and related examples). 
\end{example}

\begin{example} {\it Borel resummation and the Airy function}. In the case of the Airy function we can proceed as follows. Let us define
\be
\varphi_{\rm Ai}(z)= \sum_{n=0}^{\infty} {a_n \over z^n}, 
\ee 
where the coefficients $a_n$ are given in (\ref{anairy}). Its Borel transform can be explicitly computed as a hypergeometric function 
\be
\label{airybor}
\widehat \varphi_{\rm Ai}(\zeta) = -{5 \over 48} \, _2F_1 \Bigl( {7\over 6}, {11\over 6};2; -{3\zeta\over 4} \Bigr)
\ee
and it has a branch point singularity at $\zeta= -4/3$. The Borel resummation,  
\be
s  \left( \varphi_{\rm Ai}\right)(z)=a_0+ \int_0^{ \infty} \widehat \varphi_{\rm Ai}(\zeta) \re^{-z\zeta} \rd \zeta
\ee
is well-defined if 
\be
\label{rez}
{\rm Re}(z) \ge 0
\ee
and it reconstructs the {\it full} Airy function in the region (\ref{rez}) (see for example \cite{delabaere}). We then have, 
\be
{\rm Ai}(x)= {1\over 2x^{1/4} {\sqrt{\pi}}} \re^{-2 x^{3/2}/3} s  \left( \varphi_{\rm Ai}\right)(z), \qquad x=z^{2/3}. 
\ee
 In terms of ${\rm arg}(x)$, this representation is valid as long as
\be
\bigl| {\rm arg}\, x \bigr|<{\pi \over 3}.
\ee
\end{example} 

\begin{figure}[!ht]
\leavevmode
\begin{center}
\includegraphics[height=4cm]{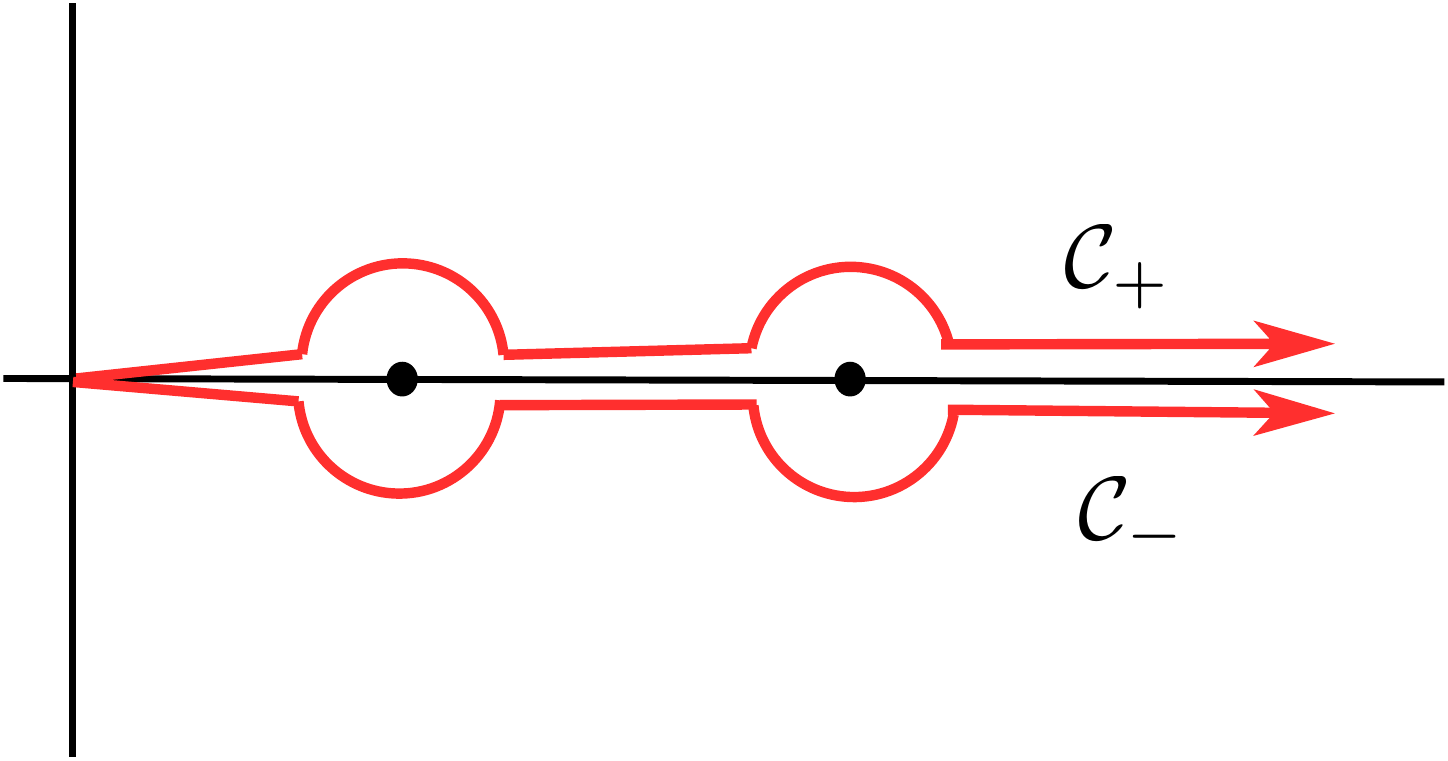}
\end{center}
\caption{The paths ${\cal C}_{\pm}$ avoiding the singularities of the Borel transform from above (respectively, below).}
\label{lateralfig}
\end{figure} 

Very often one encounters asymptotic series whose Borel transform has singularities on the positive real axis. 
In this case one needs some prescription in the integral (\ref{borelr}) to avoid the singularities. 
A standard procedure to do this is to consider {\it lateral Borel resummations}. 
Let ${\cal C}_{\theta \pm}$ be a path going from $0$ to $+\re^{\ri \theta} \infty$ and avoiding 
the singularities of $\widehat \varphi(\zeta)$ on the direction with angle $\theta$ from above (resp. below). For $\theta=0$, we will simply write the paths as $\CC_\pm$, and in this case they have the form shown in \figref{lateralfig}. The {\it lateral Borel resummations} are then defined as,
\be
\label{lateralres}
s_{  \theta \pm} \left(\varphi\right)(z)= a_0+ \int_{\CC_{ \theta\pm}} \re^{-z \zeta} \widehat \varphi(\zeta) \rd \zeta, 
\ee
provided the integral is convergent. Notice that, even if the original series has real coefficients and we choose a direction along the real axis, 
since the lateral Borel resummations are computed by integrals 
along paths in the complex plane, they lead in general to complex-valued functions. 

\begin{example} {\it Lateral Borel resummations and Euler's equation}. Let us consider the Borel transform 
(\ref{boreuler}) along the negative real axis. Since there is a singularity at $\zeta=-1$, we are forced to perform lateral Borel 
resummations:
\be
s_{\pi \pm} \left(\varphi\right)(z)=\int_{\CC_{\pi \pm} }  \re^{-z \zeta} {\rd \zeta \over 1+ \zeta}. 
\ee
These integrals give two different solutions to the original differential equation for $z<0$. Their difference can be computed as a residue, 
\be
\label{eulerjump}
s_{  \pi +} \left(\varphi\right)(z)-s_{  \pi-} \left(\varphi\right)(z)=\int_{\CC_{\pi+} -\CC_{\pi-}}{\re^{-z \zeta} \over 1+ \zeta} \rd \zeta=
2 \pi \ri \, {\rm Res}_{\zeta=-1}  {\re^{-z \zeta} \over 1+ \zeta} =2\pi \ri \, \re^z, 
\ee
and it is {\it exponentially small} along the negative real axis. In fact, it is the exponentially small term appearing in the trans-series solution (\ref{eulerts}). See \cite{ss} for 
a detailed discussion of this example. 
\end{example}

\begin{example} {\it Lateral Borel resummations for the Airy function}. Let us consider the Borel transform 
(\ref{airybor}) along the negative real axis in the $z$ variable. In terms of the $x$ variable, this corresponds to ${\rm arg}(x)=2 \pi/3$, i.e. to a 
Stokes line. Along this direction, the coefficients of the series defining the Airy function (\ref{fullai}) are no longer alternating, and this is the standard indication that the series is not Borel summable along such a direction. Indeed, since there is a singularity at $\zeta=-4/3$, we have to consider lateral Borel 
resummations:
\be
s_{\pi \pm} \left(\varphi_{\rm Ai} \right)(z)=a_0 + \int_{\CC_{\pi \pm} } \widehat \varphi_{\rm Ai}(\zeta) \re^{-z \zeta}  \rd \zeta.
\ee
We will now show that (see, for example, \cite{delabaere}) 
\be
\label{airyjump}
s_{\pi +} \left(\varphi_{\rm Ai} \right)(z)-s_{\pi -} \left(\varphi_{\rm Ai} \right)(z)=-\ri \, \re^{4z/3} s_{\pi} \left( \varphi_{\rm Bi}\right) (z)
\ee
where
\be
\varphi_{\rm Bi}(z)=\sum_{n=0}^{\infty} (-1)^n {a_n \over z^n}=\varphi_{\rm Ai}(-z).
\ee
Notice that, since $\widehat \varphi_{\rm Bi}(\zeta)$ is analytic on the negative real axis, the above Borel 
resummation is well-defined. The derivation of (\ref{airyjump}) goes as follows. By integrating by parts, changing 
variables $\zeta=-x$, and using the explicit result (\ref{airybor}), we can write down the l.h.s. of (\ref{airyjump}) as 
\be
- z\int_0^\infty \left( ~_2F_1 \left( {1\over 6}, {5\over 6}, 1; {3x \over 4} + \ri \epsilon  \right)- 
~_2F_1 \left( {1\over 6}, {5\over 6}, 1; {3x \over 4} - \ri \epsilon  \right) \right) \re^{z x } \rd x. 
\ee
The discontinuity of the hypergeometric function vanishes unless $x\ge 4/3$, and for this range of $x$ it is given by 
\be
~_2F_1 \left( {1\over 6}, {5\over 6}, 1; {3x \over 4} + \ri \epsilon  \right)- 
~_2F_1 \left( {1\over 6}, {5\over 6}, 1; {3x \over 4} - \ri \epsilon  \right)= \ri  ~_2F_1 \left( {1\over 6}, {5\over 6}, 1; 1-{3x \over 4} \right). 
\ee
After using this result and changing variables $x= u+4/3$, we find
\be
s_{\pi +} \left(\varphi_{\rm Ai} \right)(z)-s_{\pi -} \left(\varphi_{\rm Ai} \right)(z)=-\ri z \re^{4z/3} \int_0^\infty ~_2F_1 \left( {1\over 6}, {5\over 6}, 1; -{3u \over 4} \right)\re^{z u} \rd u. 
\ee
Integrating by parts again, we obtain (\ref{airyjump}). As in the previous example, the difference 
of lateral resummations is given by a trans-series 
solution. We will see in a moment that this is the way in which the Stokes phenomenon manifests itself in the context of 
Borel resummations. 
\end{example}

Let us now see in general how to apply Borel resummation to the study of ODEs. If ${\boldsymbol \varphi}_0(z)$ is a formal solution to an 
ODE, its Borel resummations (or lateral Borel resummations) 
will provide functions which solve the ODE and have the asymptotic behavior given by ${\boldsymbol \varphi}_0(z)$. 
However, we can add exponentially small corrections 
to this solution without changing the asymptotics. In general, the multi-parameter family 
\be
\label{ysols}
{\boldsymbol \varphi}_{\pm} (z; {\bf C}_\pm) =s_{\pm} \left( {\boldsymbol \varphi}_0\right)(z) +\sum_{{\bf k}} {\bf C}_{\pm}^{\bf k} z^{-{\bf k} \cdot {\boldsymbol \beta}} \re^{-{\boldsymbol \lambda} \cdot {\bf  k} z} s_{\pm} \left( {\boldsymbol \varphi}_k\right)(z), 
\ee
which is obtained by doing lateral Borel resummations on the formal trans-series solution (\ref{trans}), 
is a good solution for sufficiently large $|z|$ which asymptotes to ${\boldsymbol \varphi}_0(z)$, provided the non-vanishing terms in the trans-series are such that 
\be
{\rm Re}({\boldsymbol \lambda}  \cdot {\bf  k} z)>0.
\ee
The reciprocal is true: any solution to the ODE can be represented by such a 
Borel-resummed trans-series for an appropriate choice of ${\bf C}$'s. This is one of the main consequences of \'Ecalle's theory of resurgence, see for example \cite{ss,approche,costin} for 
detailed statements and proofs. We see that the main advantage of the Borel resummed version of asymptotic analysis is that {\it we can make sense of small exponentials}, i.e.
 we can incorporate the information encoded in the 
trans-series in a systematic way. This is not the case in classical asymptotics.

What is the interpretation of the Stokes phenomenon in the context of Borel resummation? 
In classical asymptotics, Stokes lines indicate the appearance of small exponentials, as we have seen in the analysis of the Airy function: we pass from (\ref{oneint}) to (\ref{stokesints}). However, this jump is only noticed in the classical theory when we reach the anti-Stokes line. 
Once we use Borel resummation, we can give a ``post-classical" version of the Stokes phenomenon. Let us consider a Stokes direction, 
which we take for simplicity to be ${\rm arg}(z)=0$, corresponding to the eigenvalue $A=\lambda_1>0$. Along this direction, there are two families of solutions ${\boldsymbol \varphi}_{\pm} (z; {\bf C}_\pm)$, obtained by lateral summations from below and from above. By uniqueness we 
should expect these two solutions to be related. Indeed, we have the relation
\be
\label{stokesjump}
{\boldsymbol \varphi}_{+} (z; {\bf C}) ={\boldsymbol \varphi}_{-} (z; {\bf C}+ {\bf S}),
\ee
where 
\be
{\bf S}=(S_1,0, \cdots, 0)
\ee
is called the {\it Stokes parameter} associated to the Stokes line ${\rm arg}(z)=0$, and it is an imaginary number when the coefficients of the 
trans-series are real. At leading order in the exponentially small parameter $\exp(-Az)$ we find
\be
\label{diffstokes}
{\boldsymbol \varphi}_{0;+}(z)-{\boldsymbol \varphi}_{0;-}(z) \approx S_1 z^{-\beta_1} \re^{-A z} {\boldsymbol \varphi}_{(1,0,\cdots, 0);-}(z).
\ee
The relation (\ref{stokesjump}) is the Borel-resummed version of the Stokes phenomenon. It says that the 
coefficients of the trans-series solutions have {\it discontinuous jumps} along the Stokes direction. The results (\ref{eulerjump}) and (\ref{airyjump}) are two particular examples of this general result, in which the Stokes line is the negative real axis for the $z$ variable. The Stokes parameters in 
these examples are $S=2 \pi \ri$ for the Euler equation, and $S=-\ri$ 
for the Airy function. For a pedagogical explanation of (\ref{stokesjump}) in the case of ODEs, see \cite{ss,approche}. The generalization to systems 
of ODEs is presented in \cite{costin}. 

The relationship (\ref{diffstokes}) has a very nice interpretation in terms of Borel transforms, which we will make explicit 
for simplicity in the case of a first order ODE. In this case the vector ${\boldsymbol \beta}$ has one single entry which we take to be 
$\beta=0$. Let us denote the perturbative solution by $\varphi_0 (z)$, with the form (\ref{seriesone}), and the first trans-series by 
\be
\label{phione}
\varphi_1(z)=\sum_{n\ge 0} { \varphi_{1,n}  \over z^n}.
\ee
The l.h.s. of (\ref{diffstokes}) can be written as 
\be
\label{intaround}
\int_\gamma \rd \zeta \, \re^{-z \zeta} \widehat \varphi_0 (\zeta), 
\ee
where $\gamma$ is homotopic to the contour $\CC_+ -\CC_-$, and encircles the singularities of the Borel transform. Then, (\ref{diffstokes}) tells us that the structure of 
$\widehat \varphi(\zeta)$ around the singularity at $\zeta=A$ is of the form 
\be
\label{resurg}
\widehat \varphi_0(A +\xi) = -S_1 \left(  {\varphi_{1,0} \over 2 \pi \ri\xi} + {\log(\xi) \over 2 \pi \ri} \widehat \varphi_1(\xi)+ {\rm holomorphic} \right). 
\ee
where $\widehat \varphi_1(\zeta)$ is the Borel transform of (\ref{phione}). This is easy to check: the integral (\ref{intaround}) can be evaluated by using (\ref{resurg}). The first 
term in (\ref{resurg}) gives the residue at the pole $\zeta=A$, namely 
\be
S_1 \re^{-A z} \varphi_{1,0},
\ee
while the second term in (\ref{resurg}) gives the integral of the discontinuity of the log, namely
\be
S_1 \re^{-A z} \int_0^{\infty} \rd \xi \, \re^{-z\xi} \widehat \varphi_1(\xi) = S_1 \re^{-A z} \sum_{n\ge 1} { \varphi_{1,n}  \over z^n}.
\ee
We then reconstruct the l.h.s. of (\ref{diffstokes}). 

The relationship (\ref{diffstokes}) is then telling us that the singular behavior of the Borel transform of the perturbative series is related to the first 
instanton trans-series. This shows that ``perturbative" and ``non-perturbative" phenomena are intimately related, at least in this example. In the next subsection 
we will see that this relationship 
has a powerful corollary, namely it gives an asymptotic formula for the large order behavior of the coefficients of the perturbative series. 
Equation (\ref{resurg}) is an example of the {\it resurgence relations} discovered by Jean \'Ecalle, and it forms the basis of the so-called ``alien calculus" of his theory (see \cite{cnp} for an introduction). In fact, (\ref{resurg}) is just the tip of the iceberg, since relations of this type connect all the formal power series in the trans-series, and not only the perturbative series and the first instanton. 

\begin{example} In the case of the Airy function, one can use the relation (\ref{airyjump}) to derive the following formula for the Airy function along the negative real axis \cite{delabaere}
\be
{\rm Ai}(x)= {1\over 2x^{1/4} {\sqrt{\pi}}} \biggl\{  \re^{-2 x^{3/2}/3} \, s_{\pi/2} \left( \varphi_{\rm Ai}\right)(z)+
\ri \,  \re^{2 x^{3/2}/3}\, s_{\pi/2}  \left(\varphi_{\rm Bi}\right)(z)\biggr\}, \qquad {\rm Re}(x)<0,
\ee
where $x=z^{2/3}$. This is the {\it exact} version of the oscillatory asymptotics given in (\ref{oscila}) and (\ref{stokesints}). 
\end{example}

\begin{example} \label{p2hm} {\it Painlev\'e II and the Hastings--McLeod solution}. As an example of how to reconstruct a ``true" function from the asymptotics in the case of non-linear ODEs, 
including exponentially small 
corrections, we consider the example of Painlev\'e II, whose formal structure was discussed in Example \ref{exp2}. This equation has a Stokes line at ${\rm arg}(\kappa)=0$. The PII equation has a solution, called the {\it Hastings--McLeod solution}, $u_{\rm HM}(\kappa)$ which is uniquely characterized by the following properties:
\begin{enumerate}
\item It is real for real $\kappa$. 
\item As $\kappa \rightarrow \infty$ it asymptotes
\be
 u_{\rm HM}(\kappa) \sim \kappa^{1/2}.
 \ee
 \item As $\kappa \rightarrow -\infty$ it asymptotes 
 \be
 \label{minusas}
 u(\kappa) \sim \re^{- 2 {\sqrt{2}} (-\kappa)^{3/2}/3}. 
 \ee
 \end{enumerate}
  
 \begin{figure}[!ht]
\leavevmode
\begin{center}
\includegraphics[height=4cm]{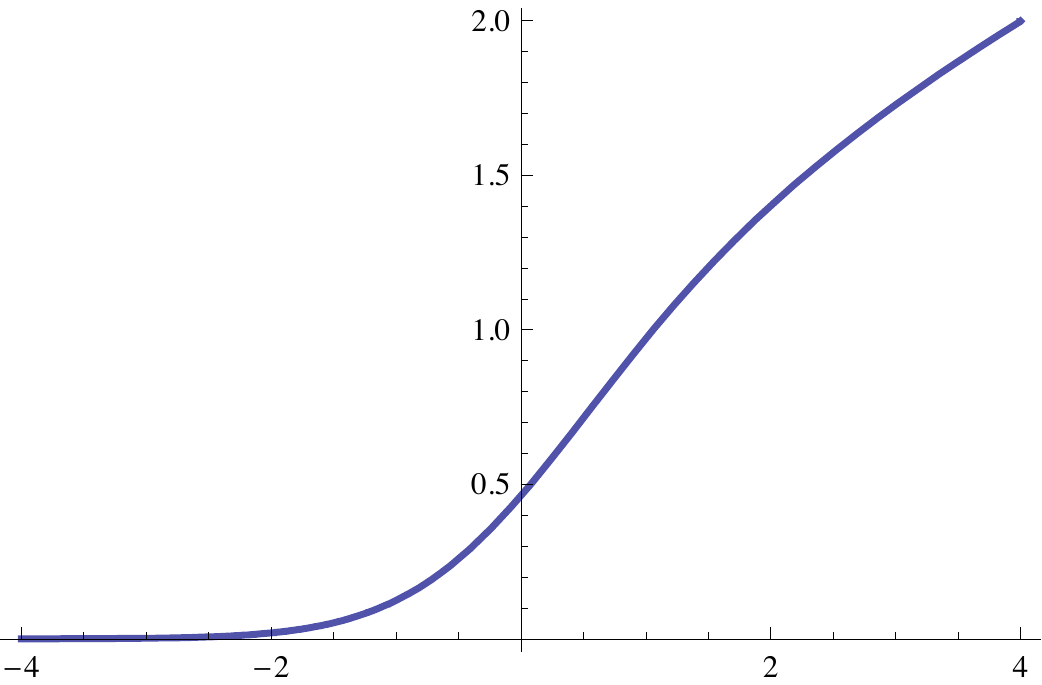}
\end{center}
\caption{The Hastings--McLeod solution to Painlev\'e II. As $\kappa \rightarrow \infty$ it asymptotes $\sqrt{\kappa}$, while as $\kappa \rightarrow -\infty$ it 
decays exponentially as indicated in (\ref{minusas}).}
\label{HMfig}
\end{figure} 

This solution of PII plays a crucial r\^ole in the Tracy--Widom law \cite{tw} and in the double-scaling limit of unitary matrix models \cite{cdm}. The Hastings--McLeod solution is shown in \figref{HMfig}. Since it is a solution to Painlev\'e II, at sufficiently large $\kappa$ one must be able to express it as the Borel resummation of the trans-series (\ref{p2trans}).The Borel transform has a singularity at $\zeta=A =4/3$, therefore we have to use lateral resummations along the positive real axis. The relation (\ref{stokesjump}) reads in this case, 
 \be
 \label{p2jump}
 u_+(\kappa;C)= u_-(\kappa;C+ S)
 \ee
 where the subscripts $\pm$ refer to the two lateral resummations of the full trans-series solution (\ref{p2trans}), and 
 \be
 S=-{\ri \over {\sqrt{2 \pi}}}
 \ee
 is the Stokes parameter. By using results on the non-linear Stokes phenomenon for the Painlev\'e II equation \cite{painlevebook} one can show that \cite{mmnp}
 \be
 u_{\rm HM}(\kappa)=u_+(\kappa;-S/2).
 \ee
Notice that this solution is real. Complex conjugation changes the sign of $S$ and also exchanges the integrals along the contours 
 $\CC_+$ and $\CC_-$, so we have 
 \be
 u_+(\kappa;-S/2)^*=u_-(\kappa;S/2)=u_+(\kappa; -S/2)
 \ee
 where we used (\ref{p2jump}). The connection to Borel resummed formal solutions gives the correct ``semiclassical" content of the Hastings--McLeod solution, and one easily shows that
 \be
  u_+(\kappa;-S/2)={1\over 2} (u^{(0)}_+ +u^{(0)}_-) - {1\over 8} S^2 (u^{(2)}_+ +u^{(2)}_-) 
 +\cdots
 \ee
\end{example}

The equation (\ref{stokesjump}) is very important conceptually, when interpreted from a physical point of view. 
As we will show in detail in these lectures, the formal power series ${\boldsymbol \varphi}_0$ 
has often the interpretation of a ``perturbative" series, while the trans-series ${\boldsymbol \varphi}_{\bf k}$ 
have the interpretation of non-perturbative effects. The coefficients ${\bf C}$ give the 
strength of these effects. But one clear implication of (\ref{stokesjump}) is that this strenght is not well-defined unless we give a prescription to perform the Borel resummation 
of the power series appearing in the formal solution. Indeed, different prescriptions lead to different coefficients, and in particular the upper and lower lateral resummations differ by a shift involving the Stokes parameters. Notice that there is a ``compensation" effect, in the sense that we can change simultaneously the resummation prescription and the strenght of the non-perturbative effects so that the solution is unchanged. This is indeed the content of (\ref{stokesjump}). This phenomenon was noticed in the literature on renormalons in gauge theories  (see for example \cite{grunberg} for a clear presentation), as well as in the study of instantons in QM \cite{zjj}, and is called in those contexts the {\it cancellation of non-perturbative ambiguities}. 

\subsection{Non-perturbative effects and large order behavior}

An important consequence of the Borel resummation technique is a relationship between the asymptotic behavior of the coefficients of a perturbative series, and the first instanton or trans-series solution. This type of relationships were anticipated in \cite{dingle} and in the work on the large 
order behavior of quantum perturbation theory \cite{largeorder}, and it will 
be important in the following lectures. We have already seen in the examples of the Euler equation and the Airy function 
that the ``action" appearing in the trans-series controls the large order behavior of the 
perturbative coefficients. This is a general phenomenon. We will now give a heuristic derivation of an asymptotic formula for the coefficients, in the case of a first order ODE with 
$\beta_1=0$ (see \cite{ck,gama} for details and rigorous proofs). We will also write down various generalizations of the result. 

We will denote the ``perturbative" series and its Borel transform by 
\be
\varphi_0(z)=\sum_{n \ge 0} {a_{n+1} \over z^{n+1}}, \qquad \widehat \varphi_0(\zeta)= \sum_{n\ge0} {a_{n+1} \over n!} \zeta^n.
\ee
Therefore
\be
{a_{n+1} \over n!} ={1\over 2\pi \ri} \oint_{\CC_0}  \rd \zeta {\widehat \varphi_0(\zeta) \over \zeta^{n+1}}.
\ee
where $\CC_0$ is a contour around the origin. We know that $\widehat \varphi_0(\zeta)$ has a singularity at $\zeta=A$, and we can deform the contour $\CC_0$ so as to enclose this 
singularity. In general, there are singularities at other points with $|\zeta|>A$, but they give exponentially small 
corrections as compared to what we are computing. We can then write 
\be
\label{per-np-odes}
{a_{n+1} \over n!} \sim {1\over 2\pi \ri} \int_{\zeta=A}  \rd \zeta {\widehat \varphi_0(\zeta) \over \zeta^{n+1}}, 
\ee
up to exponentially small corrections. We know the singularity structure of $\widehat \varphi_0(\zeta)$ near $\zeta=A$ 
thanks to the result (\ref{resurg}): there is a pole with residue $-S_1 \varphi_{1,0}/2\pi \ri $, and a logarithmic discontinuity. Changing 
variables $\zeta=A+\xi$, we obtain
\be
{a_{n+1} \over n!} \sim  {S_1\varphi_{1,0}\over 2\pi \ri} \oint_{0} {1 \over \zeta (A+\zeta)^{n+1}}{\rd \zeta \over 2\pi \ri}+ {S_1 \over 2\pi \ri} \sum_{k\ge 1} {\varphi_{1,k} \over (k-1)!} \int_0^{\infty} {\zeta^{k-1} \over (A + \zeta)^{n+1} }\rd \zeta.
\ee
Since
\be
\int_0^{\infty} {\zeta^{k-1} \over (A+ \zeta)^{n+1} }\rd \zeta =A^{k-n-1} {\Gamma(k) \Gamma(n+1-k) \over \Gamma(n+1)} 
\ee
we have the following result for the all-order asymptotics of the coefficients:
\be
\label{simpleas}
a_n \sim {S_1 \over 2\pi \ri} \Gamma(n)A^{-n}  \sum_{k\ge0} {\varphi_{1,k} A^k \over \prod_{i=1}^k (n-i)}.
\ee
This is a beautiful result. It implies that the leading asymptotics of the original (``perturbative") series is encoded in the first trans-series (or ``one-instanton") solution (there are further, exponentially suppressed contributions associated to the higher singularities). Conversely, 
all the information about the formal one-instanton series is encoded in the asymptotics of the perturbative series. Notice that the leading and next-to-leading order of the asymptotics is given by 
\be
a_n \sim n! A^{-n}
\ee
as in the examples discussed above. However, (\ref{simpleas}) contains much more information. In particular, the Stokes parameter $S_1$ plays a crucial r\^ole in the 
asymptotics, and in fact (\ref{simpleas}) provides a method to determine this parameter numerically in cases in which it is not known analytically. 

Before considering generalizations of the above equation, let us work out in some 
detail an interesting example which will illustrate most of the considerations of this first lecture. 

\begin{example} {\it A Riccati equation}. Following \cite{ss,bonet}, let us consider the ODE 
\be
\label{sric}
{\rd \varphi \over \rd z}= \varphi -{1\over z} (b^{-}+b ^+ \varphi^2).
\ee
Here, $b^\pm$ are real constans, and we assume that
\be
\beta^2=-b^- b^+
\ee
is positive. The ODE (\ref{sric}) generalizes the Euler equation (\ref{eulereq}), which is obtained (for $A=-1$) when $b^+=0$. Equation (\ref{sric}) is a particular case of the 
so-called Riccati equation, which is characterized by being quadratic in the unknown function. It is easy to see that there is a formal solution of (\ref{sric}) around $z=\infty$ which is given by 
\be
\varphi_0(z)= b^- \sum_{n\ge 1} {a_n \over z^n}, 
\ee
and the coefficients $a_k$ are obtained from the non-linear recursion
\be
a_{k+1}=-k a_k -\beta^2\sum_{\ell=1}^{k-1} a_\ell a_{k-\ell}, \quad a_0=1.
\ee
Explicitly, we find
\be
\label{particularcase}
\varphi_0(z)=b^- \Bigl\{ {1\over z} -{1\over z^2} +{2-\beta^2 \over z^3} -{6 -5\beta^2 \over z^4}+ \cdots \Bigr\}.
\ee

The Riccati ODE (\ref{sric}) has a full trans-series solution of the form 
\be
\varphi(z,C)= \sum_{n\ge 0} C^n \re^{nz} \varphi_n(z), 
\ee
therefore there is a series of ``multi-instantons" with action $nA$, and $A=-1$. The Stokes parameter for this ODE has been computed exactly in \cite{bonet}, and it is given by 
\be
S_1=-2 \pi \ri b_- \sigma(\beta), \qquad \sigma(\beta)={\sin(\pi \beta) \over \pi \beta}. 
\ee
Using (\ref{simpleas}), we obtain the following asymptotics for the coefficients $a_n$:
\be
a_n \sim  \sigma(\beta) (-1)^{n-1} (n-1)!\left( 1+ \CO\left({1\over n}\right) \right).
\ee
This is easy to test numerically. To do that, one simply studies the asymptotic behaviour of the sequence
\be
\label{snseq}
s_n=(-1)^{n-1} {a_n \over (n-1)!}
\ee
which should converge towards the constant value $\sigma(\beta)$. However, the resulting convergence is quite slow. One can accelerate it by using 
{\it Richardson transformations}. Let us assume that a sequence 
$s_n$ has the asymptotics
\be
\label{sequ}
s_n \sim \sum_{k=0}^{\infty} {c_k \over n^k}
\ee
for $n$ large. Its $N$-th Richardson transform can be defined recursively by 
\be
\label{richardson}
\ba
s^{(0)}_n&=s_n, \\
s^{(N)}_n&=s_{n+1}^{(N-1)} + {n \over N}(s_{n+1}^{(N-1)} -s_n^{(N-1)}), 
\quad N\ge 1.
\ea
\ee
The effect of this transformation is to remove subleading tails in 
(\ref{sequ}), and
\be
s^{(N)}_n \sim c_0 + \CO\Bigl({1\over n^{N+1}}\Bigr). 
\ee
 The values $s^{(N)}_n$ give numerical approximations to $c_0$, 
and these 
approximations become better as $N$, $n$ increase. Once a numerical 
approximation to $c_0$ has been obtained, the value of $c_1$ can be estimated 
by considering the sequence $n(s_n-c_0)$, and so on. In \figref{riccatif} we plot the original sequence $s_n$ and its first and second Richardson transforms, for $\beta=1/3$. The convergence towards 
\be
\sigma\left( {1\over 3}\right)\approx 0.826993...
\ee
is quite fast, and $s^{(2)}_{250}$ gives an approximate value for this constant which agrees with the right value up to the seventh decimal digit. 
 \begin{figure}[!ht]
\leavevmode
\begin{center}
\includegraphics[height=5cm]{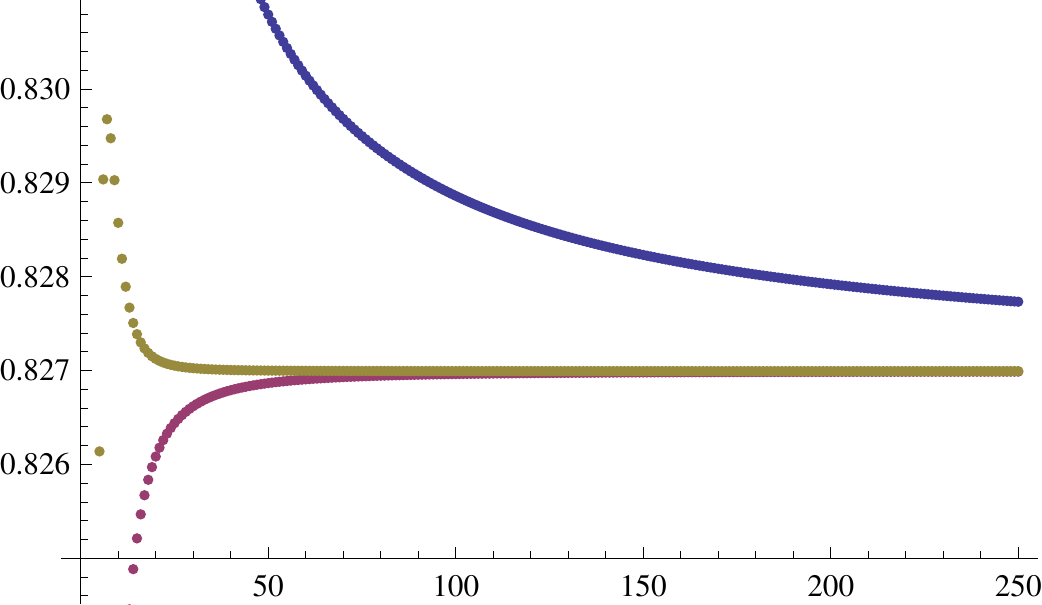}
\end{center}
\caption{The sequence $s_n$ in (\ref{snseq}) (top) and its first and second Richardson transforms (bottom).}
\label{riccatif}
\end{figure} 

\end{example}

The asymptotic result (\ref{simpleas}) has many generalizations. For example, when $\beta\not=0$, and the trans-series have the structure 
\be
\label{trans-beta}
\varphi_k (z)=z^{-\beta k} \sum_{n=0}^{\infty} \varphi_{k,n} z^{-n},
\ee
a simple generalization of the above argument shows that 
\be
\label{largeanbeta}
a_n \sim {S_1 \over 2\pi \ri} \Gamma(n-\beta)  A^{-n+\beta}  \sum_{k\ge0} {\varphi_{1,k} A^k \over \prod_{i=1}^k (n-\beta-i)}.
\ee

One can also generalize the argument to higher order ODEs. In that case, as seen in (\ref{ysols}), there are various possible instanton actions for the trans-series. The asymptotics is governed by the 
trans-series which correspond to the smallest actions, in absolute value. If there are instanton actions which have the same smallest absolute value but different phases, the asymptotics is obtained by adding their different contributions, see \cite{ck} for a precise mathematical statement. For example, for the coefficients $u_{n,0}$ of the perturbative solution (\ref{asymp1}) of PI, one has the following asymptotics \cite{kapaev,jk}
\be
u_{0,n} 
\sim A^{-2n+{1\over 2}}  \Gamma\Bigl(2n-{1\over 2} \Bigr)\, 
{S_1 \over \pi \ri} \biggl\{1 + \sum_{l=1}^{\infty} {u_{1,l} A^{l} 
\over \prod_{k=1}^{l} (2n-1/2 -k)} \biggr\},
\ee
which comes from two instanton actions $\pm A$. In this expression, $u_{1,l}$ are the coefficients of the $1$-instanton 
series $\epsilon^{(1)}(\kappa)$ appearing in (\ref{el}), $A$ is the instanton action (\ref{aaction}), and
\be
S_1 = -\ri {3^{1\over 4} \over 2{\sqrt{\pi}}}
\ee
is a Stokes parameter. 

Since the trans-series solutions are also formal, asymptotic series, one can ask what is the asymptotic behavior of their coefficients. 
In the same way that the asymptotics of the perturbative coefficients is encoded in the first trans-series, it turns out that the asymptotics of the coefficients of a given trans-series is encoded in higher trans-series solutions. The study of this question requires the full machinery of resurgent analysis, see 
\cite{gama,gikm,asv} for various results along this direction. 

\subsection{Lessons}

We can now summarize some of the results that can be learned from the study of asymptotic series appearing in ODEs. All of these results will have a counterpart when we look at 
instanton corrections in quantum theories and string theories, and therefore they constitute a sort of ``r\^ole model" for their study. 

\begin{enumerate}

\item The standard perturbative contribution (${\bf y}_0(x)$, in the context of ODEs) is a {\it factorially divergent, asymptotic series}. At the formal level, one 
can also obtain trans-series solutions. These will correspond to perturbation theory around instanton solutions, i.e. to {\it non-perturbative effects}. 

\item The weights of the instanton solutions are {\it a priori} undetermined. Therefore, the general trans-series solution gives a multi-parameter family of 
formal solutions. This is the {\it non-perturbative ambiguity}.

\item By Borel resummation, the family of formal solutions becomes a family of ``true" solutions. Therefore, we obtain 
a family of {\it non-perturbative completions}. 
If we have a {\it non-perturbative definition}  
of the theory we can fix the non-perturbative ambiguity by choosing the values of the parameters that reproduce the non-perturbative definition. Along directions where the series is Borel summable, the original solution is typically reconstructed by Borel resummation of the perturbative series solely.

\item Along Stokes lines there are different prescriptions for resummation, due to singularities in the Borel transform. Their difference is purely non-perturbative and defines the Stokes parameter. This is the ``resurgent" version of the Stokes phenomenon. The reconstruction of the non-perturbative solution involves in a crucial way the Borel-resummed non-perturbative effects, as we showed in Example \ref{p2hm} for the 
Hastings--McLeod solution of PII .

\item The large order behavior of the perturbative expansion along a Stokes line encodes the action of the instanton, the Stokes parameter, and the coefficients of the 
first instanton correction. This relation is extremely powerful, since it says that the large order behavior of perturbation theory knows about non-perturbative corrections. 
In cases where there is no clear technique (or even framework!) to address the computation of non-perturbative effects, the large order behavior of the perturbative series gives an important hint about 
their structure. 

\end{enumerate}

\sectiono{Non-perturbative effects in Quantum Mechanics and Quantum Field Theory}

\subsection{Trans-series in Quantum Mechanics}

Before discussing non-perturbative effects in QFT and matrix models, 
it is instructive to  first consider simple quantum-mechanical examples, where the analysis of non-perturbative effects can be made in detail. We will focus on the ground state energy of 
one-dimensional particles with Hamiltonian 
\be
H={p^2 \over 2} +V(q). 
\ee
We will assume that the potential $V(q)$ is of the form 
 \be
 \label{wint}
 V(q)={1 \over 2} q^2 + g V_{\rm int}(q),
 \ee
 where $g V_{\rm int}(q)$ is the interaction term. A typical example is the quantum anharmonic oscillator, where
\be
V(q) ={q^2 \over 2} + {g \over 4} q^4.
\ee

The ground 
state energy of a quantum mechanical system in a potential $V(q)$ can be computed in a variety of ways. The most elementary method is of course 
Rayleigh--Schr\"odinger perturbation theory,  and the resulting series has the form 
\be
\label{gsseries}
E= \sum_{n=0}^\infty a_n g^n,
\ee
where (in units where $\hbar=1$) 
\be
a_0={1\over 2}
\ee
is the ground state energy of the harmonic oscillator, and we have an infinite series of corrections due to the interaction term in (\ref{wint}). 
In order to make contact with QFT it is instructive to calculate the series (\ref{gsseries}) in terms of diagrams. 
To do this, we first notice that the ground state energy can be extracted from the small temperature behavior of the 
thermal partition function, 
\be
Z(\beta) = {\rm tr} \, \re^{-\beta H(\beta)}, 
\ee
as
\be
\label{grounde}
E= -\lim_{\beta \to \infty} {1\over \beta } \log \, Z(\beta). 
\ee
In the path integral formulation one has
\be
\label{pathintegral}
Z(\beta) =\int \, {\cal D}[q(t)]  \re^{-S(q)}, 
\ee
where $S(q)$ is the action of the Euclidean theory,
\be
S(q)=\int_{-\beta/2}^{\beta/2} \rd t \,  \biggl[ {1\over 2} (\dot q (t))^2 + V(q(t))\biggr], 
\ee
and the path integral is over periodic trajectories
\be
\label{periodic}
q(-\beta/2)=q(\beta/2).
\ee
The path integral defining $Z$ can be computed in standard Feynman perturbation theory  by expanding in $V_{\rm int}(q)$. We will actually work in the limit in which 
 $\beta \rightarrow \infty$, since in this limit many features are simpler, like for example the form of the propagator. In this limit, the free energy will be given by $\beta$ times a $\beta$-independent constant, as follows from (\ref{grounde}). In order to extract the ground state 
 energy we have to take into account the following facts:
 \begin{enumerate}
 
 \item Since we have to consider $F(\beta)=\log Z(\beta)$, only connected bubble diagrams contribute. 
 
 \item The standard Feynman rules in position space 
 will lead to $n$ integrations, where $n$ is the number of vertices in the diagram. One of these integrations just gives as an overall factor the ``volume" of spacetime, i.e. the factor $\beta$ that we just mentioned. Therefore, in order to extract $E(g)$ we can just perform $n-1$ integrations over $\IR$. 
 
 \end{enumerate}

\begin{figure}[!ht]
\leavevmode
\begin{center}
\includegraphics[height=5cm]{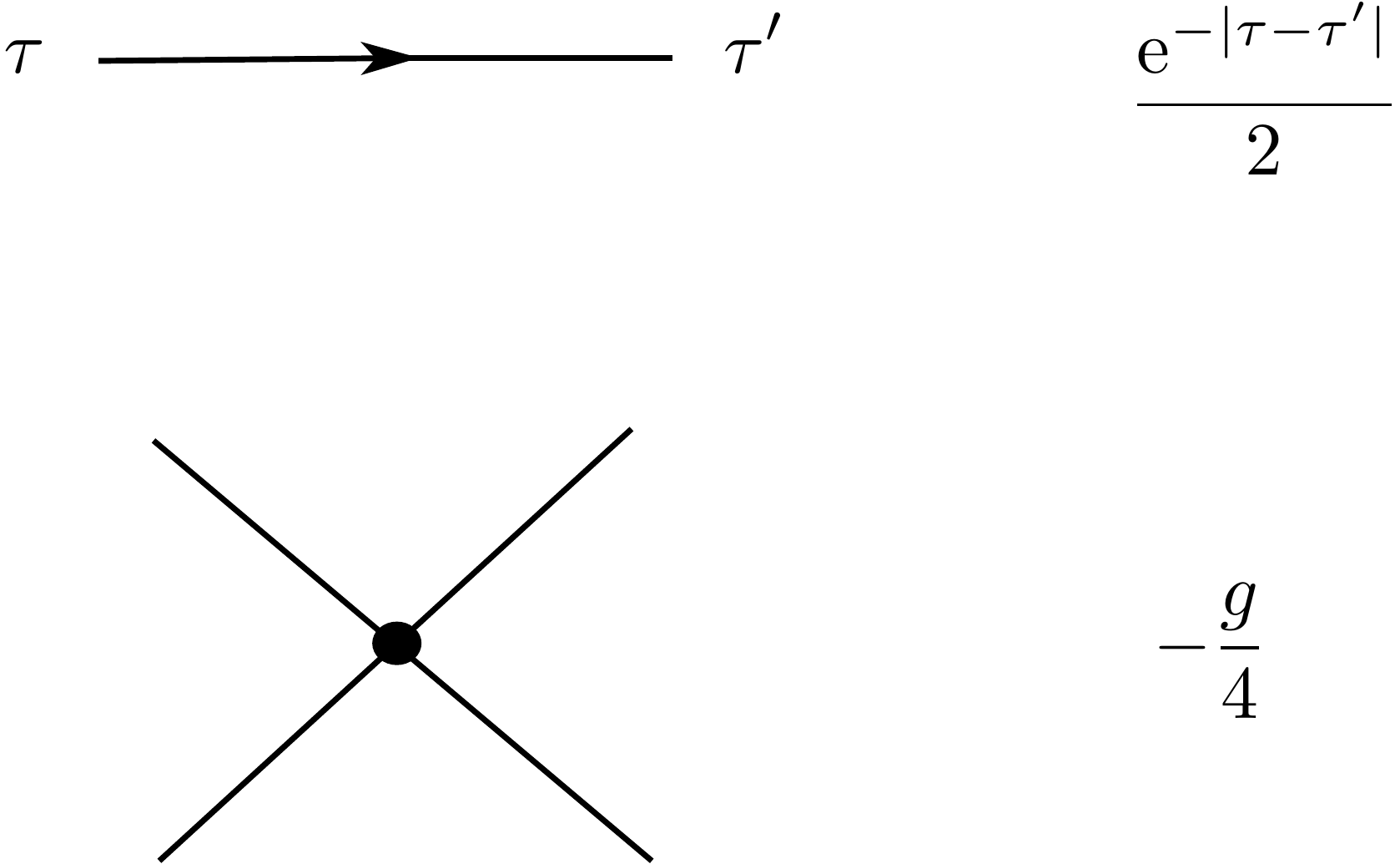}
\end{center}
\caption{Feynman rules for the quantum mechanical quartic oscillator.}
\label{qmf}
\end{figure}

\begin{figure}[!ht]
\leavevmode
\begin{center}
\includegraphics[height=8cm]{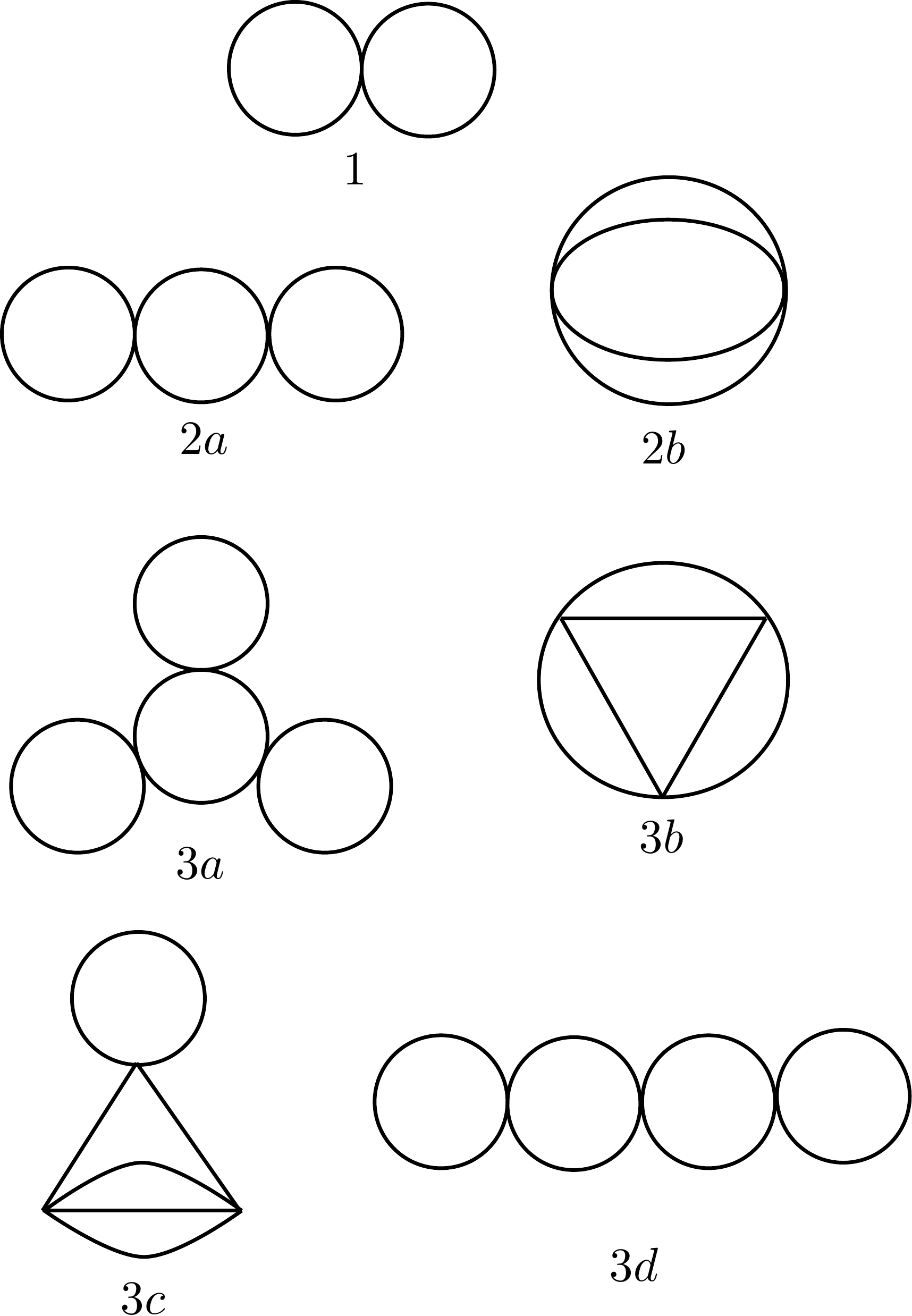}
\end{center}
\caption{Feynman diagrams contributing to the ground state energy of the 
quartic oscillator up to order $g^3$. }
\label{feynmanquartic}
\end{figure}

 For $\beta \rightarrow \infty$ the propagator of this one-dimensional field theory is simply
 \be
 \int {\rd p \over 2 \pi} {\re^{\ri p \tau}\over p^2 + 1}={\re^{- |\tau|} \over 2}. 
 \ee
 For a theory with a quartic interaction (i.e. the anharmonic quartic oscillator) 
 \be
 V_{\rm int}(q)={g \over 4} q^4
 \ee
 the Feynman rules are illustrated in \figref{qmf}.  One can use these rules to compute the perturbation series of the ground energy of 
 the quartic oscillator (see Appendix B of \cite{oldbw} for some additional details). We have, schematically, 
\be
a_n =\sum \left( {\text{connected vacuum bubbles}}\right).
\ee
For example, the diagrams contributing up to order $g^3$ are shown in 
 \figref{feynmanquartic}, and after performing the integrals over the propagators one finds,
 \be
 E={1\over 2} +{3\over 4} \Bigl( {g \over 4} \Bigr)  -{21\over 8} \Bigl( {g \over 4} \Bigr)^2 + {333 \over 16} \Bigl( {g \over 4} \Bigr)^3 +\CO(g^4), 
 \ee
which agrees with the result of standard Rayleigh--Schr\"odinger perturbation theory. 

A basic property of the above perturbative series is that the coefficients $a_n$ grow {\it factorially} when $n$ is large. 
Moreover, this behavior is due to the {\it factorial growth in the number of diagrams} (the Feynman integrals over products of propagators only grow exponentially). 
To see this, remember that $a_n$ can be computed as a sum over connected quartic graphs. The total number of connected graphs with $n$ quartic vertices is given by
\be
{1\over n!} \langle (x^4)^n \rangle ^{(c)},
\ee
where
\be
 \langle (x^4)^n \rangle={\int_{-\infty}^{\infty} \rd x \, \re^{-x^2/2} x^{4n} \over \int_{-\infty}^{\infty} \rd x \, \re^{-x^2/2}}
 \ee
 is the Gaussian average. By Wick's theorem, the average 
 counts all possible pairings among $n$ four-vertices. The superscript $(c)$ means that we take the connected part of the average. Since
 \be
 \label{simplewick}
  \langle x^{2k} \rangle=(2k-1)!! ={(2k)! \over 2^k k!}
  \ee
  we find 
  \be
  \label{x4corr}
  {1\over n!} \langle (x^4)^n \rangle = {(4n-1)!!\over n!}= {(4n)! \over 4^n n! (2n)!}.
  \ee
As $n \rightarrow \infty$ this behaves like
\be
4^{2n} n!,
\ee
i.e. there is a factorial growth in the number of disconnected diagrams. One could think that there might be a substantial reduction in this number 
when we consider connected diagrams, but a careful analysis \cite{bendercaswell} shows that this is not the case: at large $n$, the quotient of the number of 
connected and disconnected diagrams differs from $1$ only in $\CO(1/n)$ corrections. We conclude that there are $\sim n!$ diagrams that contribute 
to $a_n$. The resulting factorial behavior of the perturbative series of the quartic oscillator can be verified by a detailed consideration of Feynman diagrams \cite{bwstat} (see \cite{bender} for a review of these early developments). Therefore, we conclude that the perturbative series for the ground state energy is a formal, divergent power series. This series gives at best an asymptotic expansion of the true non-perturbative ground-state energy, defined in terms of the exact spectrum of the Schr\"odinger operator.

\begin{figure}[!ht]
\leavevmode
\begin{center}
\includegraphics[height=4cm]{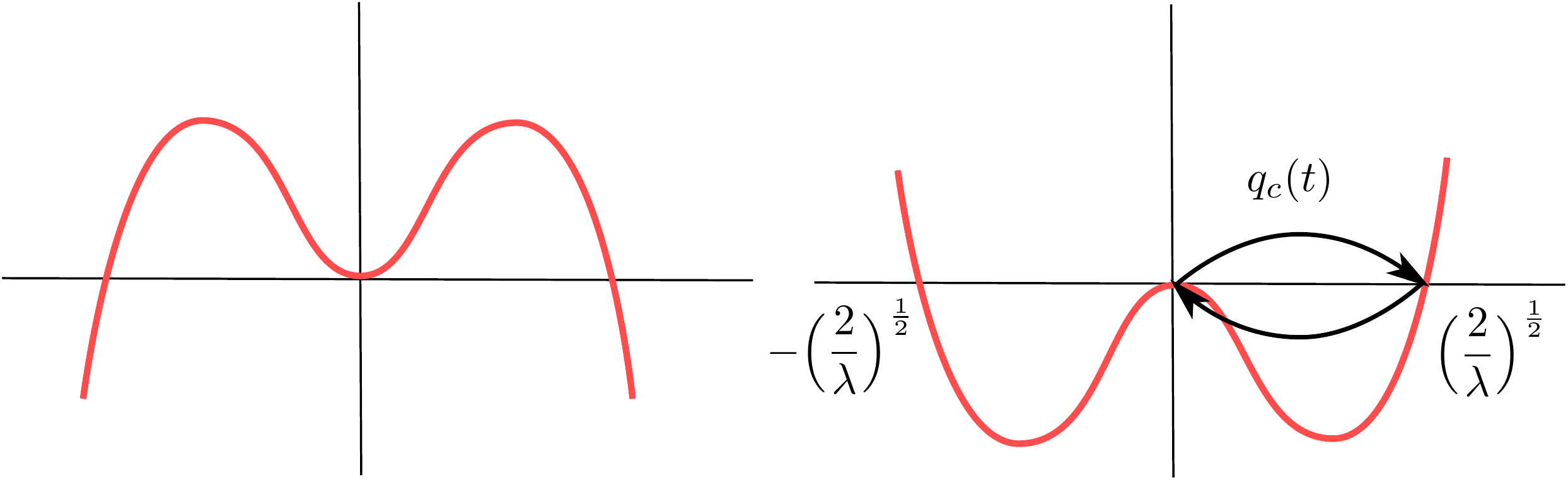}
\end{center}
\caption{The inverted potential relevant for instanton calculus in the quartic case. The 
instanton or bounce configuration $q_c(t)$ leaves the origin at $t=-\infty$, reaches the zero $(2/\lambda)^{1\over2}$ at $t=t_0$, 
and comes back to the origin at $t=\infty$. }
\label{qinst}
\end{figure}

We can now ask what is the analogue of the trans-series for this type of problems. As it is well-known (see for example the discussion in the textbook \cite{zj}), 
there are instanton contributions to the thermal partition function. For concreteness, let us consider again the quartic oscillator 
and let us suppose that the coupling constant is negative, i.e. $g=-\lambda$, with $\lambda>0$, 
so that we have a potential of the form 
shown in the left hand side of \figref{qinst}. In this case, the Euclidean action has non-trivial saddle-points. The EOM reads 
\be
\label{aneom}
-\ddot q (t) + q(t) -\lambda q^3(t) =0.
\ee
In the limit $\beta \rightarrow \infty$ one finds the following trajectory with $E=0$, 
\be
\label{qcsaddle}
 q_c(t) =\pm \Bigl( {2\over \lambda}\Bigr)^{1\over 2} {1\over \cosh (t-t_0)},
\ee
where $t_0$ is an integration constant or modulus of the solution. When $\beta \rightarrow \infty$, such a trajectory starts at the origin in the infinite past, 
reaches the zero of the potential $V(q)$ at $t=t_0$, and returns to the origin in the infinite future. 
As is well-known, the Euclidean action can be regarded as an action in Lagrangian mechanics with an ``inverted" potential $-V(q)$, 
and the non-trivial saddle-point described above is simply a trajectory of zero energy in this inverted potential.

The partition function around this non-trivial saddle can be computed at one-loop by 
using standard techniques, which we will not review in here (a very complete and updated discussion can be found in chapter 39 of \cite{zj}). 
It can be seen that this saddle-point is unstable: it has one, and exactly one, negative mode. This means that the instanton contribution is imaginary. 
A detailed analysis, which can be found in for example \cite{cs,zj}, shows that this one-instanton calculation determines the 
{\it discontinuity} of the partition function for negative values of the coupling:
\be
\label{reson}
{\rm disc}\, Z(-\lambda)=Z(-\lambda + \ri \epsilon)-Z(-\lambda -\ri \epsilon) =2 \ri \, 
 {\rm Im}\, Z(-\lambda).
\ee
This leads to a discontinuity in the ground-state energy, as a function of the coupling. In the case of the quartic oscillator one finds, at one loop, 
\be
\label{corrE}
{\rm disc}\, E(-\lambda)=2 \ri \, 
 {\rm Im}\, E(-\lambda)\approx {8 \ri \over  {\sqrt {2 \pi \lambda}}} \re^{-{A/\lambda}},
\ee
where
\be
\label{bounceaction}
A={4\over 3} 
\ee
is the action of the saddle (\ref{qcsaddle}) for $\lambda=1$. 
This imaginary correction to the energy has a clear physical interpretation: since for negative coupling the potential is unstable, a 
particle in its ground state will eventually tunnel. The width of the ground state energy
\be
\Gamma=2|{\rm Im} \, E|, 
\ee
is inversely proportional to the life-time of the ground state. 

The above calculation is just the one-loop approximation to the one-instanton sector. But if we consider multi-instanton expansions at all loops, we expect 
to find for the ground state energy a trans-series structure of the form
\be
E(g) =\sum_{\ell=0}^{\infty} C^{\ell} E^{(\ell)}(g), 
\ee
where $E^{(0)}$ is the asymptotic, perturbative series (\ref{gsseries}), and 
\be
E^{(\ell)}(g)\propto \re^{-\ell A/g}
\ee
are the $\ell$-instanton corrections, themselves asymptotic expansions in the coupling constant $z$. In the case of the quartic oscillator they have the structure 
\be
E^{(\ell)}(-z)=z^{\ell  \beta}   \re^{-\ell A/z}\sum_{n=0}^\infty a_{\ell, n} z^n, 
\ee
which is identical to (\ref{trans-beta}) (here the expansion is around $z=0$, while in (\ref{trans-beta}) we do the expansion around $z=\infty$). 
In other cases, like the double-well potential analyzed in \cite{zj-first,zj,zjj} they are more complicated and include terms of the form $z^n \log z$. 

The structure of the trans-series in QM suggests that the instanton action determines the positions of the singularities of the Borel 
transform, and that the large order behavior of the coefficients in the perturbative series (\ref{gsseries}) is controlled by the first instanton contribution. These expectations are indeed true, and one can show that much of the structure appearing in ODEs can be extended to the analysis of quantum-mechanical potentials in one dimension. In particular, the ``resurgent" structure of the formal trans-series calculating the energies of bound states in 
QM has been established, following the work of Voros \cite{voros}, in \cite{delabaerepham}.  

In the case of the quartic oscillator, the Borel transform of the formal power series (\ref{gsseries}) has a singularity at $\zeta=-4/3$. 
It is therefore Borel-summable along the positive real axis (i.e. for $g>0$), and its Borel resummation is indeed the exact ground-state energy, 
as defined by the Schr\"odinger operator. 
This was originally proved in \cite{ggs}, and a proof using the theory of resurgence can be found in \cite{delabaerepham}. 
To understand the large order behavior of the series (\ref{gsseries}) we have to take into account the presence of the instanton at negative 
$g=-\lambda<0$. In this case, one has to consider lateral resummations of $E(g)$ along the negative real axis. The discontinuity ${\rm disc}\, E(-\lambda)$ gives then the difference between lateral Borel resummations, and the result (\ref{corrE}) can be interpreted as the analogue of (\ref{diffstokes}) in the theory of ODEs: the asymptotic expansion of this difference is 
given (at leading order) by the first instanton correction to the energy, which has the general structure 
\be
s_+ (E)(-z)-s_-(E)(-z) \approx S_1 E^{(1)}(z),
\ee
where $S_1$ is a Stokes parameter. At one-loop we have the asymptotic result, 
\be
s_+ (E)(-z)-s_-(E)(-z) \approx {8 \ri \over  {\sqrt {2 \pi z}}} \re^{-{4\over 3 z}}.
\ee
Notice, in particular, that the coefficient of the one-loop calculation of the instanton partition function gives the Stokes parameter of the problem. 

\begin{figure}[!ht]
\leavevmode
\begin{center}
\includegraphics[height=5cm]{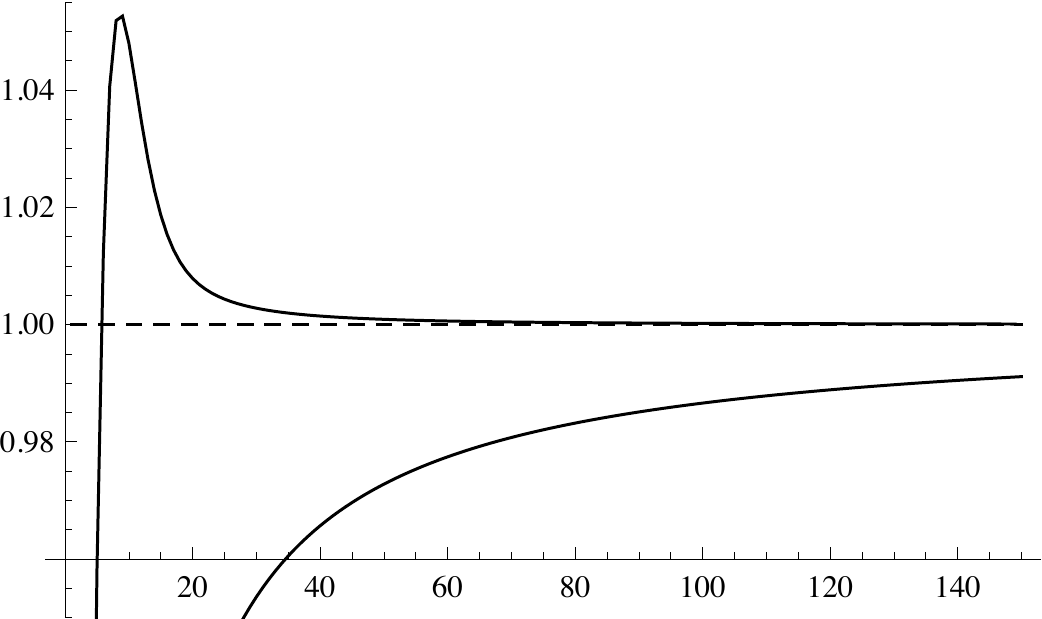}
\end{center}
\caption{The bottom line shows the joined points of the sequence $Q_n$, which is defined in (\ref{qn}) from the coefficients $a_n$ of the perturbative series of the ground state energy for the quartic oscillator. The top line is its first Richardson transform, with accelerated convergence to the expected value $1$. }
\label{rt-quartic}
\end{figure}

As in the theory of ODEs, one can use this result to derive the large order behavior of the coefficients $a_n$ in (\ref{gsseries}). Writing 
\be
\label{ol-qm}
S_1 E^{(1)}(-z)=\ri z^{ \beta}   \re^{-A/z}\sum_{n=0}^\infty c_n z^n,
\ee
we find the asymptotic growth
\be
\label{akgen}
a_n \sim { (-1)^{n+1}A^{-n+\beta} \over 2 \pi} \Gamma(n-\beta) \left\{ c_0 + \sum_{l=1}^\infty{ c_{l+1} A^l \over \prod_{m=1}^l (n-\beta-m)} \right\}. 
\ee
This formula is exactly like the one in (\ref{largeanbeta}), with the only difference of the extra factor $(-1)^{n+1}$ which is due to the fact that we do the perturbative expansion in the variable $g=-z$. Plugging in the concrete values of the quartic oscillator for the different quantities, i.e. 
\be
\beta=-{1\over 2}, \quad c_0=4 {\sqrt {2 \over \pi}}, \quad A=4/3.
\ee
we find for the large-order behavior
\be
\label{famousbw}
a_n \sim (-1)^{n+1}{ {\sqrt 6} \over \pi^{3/ 2}} \Bigl( {3 \over 4} \Bigr)^n \Gamma\Bigl(n +{1\over 2}\Bigr).
\ee
This can be tested against an explicit study of the behavior of the coefficients $a_n$ as $n$ grows large. These coefficients can be computed explicitly at for large values of $n$ 
by using a recursion relation found in \cite{oldbw}. In \figref{rt-quartic} we plot the quotient
\be
\label{qn}
Q_n= (-1)^{n+1}{\pi^{3/ 2}  \over  {\sqrt 6}}  \left( {3 \over 4} \right)^{-n} { a_n  \over  \Gamma\left(n +{1\over 2}\right)}
\ee
which should behave, at large $n$, as 
\be
Q_n =1 + \CO\left({1\over n}\right). 
\ee
We also plot its first Richardson transform to eliminate subleading tails. The ``prediction" (\ref{famousbw}) is indeed verified experimentally. 

The story of the famous result (\ref{famousbw}) is a fascinating chapter of modern mathematical physics 
(see \cite{simonreview}). The behavior of the $a_n$ at large $n$ 
was first obtained by Bender and Wu in \cite{oldbw} 
by studying numerically the sequence of the first seventy-five coefficients. 
They were even able to guess, from these numerical experiments,  the exact form of the prefactor in (\ref{famousbw}). 
In a subsequent (and classic) paper \cite{bw}, 
they showed that the result (\ref{famousbw}) could be derived analytically by looking at the one-instanton sector.

In the case of the quartic oscillator with positive coupling $g$, the origin leads to a stable quantum-mechanical ground state. The instanton solution only appears 
when one inverts the sign of the coupling, and as a result the perturbative series for positive $g$ is alternating and Borel summable. In other cases, like for 
example the cubic oscillator 
\be
V(x)={1\over 2} x^2 -g x^3,
\ee
the origin is always unstable quantum-mechanically for any real value of the coupling constant $g$. An elementary calculation shows that the action of the instanton mediating the tunneling is 
\be
\label{cubicaction}
S_c= {A \over g^2}, \qquad  A={2\over 15}.
\ee
The large order behavior of the coefficients in the formal power series for the ground state energy 
\be
E(g)=\sum_{n=0}^\infty a_n g^{2n}
\ee
can be computed by using a small modification of (\ref{akgen}) -essentially, one has to consider the discontinuity along the positive real axis of $g^2$ and then there is no sign alternating factor. Therefore, 
\be
a_n \sim A^{-n} \Gamma\left( n +1/2\right)
\ee
where $A$ is given in (\ref{cubicaction}), see for example \cite{alvarezcubic} for a derivation of this result. 
In this case the series is not Borel summable, reflecting the instability of the perturbative ground state. 

\begin{figure}[!ht]
\leavevmode
\begin{center}
\includegraphics[height=4cm]{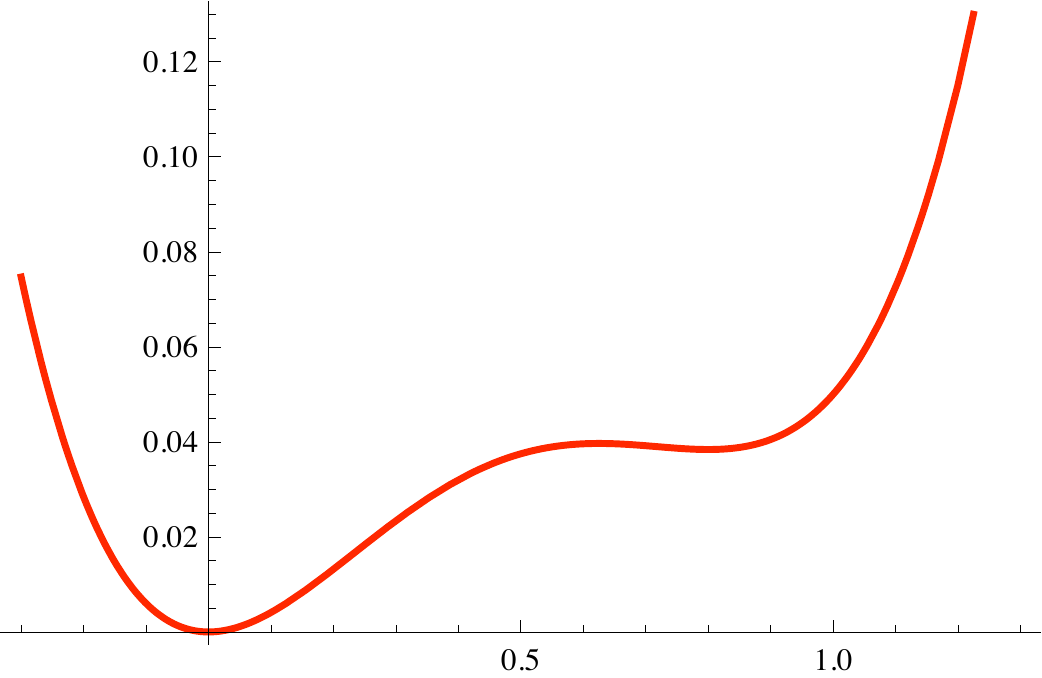} \qquad  \qquad \includegraphics[height=4cm]{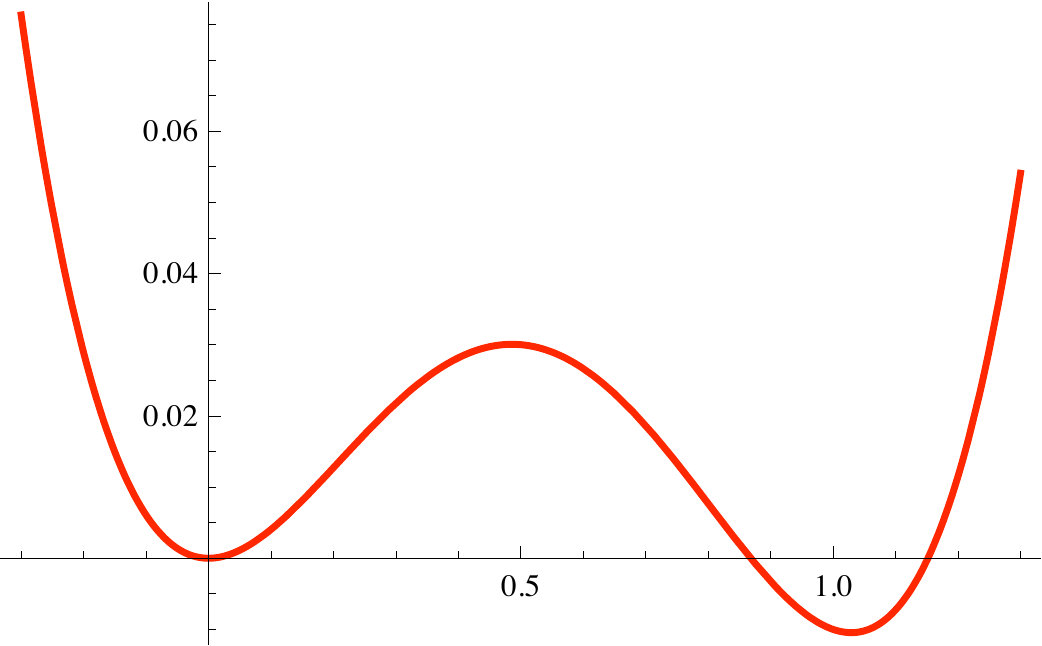}
\end{center}
\caption{On the left: the potential $V(x)$ in (\ref{gammapot}) for $\gamma=0.95$, where the origin is the absolute minimum. On the right: the potential (\ref{gammapot}) for $\gamma=1.01$; the origin is now unstable to quantum tunneling.}
\label{gammapotplots}
\end{figure}

In general, in one-dimensional quantum-mechanical problems, we will have {\it complex} instanton solutions with complex actions. They lead to perturbative series 
which are Borel summable and have an oscillatory character. As we mentioned in the case of ODEs, the large order behavior is controlled by 
the instantons with the smallest action in absolute value, and the phase of the action determines the oscillation period of the series. 
Let us analyze in some detail a very instructive example, following \cite{blgzj} (a useful discussion can be also found in chapter 42 of \cite{zj}). Let us consider a particle situated at the origin of the potential 
\be
\label{gammapot}
{1\over g^2} V(gx) , \qquad V(x)={1\over 2} x^2 -\gamma x^3 +{1\over 2} x^4.
\ee
The ground state energy has the expansion 
\be
E(g)=\sum_{n\ge 0} a_n g^{2n}. 
\ee
In this example there are two different situations (see \figref{gammapotplots}):
\begin{enumerate}
\item For $|\gamma| >1$, the origin is not an absolute minimum, which is in fact at 
\be
x_0=\frac{3\,\gamma+ {\sqrt{-8 + 9\,\gamma^2}}}{4}.
\ee

\item For $|\gamma|<1$, the origin is the absolute minimum. 

\end{enumerate}

In the first case $|\gamma| >1$, the vacuum located at the origin is quantum--mechanically unstable, and there is a real instanton given by a trajectory from $x=0$ to the 
turning point 
\be
x_+=\gamma-{\sqrt{\gamma^2-1}}.
\ee
The action of this instanton can be written as 
\be
S_c={A\over g^2}, \qquad A=2 \int_0^{x_+} \rd x \, (2 V(x))^{1\over 2} =-{2\over 3} +\gamma^2  -{1\over 2} \gamma(\gamma^2-1)\log {\gamma +1\over\gamma-1}. 
\ee
The one-loop prefactor for this instanton, appearing in (\ref{ol-qm}), is given by 
\be
c_0= {2\over \pi^{1/2}}(\gamma^2-1)^{-1/2}.
\ee

The behavior when $|\gamma|<1$ is obtained by analytic continuation of this instanton configuration, which is 
now complex. In fact, there are two complex conjugate instantons described by a particle which goes from $x=0$ to 
\be
x=\gamma \pm \ri  {\sqrt{1-\gamma^2}}.
\ee
We have then to {\it add} the contributions of both instantons. Since $c_0$ becomes imaginary when $|\gamma|<1$, adding the complex-conjugate contributions of the two instantons gives 
\be
a_k  \sim \Gamma(k+1/2)  {\rm Im}\, A^{-k-1/2}.
\ee
More generally, if we have a quantum-mechanical problem involving a complex instanton and its complex conjugate, and 
\be
A=|A| \re^{-\ri \theta_A}, \qquad 
c_0=|c_0| \re^{\ri \theta_c}, 
\ee
the large order behavior, obtained by adding the contribution of the two instantons, is oscillatory
\be
a_k \sim \Gamma(k-\beta) |A|^{-k+\beta} \cos\left( (k-\beta)\theta_A + \theta_c\right).
\ee

As in the case of ODEs analyzed in section \ref{ODEs}, when a perturbative series is Borel summable (like in the case of the quartic oscillator with $g>0$ or in the potentials with complex instantons), the Borel resummation of the perturbative series reconstructs the non-perturbative answer. There are two types of situations where there is no Borel summability: the first case corresponds to perturbative series around unstable minima, like the quartic oscillator with $g<0$ or the cubic oscillator. A different situation occurs in the case of the double-well potential. In that case, there is a stable ground state but the perturbative series is not Borel summable, and one has to consider lateral Borel resummations. The ground state energy can be reconstructed from the Borel-resummed perturbative series and the Borel-resummed instanton or trans-series solutions, in a way which is similar to the analysis of the Hastings--McLeod solution of Painlev\'e II in Example \ref{p2hm}, see \cite{zj-first,delabaerepham,zjj} for more details on this quantum-mechanical problem.

\subsection{Non-perturbative effects in Chern--Simons theory}
\label{nonpert-cs}

We have seen that many of the structures found in the study of ODEs reappear in QM: perturbative series are asymptotic series, 
and expansions around non-trivial saddle points or 
instantons are the analogues of trans-series. In particular, the singularity structure of the Borel transform of the perturbative 
series is governed by the non-trivial saddles. In principle, the extension of these ideas to QFT should be straightforward: the analogue of a trans-series would be 
the perturbative expansion around instanton configurations, and could think that these trans-series control the Borel transform of the perturbative series, and therefore 
its large order behavior. However, the extension of the above ideas to realistic QFTs is plagued with serious difficulties. Probably, 
the most important one is the fact that in renormalizable QFTs 
there are other sources of factorial divergence in perturbative series, namely {\it renormalons} (see \cite{beneke}). Renormalons are particular types of 
diagrams which diverge factorially due to the integration over momenta in the Feynman integral. Due to the existence of renormalons, 
the analysis of the large order behavior of 
perturbation theory inspired by QM does not extend straightforwardly to standard QFTs. 
There are however QFTs where renormalon effects are absent, like Chern--Simons (CS) theory and many supersymmetric QFTs, and we will 
focus here on this ``toy" QFTs, and more particularly on CS theory, where many different aspects of non-perturbative effects are relatively well understood.

CS theory is a QFT defined by the action
\be
S=-{k \over 4\pi} \int_M {\rm Tr} \Bigl( \CA\wedge \rd \CA + {2 \ri \over 3} \CA
\wedge \CA \wedge \CA \Bigr).
\label{csact}
\ee
Here, $\CA$ is a $G$-connection on the three-manifold $M$, where $G$ is a gauge group. We will mostly consider $G=U(N)$, and in this case our conventions are such that $\CA$ is a Hermitian $N \times N$ matrix-valued one-form. Gauge invariance of the action requires \cite{deser}
\be
k \in \IZ. 
\ee
The 3d QFT defined by this action is a remarkable one: it is exactly solvable, 
yet highly nontrivial, and provides a QFT interpretation of quantum invariants of knots and three-manifolds \cite{wittencs}. 

The partition function of the theory on $M$ is defined by the path integral 
\be
\label{zcs}
Z(M) =\int {\mathcal D}\CA\,  \re^{\ri S(\CA)}
\ee
and in principle it can be computed by standard perturbative techniques (see \cite{mmcslectures} for a review). 
The saddle-points of the CS action are just flat connections  
\be
F(\CA)=0. 
\ee
These are in one-to-one correspondence with embeddings 
\be
\pi_1(M) \rightarrow G, 
\ee
where $G$ is the gauge group. In principle, the path integral (\ref{zcs}) has various contributions coming from 
expansions around the different saddle-points. The perturbative sector is defined by expanding around the trivial flat connection 
\be
\CA=0,
\ee
while instanton sectors are associated to non-trivial flat connections. Formal expansions around these instanton sectors define the analogue of trans-series for this QFT.  

\begin{example} {\it Lens spaces}. The lens space $L(p,1)$ has fundamental group $\pi_1(L(p,1)) =\IZ_p$. The set of $U(N)$ flat connections is 
given by homomorphisms
\be
\IZ_p \rightarrow U(N),
\ee
modulo gauge transformations. These are in turn given by splittings of $U(N)$ into $p$ factors
\be
\label{lens-split}
U(N)  \rightarrow U(N_1) \times U(N_2) \times \cdots \times U(N_p),
\ee
corresponding to the homomorphism 
\be
\xi \rightarrow {\rm diag}\left(\underbrace{1, \cdots, 1}_{N_1}, \underbrace{\xi, \cdots, \xi}_{N_2}, \cdots, \underbrace{\xi^{p-1}, \cdots, \xi^{p-1}}_{N_p} \right),
\ee
where
\be
\xi=\exp\left( {2 \pi \ri \over p} \right). 
\ee
Therefore, instanton sectors are in one-to-one correspondence with partitions of $N$ into $p$ nonzero integers. 
The CS action evaluated at the flat connection labelled by $\{ N_j\}_{j=1, \cdots, p}$ is given by 
\be
\label{csia}
A={\pi \ri k \over p} \sum_{j=1}^p (j-1)^2 N_j. 
\ee
\end{example}

\begin{figure}[!ht]
\leavevmode
\begin{center}
\includegraphics[height=2.5cm]{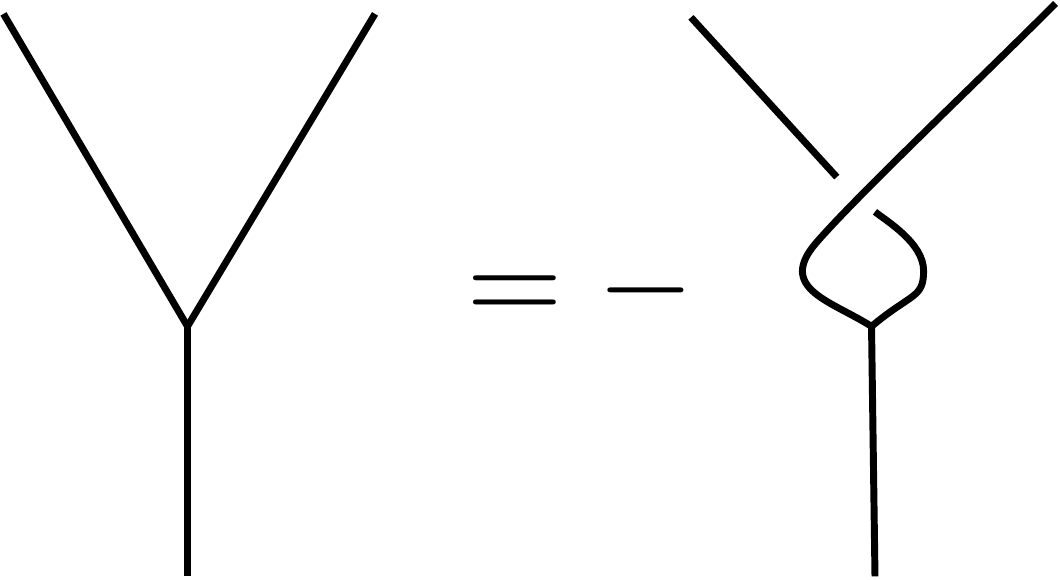} \\ \vskip .75cm\includegraphics[height=2cm]{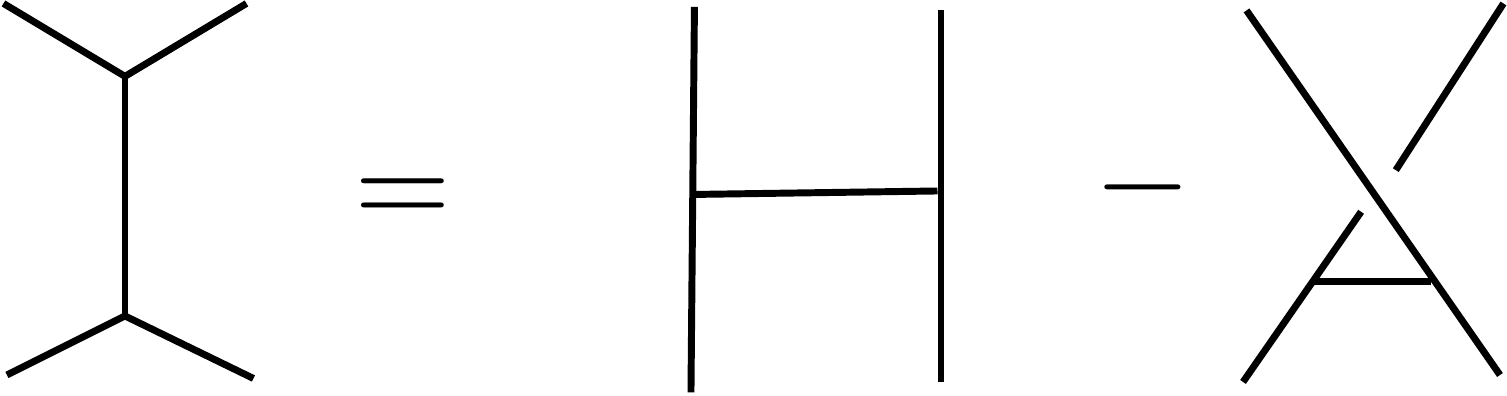}
\end{center}
\caption{AS and IHX relations. }
\label{rels}
\end{figure}

We can now ask what is the nature of the perturbative series appearing in CS theory. It turns out that, generically, perturbation theory around the trivial connection $\CA=0$ 
is factorially divergent. The reason for this is the same as in QM, namely, the factorial growth in the number of diagrams. Let us see this in some detail. We will denote by 
\be
F(M, {\bf g}, g_s)
\ee
the contribution of the trivial connection to 
the free energy $\log \, Z$ of CS theory. Here, ${\bf g}$ is the Lie algebra associated to $G$, and 
\be
\label{gscs}
g_s={2\pi \ri \over k}. 
\ee
Using standard perturbative techniques, it is easy to see that the free energy can be written as a formal power series of the form 
\be
\label{fM}
F(M, {\bf g}, g_s) =\sum_{n=1}^{\infty} \sum_{\Gamma \in \CA^{(c)}_n(\emptyset)} c_\Gamma(M) W_{\bf g}(\Gamma) g_s^n. 
\ee
Let us spell out in detail the ingredients in this formula. We first construct the space of Feynman diagrams, 
$\CA^{(c)}(\emptyset)$. This is the space of connected, trivalent diagrams with no 
external legs (i.e. connected vacuum bubbles) modulo 
the so-called IHX and AS relations, shown in \figref{rels}. It is graded by the {\it degree of the diagram} $n$, which is half the number of vertices (and 
also equals the number of loops minus one):
\be
\CA^{(c)}(\emptyset) =\oplus_{n=1}^{\infty} \CA^{(c)}_n(\emptyset).
\ee
A basic fact is that, for each $n$, this space has finite dimension. The very first dimensions are listed in Table \ref{dims}. 

\begin{table}[htbp] 
\centering 
\begin{tabular}{|c||c|c|c|c|c|c|c|c|c|c|}\hline
 $n$ & 1 & 2 & 3 & 4 & 5 & 6 & 7 & 8 & 9 & 10  \\  \hline
 $d(n)$ & 1 & 1 & 1 & 2 & 2 & 3 & 4 & 5 & 6 & 8  \\ 
\hline
\end{tabular}
\caption{Dimensions $d(n)$ of ${\cal A}^{(c)}_n(\emptyset)$ up
to $n=10$.}
\label{dims}
\end{table}
An explicit choice of basis up to $n=5$ is shown below:

\begin{eqnarray}
\label{graphi}
n=1: & & \,\,\,\,\, \twoVgraph \nonumber\\
n=2: & & \,\,\,\,\, \fourVgraph \nonumber\\
n=3: & & \,\,\,\,\, \sixVgraph \nonumber\\
n=4: & & \,\,\,\,\, \eightVgraphI \,\,\,\,\, 
\eightVgraphII \nonumber\\  
n=5: & & \,\,\,\,\,  \tenVgraphI \,\,\,\,\, 
\tenVgraphII \nonumber\\
\end{eqnarray}

\begin{figure}[!ht]
\leavevmode
\begin{center}
\includegraphics[height=2cm]{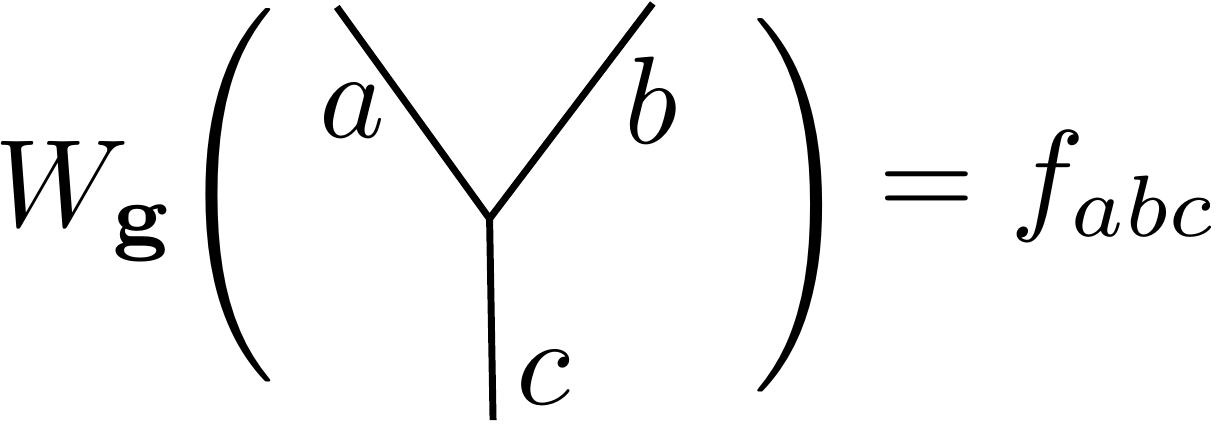} \end{center}
\caption{Weight system. }
\label{weight}
\end{figure}

The second ingredient is the {\it weight system}. This is an instruction to produce a number for each diagram, given the data of a Lie algebra with structure constants and 
a Killing form, 
\be
[T_a, T_b]=f_{abc}. 
\ee
To each trivalent vertex we associate the structure constant $f_{abc}$ as shown in \figref{weight}. In QFT we call 
this ``computing the group factor of the diagram $\Gamma$." 
The final ingredient is $c_\Gamma(M)$. It is simply given by the Feynman integral associated to the graph $\Gamma$. It is possible to show that each $c_\Gamma (M)$ is a 
topological invariant of $M$. For example,
\be
c_{\theta} (M)=\text{Casson invariant of $M$}. 
\ee

Based on our experience with the quantum anharmonic oscillator, we should ask 
how many diagrams we have at each loop order $n+1$. It has been shown by Garoufalidis and Le in \cite{gl} that
\be
{\rm dim}\left( \CA^{(c)}_n(\emptyset) \right) \sim n!, \qquad n\gg 1. 
\ee
Therefore, the series (\ref{fM}) will be factorially divergent. Interestingly, there is no other source of factorial divergences. These divergences could come from the weight factors, or from the Feynman integrals. However, it is easy to see that the weight factors 
can only grow exponentially. It is also shown in \cite{gl} that the Feynman integrals grow with the degree as 
\be
\left| c_\Gamma (M) \right| \sim C^n_M,
\ee
where $C_M$ is a constant that depends on the three-manifold under consideration. In QFT terms this means that Feynman integrals grow at most exponentially, i.e. that there are no renormalons (since in a renormalon diagram with $n$ loops, the Feynman integral diverges itself factorially, as $n!$).

\begin{example} \label{cs-example} {\it Chern--Simons theory on Seifert spheres}. The general prediction of factorial divergence can be verified in detail for CS theory on Seifert homology spheres. A Seifert homology sphere is specified by 
$r$ pairs of coprime integers $(p_j, q_j)$, $j=1, \cdots, r$, and is denoted by 
\be
M=X\left( {p_1 \over q_1}, \cdots, {p_r \over q_r}\right).
\ee
One also defines
\be
P =\prod_{s=1}^r p_s, \qquad H=P \sum_{s=1}^r {q_s \over p_s}. 
\ee
$H$ is the order of the first homology group $H_1(M, \IZ)$. 
When $r=1$, Seifert spaces are just lens spaces, 
\be
L(p,q)=X(q/p). 
\ee
The partition function of CS theory on a Seifert space can be written as a matrix integral. This was shown for $G=SU(2)$ by Lawrence and Rozansky \cite{lr}, 
and it was extended to arbitrary gauge groups in \cite{mm}. There are two types of non-trivial flat connections: the reducible ones, and the irreducible ones. 
In a Seifert space and when $G$ is a simply-laced group, 
reducible flat connections are labelled by elements in 
\be
t \in \Lambda_r/H \Lambda_r, 
\ee
where $\Lambda_r$ is the root lattice. The contribution of such a connection to the partition function is given by (up to an overall normalization, see \cite{mm} for the details) 
\be
\label{betain}
Z(M) \propto \int \rd \lambda \,  {\rm e}^{ -{\lambda^2/2 \hat g_s } - k t\cdot \lambda}{ 
\prod_{s=1}^r \prod_{\alpha>0} 2 \sinh {\lambda\cdot \alpha \over
2 p_s} \over \prod_{\alpha>0} 
\Bigl( 2 \sinh {\lambda\cdot \alpha  \over
2} \Bigr)^{r-2}},
\ee
where 
\be
\label{hatg}
\hat g_s  ={P\over H} g_s,
\ee
$\lambda$ belongs to the weight lattice $\Lambda_w$, $\alpha>0$ are the positive roots, and the products are computed with the standard Cartan--Killing form. In the case of 
$U(N)$ we have 
\be
t \in \IZ^N/ H \IZ^N, 
\ee
and we write 
\be
t=\sum_{i=1}^N t_i e_i, \qquad \lambda =\sum_{i=1}^N \lambda_i e_i, \qquad \{ \alpha\}=\{ e_i -e_j\}_{1\le i<j\le N}, 
\ee
where $e_i$ is the orthonormal basis of the weight lattice, and $0\le t_i \le H-1$. We then find, 
\be
\label{unbetain}
Z(M)\propto \int  \prod_{i=1}^N \rd \lambda_i  \,  {\rm e}^{ -{1\over 2 \hat g_s} \sum_{i=1}^N \lambda_i^2- k \sum_{i=1}^N t_i \lambda_i}  
{\prod_{s=1}^r \prod_{i<j} 2 \sinh {\lambda_i-\lambda_j  \over
2 p_s} \over \prod_{i<j} 
\Bigl( 2 \sinh { \lambda_i-\lambda_j  \over
2} \Bigr)^{r-2}}.
\ee
The integration contour in (\ref{betain}), (\ref{unbetain}) is chosen in such a way that the Gaussian integral 
converges. 

The case of $SU(2)$ is particularly simple. Up to an overall constant, the contribution of the trivial connection to the partition function is just an integral, 
\be
Z\propto \int \rd \lambda \,  {\rm e}^{ -{\lambda^2/4 \hat g_s } }  f(\lambda; p_s), 
\qquad f(\lambda;p_j)=
\left( 2 \sinh { \lambda  \over
2} \right)^{2-r} \prod_{s=1}^r \left( 2 \sinh {\lambda  \over
2 p_s} \right).
\ee
This can be expanded in power series in $g_s$,
\be
Z \propto \sum_{n=0}^{\infty} a_n g_s^n, \qquad a_n ={f^{(2n)}(0)\over n!},
\ee
and one finds \cite{lr}
\be
a_n \sim  \left( -{P \over \pi^2 H}\right)^n n!, 
\ee
i.e. we have a factorial divergence, as expected. The growth is controlled by 
\be
A =-{\pi^2 H \over P}. 
\ee
We would expect this quantity to be the action of a non-trivial saddle-point of the theory, and indeed this is the action of an irreducible flat connection on the Seifert manifold.

\end{example}

The results of \cite{gl} about the structure of the CS perturbative series rely on a mathematical construction for this series 
called the LMO invariant \cite{lmo}, which has been 
much studied during the last years. The structure of the general trans-series, which correspond to the perturbative expansion around a non-trivial instanton solution, is less understood, and no mathematical construction has been proposed so far. Questions on classical asymptotics and Borel summability in CS theory have started to be addressed only recently, see  \cite{garoufalidis,wittenanalytic} for some results and/or conjectures. 

\subsection{The $1/N$ expansion}

The problem of non-perturbative effects and asymptotics becomes much more interesting when we look at gauge theories in the $1/N$ expansion \cite{thooft}. In this expansion, the free energy and correlation functions of the gauge theory are expanded in powers of $N$ or of the coupling constant $g_s$, but keeping the 't Hooft parameter
\be
t=g_s N
\ee
fixed. For example, the expansion of the free energy around the trivial connection (i.e. what we have called the perturbative series) is re-organized as
\be
\label{genusex}
F=\sum_{g=0}^{\infty} F_g(t) g_s^{2g-2}, 
\ee
where $F_g(t)$ is a sum over {\it double-line graphs} or {\it fatgraphs} of genus $g$. In this reorganization of the theory, the dominant contribution comes from the genus zero or planar 
diagrams. 

\begin{figure}[!ht]
\leavevmode
\begin{center}
\includegraphics[height=.8cm]{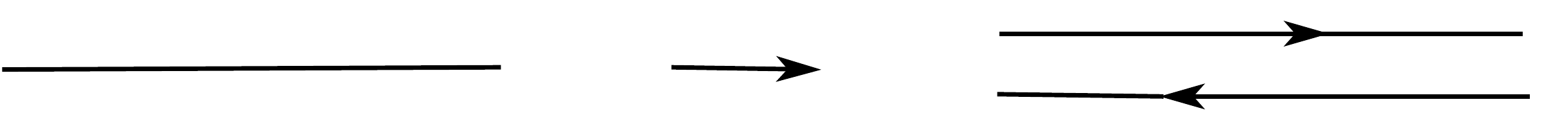}
\end{center}
\caption{Thickening an edge.}
\label{edgepropa}
\end{figure}
\begin{figure}[!ht]
\leavevmode
\begin{center}
\includegraphics[height=3cm]{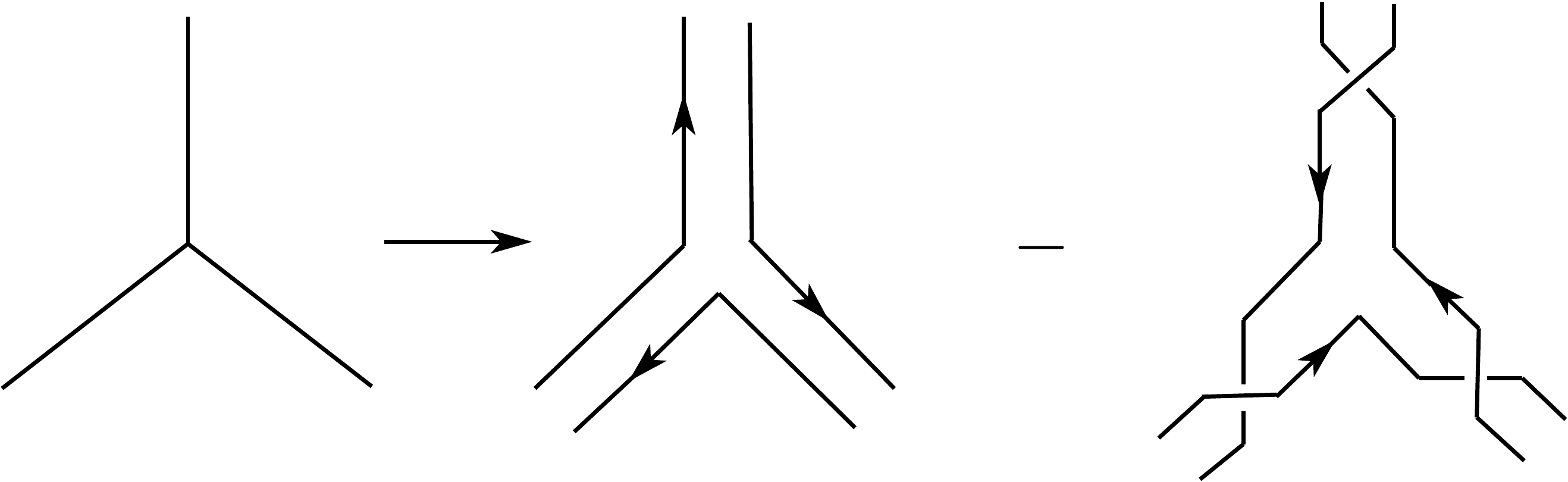}
\end{center}
\caption{Thickening a marked vertex.}
\label{resolution}
\end{figure}

\begin{figure}[!ht]
\leavevmode
\begin{center}
\includegraphics[height=3cm]{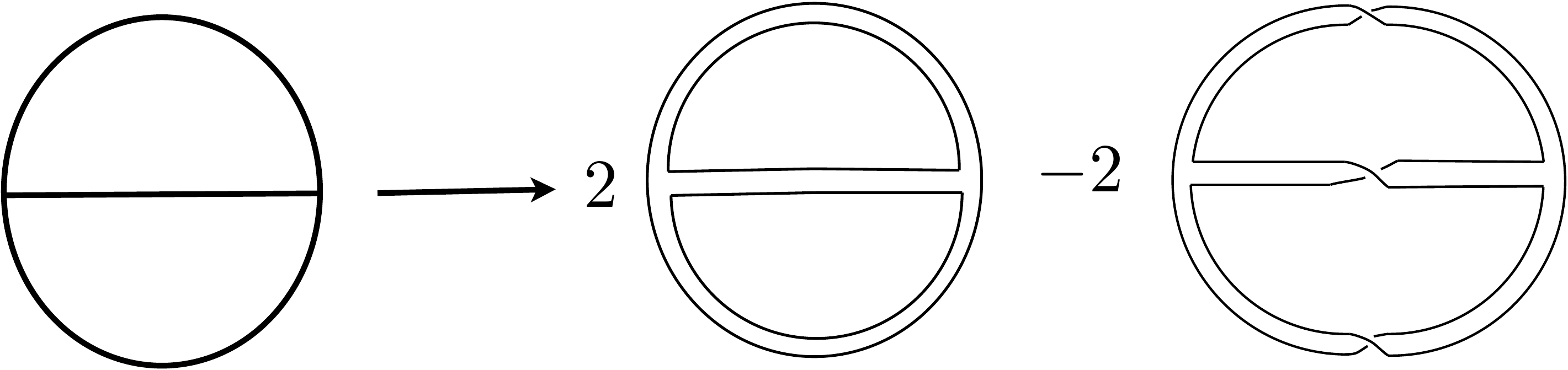}
\end{center}
\caption{Fatgraphs obtained from the theta diagram.}
\label{thetares}
\end{figure}

In the case of CS theory, the structure of the $1/N$ expansion can be made very explicit, as follows. Consider a graph 
$\Gamma$ in $\CA^{(c)}_n(\emptyset)$, and apply the {\it thickening rules} depicted in \figref{edgepropa} and \figref{resolution}. 
The thickening rules can be regarded as a map that associates to each diagram $\Gamma$ a formal linear combination of {\it fatgraphs} $\Gamma_{g,h}$, 
which are Riemann surfaces with boundaries and are classified topologically by their genus $g$ and number of boundaries $h$:
\be
\Gamma \rightarrow \sum_{g,h} p_{g,h}(\Gamma) \Gamma_{g,h}.
\ee
It is easy to see that the weight system of $U(N)$ can be written in terms of fatgraphs \cite{cvitanovic,bar-natan}, 
\be
W_{{\rm u}(N)}(\Gamma)=\sum_{g,h} p_{g,h}(\Gamma) N^h. 
\ee
An example is shown in \figref{thetares}. One then finds the following expression for the free energy around the trivial connection: 
\be
F(M, {\rm u}(N), g_s) = \sum_{g=0}^{\infty} \sum_{\Gamma_{g,h}} c_\Gamma (M)  p_{g,h}(\Gamma) N^h g_s^{E(\Gamma)-V(\Gamma)},
\ee
where $E(\Gamma), V(\Gamma)$ are the number of edges and vertices in $\Gamma$ (these topological data do not depend on the fattening of the graph). 
If we now use Euler's relation, 
\be
E(\Gamma)-V(\Gamma)=2g-2+h,
\ee
we see that $F(M, {\rm u}(N), g_s)$ is given by the formal series (\ref{genusex}), where 
$F_g(t)$ is defined as a formal infinite sum over all fatgraphs $\Gamma_{g,h}$ with {\it fixed} $g$ 
\be
\label{aght}
F_g(t)=\sum_{h\ge 0} a_{g,h} t^h, \qquad  a_{g,h}=\sum_{\Gamma_{g,h}} c_{\Gamma}(M) p_{g,h}(\Gamma).
\ee

As is well-known (see for example the classic review in \cite{coleman}) the $1/N$ expansion described above can be implemented in any $U(N)$ 
gauge theory where the fields transform in the adjoint representation of $U(N)$, and it can be applied to any gauge-invariant observable of the 
theory (when one expands around the trivial connection). 
The structure of the free energy as a double power series, 
see (\ref{genusex}) and (\ref{aght}), which we have written above based in the analysis of CS theory, can be easily seen to hold in any theory with fields in the adjoint of $U(N)$. The $1/N$ expansion is particularly clean in theories where the coupling constant $g_s$ does not run, i.e. in conformal field theories and topological field theories.

The first question that we have to ask in the search for non-perturbative effects is: what is the nature of the formal power series appearing in the theory, like for example in the series defining the free energy of the theory expanded around the trivial connection? In the case of the $1/N$ expansion, since there are two parameters, we have two different questions to ask: 

\begin{enumerate} 

\item What is the nature of the formal power series in $t$ appearing in (\ref{aght}), defining $F_g(t)$?

\item For a fixed $t$, what is the nature of the power series in $g_s$ appearing in (\ref{genusex})?

\end{enumerate}

The answer to these questions is the following: in theories with no renormalons, the functions $F_g(t)$ are analytic 
at the origin, i.e. the power series (\ref{aght}) have a finite radius of convergence which, moreover, 
is common to all of the $F_g(t)$. However, for fixed $t$, the functions $F_g(t)$ grow like 
\be
F_g(t) \sim (2g)! 
\ee
and the series (\ref{genusex}) diverges factorially. 

The analyticity of the $1/N$ expansion at fixed genus was first noticed in \cite{knn} and analyzed in 
some models by 't Hooft \cite{thooft}. It can be proved in detail in simple $U(N)$ gauge theories, such 
as matrix models (see \cite{gp} for a recent study) and CS theory \cite{glm}. The factorial growth of the $1/N$ expansion was pointed out, in a slightly different context, in \cite{shenker}.  

\begin{figure}[!ht]
\leavevmode
\begin{center}
\includegraphics[height=3.5cm]{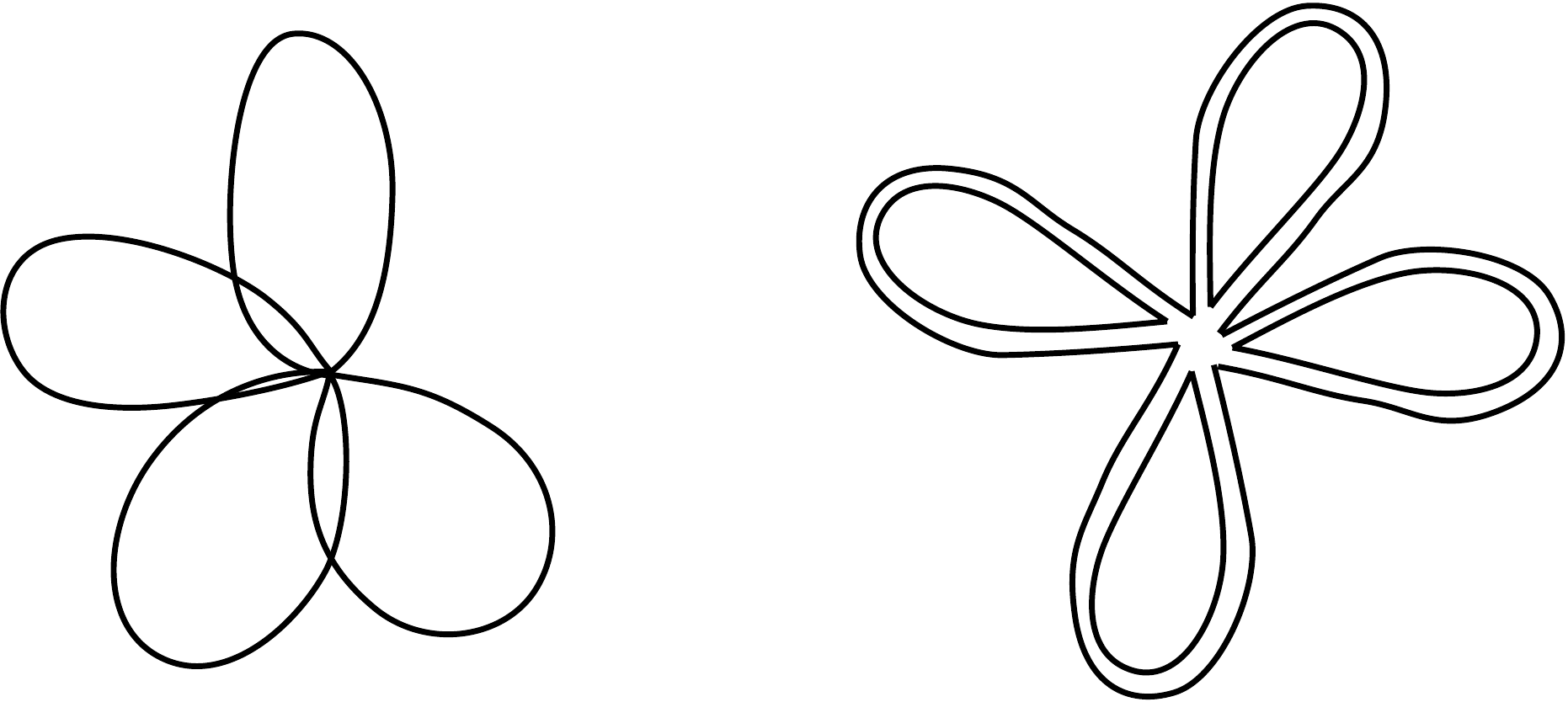} 
\end{center}
\caption{Counting flowers with $n$ petals.}
\label{petals}
\end{figure}

The basic reason for the analyticity of the $F_g(t)$ is that the number of fatgraphs of {\it fixed} genus grows only {\it exponentially}, and not factorially. A nice example of 
this contrasting behaviour is the counting of flower-like graphs. The number of flowers made out of  $n$ ``thin petals",  like the one shown on the left in \figref{petals}, is given by 
the number of possible contractions in a vertex with $2n$ legs, and equals $(2n-1)!!$, which grows factorially for $n\gg 1$. In contrast, the number of ``genus zero" flowers with $n$ ``thick petals," like the one shown on the right in \figref{petals}, is given by the Catalan number (see for example \cite{difrancesco})
\be
C_n ={(2n)! \over (n+1)! n!},
\ee
which at large $n$ grows only exponentially, like $4^n$. 

We will now give an argument for the analyticity of the free energy in theories without renormalons, following \cite{glm}. 
Let us come back to the formal power series appearing in (\ref{aght}). It is easy to see that 
\be
p_{g,h}(\Gamma) \le C_p^{2g-2+h}, 
\ee
where $C_p$ is a constant. For example, in a theory with a purely cubic interaction, like CS theory, 
each vertex gives two resolutions (see \figref{resolution}), and the maximum number of terms is $2^V$. Since in a theory with a cubic interaction we have
\be
3V=2 E,
\ee
we deduce
\be
p_{g,h}(\Gamma) \le 2^V=4^{E-V}=4^{2g-2+h}.
\ee
In Yang--Mills theory there are also quartic vertices, which from the point of view of this counting can be regarded as two cubic vertices joined by an edge, leading to a similar estimate. The next step is to analyze the Feynman integrals, $c_\Gamma$. If the theory has renormalons, they can grow factorially with the number of vertices. But in a theory without renormalons they grow only exponentially in the number of vertices, and we can write
\be
\label{norenormalon}
|c_\Gamma|\sim C_F^{2g-2+h},
\ee
where $C_F$ is another constant. This has been shown to be the case for a large class of diagrams in 
QM \cite{bender}, and it has been proved in CS theory, by using the formulation in terms of the LMO invariant \cite{gl}. We then have, 
\be
a_{g,h} \sim (C_p C_F)^{2g-2+h} N_{g,h},
\ee
where $N_{g,h}$ is the number of double-line diagrams with genus $g$ and $h$ holes, counted with the appropriate weight. 
For example, if we normalize all vertices of degree $p$ with a factor $1/p!$, the weight of a diagram $\Gamma_{g,h}$ is given by 
\be
{1\over \left|  {\rm Aut}(\Gamma_{g,h}) \right|},
\ee
i.e. the inverse of the order of its automorphism group, see \cite{difrancesco}. 
The counting of fatgraphs (weighted by their automorphism group, as above) has been developed very much both in 
combinatorics and in mathematical physics. The main result we have in this respect is that
\be
N_{g,h} \sim C_D^V C_G^g (2g)!, 
\ee
see for example \cite{glm}. We conclude that
\be
a_{g,h}\sim (2g)! C_1^g C_2^h. 
\ee
Therefore, for {\it fixed genus}, and assuming the growth (\ref{norenormalon}), the functions $F_g(t)$ defined by the power series (\ref{aght}) are analytic at $t=0$ with a finite radius of convergence $\rho$ common to all $g$. This is the analiticity result 
we wanted to establish. Generically $\rho<\infty$, and there is typically a singularity $t_c$ in the $t$-plane somewhere in the circle of radius $\rho$,
\be
\label{singrho}
|t_c|=\rho.
\ee
This argument shows as well that {\it for fixed $t$} (and inside the common domain of convergence) the sequence $F_g(t)$ will diverge like $(2g)!$. This was our second claim above.
 
 The analyticity of the genus $g$ free energies $F_g(t)$ suggests that $t$ should be regarded naturally as an arbitrary, 
{\it complex} variable. In the same way, $g_s$ should be also thought of as a complex variable, although in the original gauge theory it can only take special 
values: in Yang--Mills theory $g_s$ is the square of the Yang--Mills coupling constant 
$g^2_{\rm YM}$ and it is real and positive, while in CS theory it is of the form (\ref{gscs}). The complexification of these variables is natural in 
the context of some large $N$ dualities. For example, in dualities between large $N$ CS theory and topological string theory \cite{gv}, the 't Hooft parameter is identified with 
a {\it complexified} K\"ahler parameter of a Calabi--Yau manifold. We will then take the point of view that all gauge theory parameters belong to a complex {\it moduli space}.

\subsection{Large $N$ instantons}

We have seen that, in theories without renormalons, 
the power series defining free energies $F_g(t)$ at {\it fixed} genus are analytic at $t=0$. Therefore, there is in principle no indication of non-perturbative 
effects to be taken into account. Moreover, in many interesting QFTs, one can perform an analytic continuation of the functions $F_g(t)$ 
to a region in the complex plane which includes all physical values of the 
't Hooft coupling, providing in this way a complete description of the theory order by order in the $1/N$ expansion. 
This procedure is sometimes called {\it weak-strong coupling interpolation}, since one starts with a function defined at weak 't Hooft coupling 
(i.e. in a neighbourhood of $t=0$) and ends up with a function defined for some angular region in the complex plane where $|t|\gg 1$. 
We will see below an example of this analytic continuation, concerning the free energy of ABJM theory).  

However, as we have also seen, the price to pay for the analyticity of $F_g(t)$ for fixed $g$ is that, 
at fixed $t$, the sequence of free energies grows doubly-factorially with the genus. More precisely, 
one has the growth 
\be
\label{largeg}
F_g(t) \sim (2g)! (A(t))^{-2g}, \qquad g \gg1,
\ee
where $A(t)$ is a function of $t$. This behavior has been found in many simple models of the $1/N$ expansion, 
like matrix models (and their double-scaling limits), and in some supersymmetric gauge theories. 

As we have seen, in theories without renormalons, the standard factorial growth of the perturbative expansion is typically 
related to the existence of instantons in the theory. It is then natural to suspect that the growth (\ref{largeg}) of the 
$1/N$ expansion should be also due to instanton-like objects which we will call {\it large $N$ instantons}, and that $A(t)$ is the action of such an object. 

What is a large $N$ instanton? In general, large $N$ instantons are built upon classical instantons. 
To see this, consider a gauge theory with coupling constant $g_s$ and 't Hooft parameter $t$, as well as an instanton 
solution whose action (including the coupling) is given by 
\be
S_c={A \over g_s}.
\ee
We will also assume that $A$ is of order one at large $N$:
\be
A \sim \CO(1).
\ee
For example, in Yang--Mills theory the usual instantons with low instanton number satisfy this property. This is due to the fact we can build an instanton 
by using just an $SU(2)$ subgroup of $U(N)$. This is also the case in matrix models, where, as we will see in detail, instantons are obtained by eigenvalue 
tunneling of one eigenvalue out of $N$. There are instanton configurations whose action is of order $N$ (``giant instantons"), but we will not consider them here.
 
Let us now evaluate the free energy of a gauge theory around such an instanton configuration, in perturbation theory. The one-loop fluctuations give a term 
with the generic form (at large $N$)
\be
\left( {c_0  \over g_s}\right)^{c_1 N},
\ee
where $c_1 N$ is the number of zero modes, or collective coordinates of the instanton, at large $N$. This factor comes from the canonical normalization of the 
modes in the path integral, since we can always normalize the fields in such a way that the action has an overall power of $1/g_s$. On top 
of the classical action and the one-loop fluctuations, at large $N$ we have to consider as well all vacuum, connected planar 
diagrams (at all loops) in the background of the classical instanton configuration. To see how these appear, 
let us focus for simplicity on an interaction given by a cubic vertex, as in \figref{resolution}. Let us consider fluctuations around 
the instanton solution $\overline{\CA}$
\be
\CA=\overline{\CA} +\CA'.
\ee
\begin{figure}[!ht]
\leavevmode
\begin{center}
\includegraphics[height=3cm]{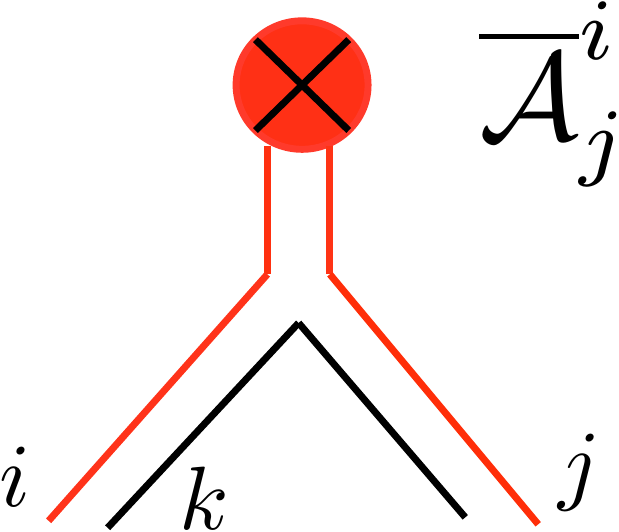} \qquad \qquad \includegraphics[height=3.5cm]{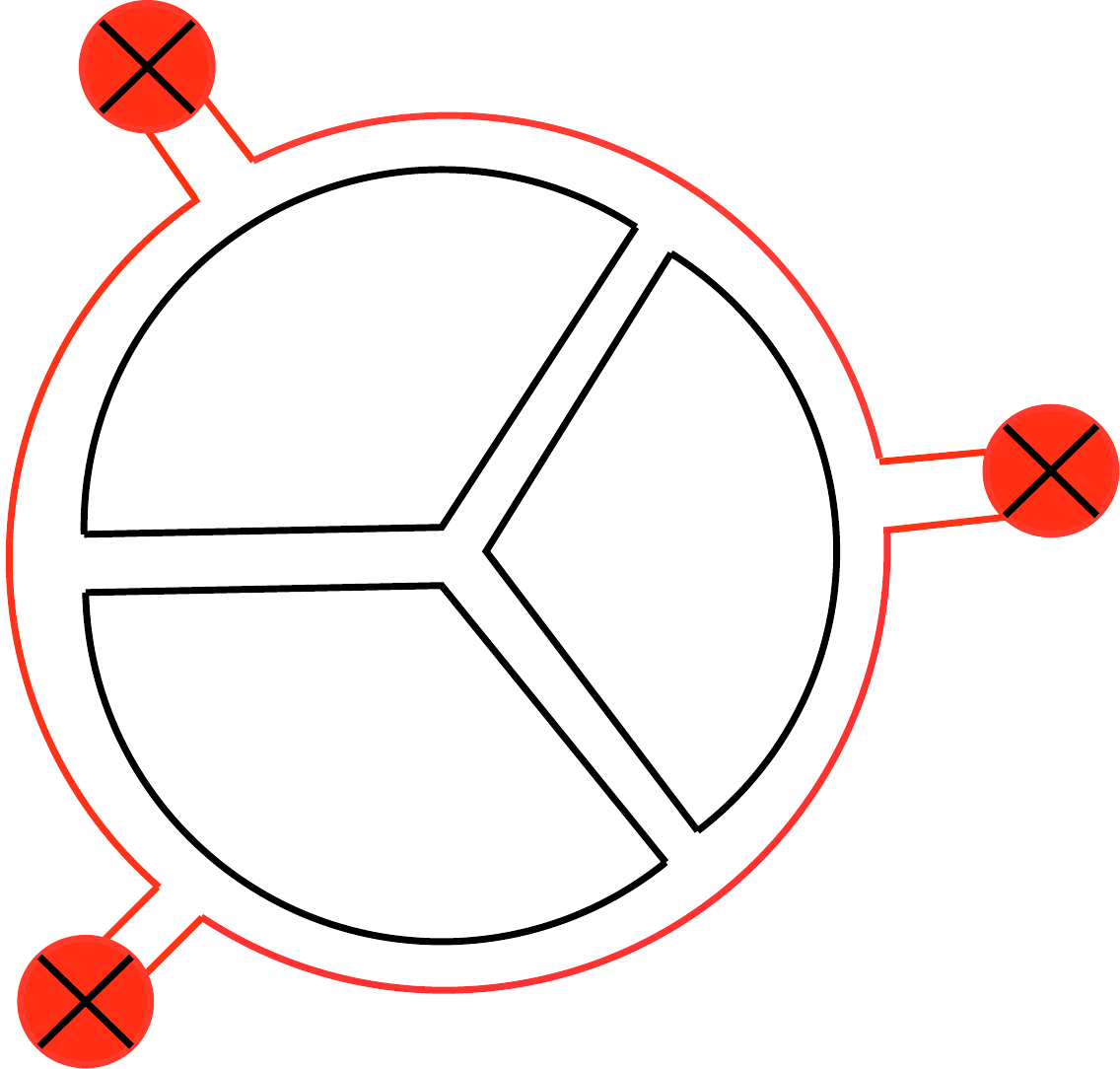} 
\end{center}
\caption{The instanton vertex (\ref{ivertex}) (left) and a planar diagram contributing to the large $N$ instanton action $A(t)$ a term of order $t^3$ (right).}
\label{instvertex}
\end{figure}
The action for the fluctuations will include a vertex of the form 
\be
\label{ivertex}
\sum_{i,j,k} (\overline{\CA}_\mu)^i_{~j} \,  (\CA_\nu)^j_{~k} \,  (\CA_\rho)^k_{~i}
\ee
and involving the instanton background. We can represent this vertex in the double-line notation as in \figref{instvertex}, where the red line ending on the blob 
corresponds to the 
instanton background. It gives a factor 
of $g_s$, but only the interior line gives a factor of $N$ after tracing over. A simple example of a diagram 
contributing to the instanton action is the one depicted on the r.h.s. of \figref{instvertex}. The inner closed lines gives a factor of $N^3$, 
and the diagram is proportional to 
\be
\tr\bigl( {\overline{\CA}}^3 \bigr) N^3 g_s^2 = {1\over g_s} t^3 \tr\bigl( {\overline{\CA}}^3 \bigr), 
\ee
since there are nine edges $E=9$ and seven vertices $V=7$, so the power of $g_s^{E-V}$ is two. Therefore, this diagram gives a correction of order $t^3$.  

We conclude that, at large $N$, the contribution of an instanton to the free energy is of the form 
\be
\exp\left( -{A(t) \over g_s} \right),
\ee
where
\be
\label{exlna}
A(t)= A - c_1 t \log \left( {c  \over t} \right)+ \CO(t).
\ee
The first term in this equation comes from the classical action of the instanton. The second term, which is logarithmic in $t$, incorporates the one-loop correction. 
Higher loop corrections lead to a series in $t$. 
We then see that, from the point of view of the $1/N$ expansion, 
the instanton action is promoted to a non-trivial function $A(t)$ of the 't Hooft parameter, which we calll the {\it large $N$ instanton action}. 

The calculation of large $N$ instanton actions in realistic theories is of course difficult, since we have to sum up an infinite number of planar diagrams (in the same 
way that calculating the genus zero free energy involves adding up an infinite number of diagrams, at all loops). 
An alternative, more general way to think about large $N$ instantons is in terms of large $N$ effective actions. 
It is believed that gauge theories at large $N$ can be reformulated
 in terms of a ``large $N$ effective action" with coupling constant (or $\hbar$ constant) equal to $1/N$, and involving a field 
sometimes called the ``master field" \cite{wittenmaster}. In this effective theory, correlation functions 
at large $N$ are obtained simply by solving the classical equations of motion of the effective action in the presence of sources. 
A large $N$ instanton is an instanton solution of this large $N$ effective theory, i.e. a saddle point of the Euclidean version of the theory, with finite action. 
This is in general different from the usual instanton configurations, which are saddle point of the {\it classical} Euclidean action. 
However, as we have seen, large $N$ instantons can often be thought of as deformations of the classical instantons, 
where the deformation parameter is the 't Hooft parameter:  as it is manifest in (\ref{exlna}), when $t\rightarrow 0$ we recover the action of the ``classical" gauge theory instanton. 
Explicit examples of large $N$ instantons were obtained in the $\IC\IP^N$ model in 
\cite{affleck,munster}, as deformations of classical instantons. A particularly beautiful example is the large $N$ 
instanton of two-dimensional Yang--Mills theory obtained in \cite{gm}. In the next section we will see 
an example of a large $N$ instanton in a simple toy model, namely matrix quantum mechanics. 

Large $N$ instantons play an important r\^ole in the structure of the $1/N$ expansion. First of all, 
in the calculation of physical quantities in a large $N$ theory, we expect to have a trans-series structure 
with a perturbative sector around the trivial instanton sector, and then a series of sectors corresponding to 
$1/N$ expansions around large $N$ instantons. The instanton sectors are weighted by 
\be
\re^{-A(t)/g_s}.
\ee
As it will be obvious in examples, $A(t)$, the large $N$ instanton action, is in general a non-trivial function of the 't Hooft parameter $t$. 
As expected from the examples developed in these lectures, this is the same quantity appearing in (\ref{largeg}) and controlling the subleading large $g$ asymptotics. 

The second important property of large $N$ instantons is their dynamical r\^ole in triggering phase transitions. 
If ${\rm Re}(A(t)/g_s)>0$, large $N$ instantons are suppressed exponentially at large $N$ (or small $g_s$). This might lead to think 
that ``instantons are suppressed at large $N$," but as Neuberger pointed out in \cite{neuberger}, 
this is not necessarily the case. It might happen for example that $A(t)$ 
vanishes at a particular value of $t$, and in this case the contribution of instantons become as important as the perturbative contributions. 
The value of the 't Hooft parameter for which $A(t)$ vanishes signals very often a {\it large $N$ phase transition}, or a critical point, in the theory. 
This behavior is nothing but the large $N$ version of the jump in the asymptotics occurring along 
an anti-Stokes line. It turns out that the critical value of the 't Hooft parameter is also, in many cases, the first singularity $t_c$ in the $t$-plane which 
we found in (\ref{singrho}). One of the first examples of such a transition was found in \cite{gw,wadia,wadia2}, and it was argued in \cite{neuberger} 
that this should be due to the vanishing of the instanton action, see \cite{mmnp} for a detailed verification for the model analyzed in \cite{gw,wadia}.

\begin{remark} {\it Large $N$ instantons as D-branes}. Notice that the diagrams contributing to the large $N$ instanton action 
are similar to the diagrams involving an external boundary, coming from a Wilson loop for example. This means that, if the large $N$ theory has a string dual, 
the large $N$ instanton should be associated to a hole in the worldsheet --in other words, to a D-brane. The relation between large $N$ instantons and D-branes 
in string theory goes back to the observation by Shenker \cite{shenker} that the large order behavior (\ref{largeg}) should be typical of genus expansions 
in string theory \cite{shenker}. This observation was based on the duality between doubly-scaled large $N$ matrix models (which also display this $(2g)!$ growth) 
and noncritical strings. According to Shenker, however, this growth should be a universal feature of string perturbation theory. The non-perturbative effects associated 
to this stringy behavior were later on identified with D-brane and membrane effects in \cite{polchinski,bbs}. 
\end{remark}

\begin{example} {\it A large $N$ instanton in CS theory}. Let us consider $U(N)$ CS theory on the lens space $L(2,1)$, and the instanton (i.e. the non-trivial 
flat connection) characterized by the splitting 
\be
(N-1, 1). 
\ee
The path integral around this instanton configuration is weighted by the exponentially small factor $\re^{-A/\hat g_s}$, 
where $\hat g_s=2 g_s$ (see (\ref{hatg})), and one finds from (\ref{csia}) that
\be
A= {\pi^2 \over 2}. 
\ee
Using the matrix model representation of the partition function given in (\ref{unbetain}), it is possible to calculate the large $N$ instanton action built on this classical instanton \cite{mpp}. The result is $A(t)/\hat g_s$, where
\be
\label{lensaction}
A(t) =2 \, {\rm Li}_2(\re^{-t/2}) -2\,  {\rm Li}_2(-\re^{-t/2})= {\pi^2 \over 2} -t \log \left( {4 \re \over t}\right)  +\CO(t^3),
\ee
which has indeed the expected structure (\ref{exlna}). 
\end{example}

In general, it is difficult to obtain explicit expressions for the $1/N$ expansion around large $N$ instanton configurations, hence our understanding of non-perturbative 
effects in general large $N$ theories is limited. In the next section we will make a detailed study on large $N$ instantons in matrix models, where one has a lot of analytic control and many 
results are available. We will now present another instructive solvable example, namely large $N$ instantons in Matrix Quantum Mechanics. 

\subsection{Large $N$ instantons in Matrix Quantum Mechanics}

In order to illustrate the general considerations on $1/N$ expansions and large $N$ instantons explained above, we will now discuss Matrix Quantum Mechanics (MQM), introduced and first studied in \cite{bipz}. This is a quantum-mechanical model where the degrees of freedom are the entries of a Hermitian $N\times N$ matrix $M$ 
and the Euclidean Lagrangian is given by
\be
\label{LagM}
L_M  =\tr \Bigl[ {1\over 2}\dot M^2 + V(M) \Bigr], 
\ee
where $V(M)$ is a polynomial in $M$. Notice that this problem has a $U(N)$ symmetry 
\be
M \rightarrow U M U^{\dagger}
\ee
where $U$ is a constant unitary matrix, and it promotes the standard one-dimensional QM problem to a problem where fields are in the adjoint 
representation of a $U(N)$ symmetry group and can then be studied in the $1/N$ expansion. We will 
 assume that the potential $V(M)$ is of the form 
 \be
 V(M)={1\over 2} M^2 + V_{\rm int}(M)
 \ee
 where $V_{\rm int}(M)$ is the interaction term.
The Feynman rules 
 are the same as in the case of QM, with the only difference that we will 
 now have ``group factors" due to the fact that $M$ is matrix valued.

\begin{figure}[!ht]
\leavevmode
\begin{center}
\includegraphics[height=5cm]{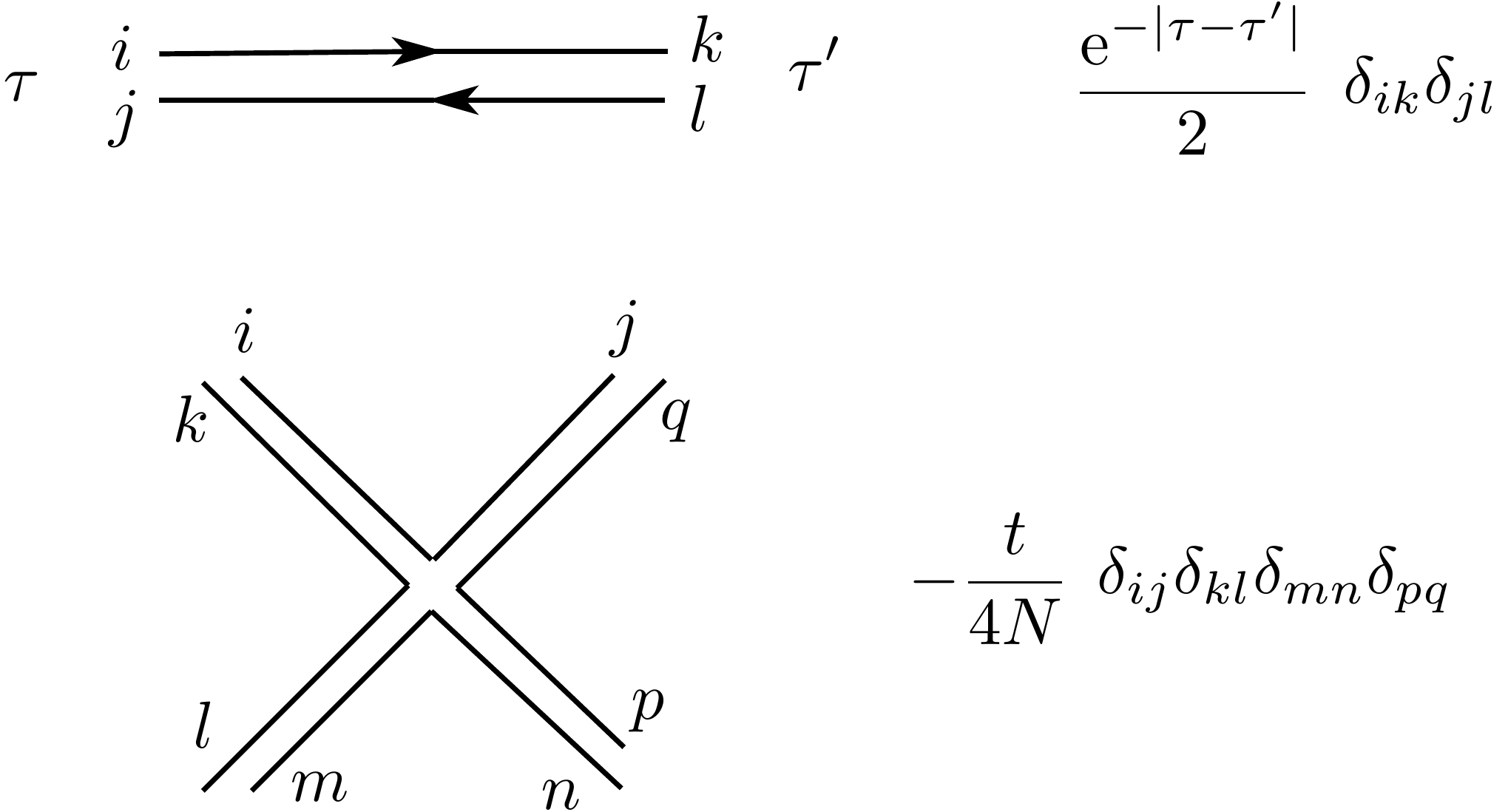}
\end{center}
\caption{Feynman rules for matrix quantum mechanics.}
\label{mqmf}
\end{figure}  

 The propagator of MQM is
 \be
{\re^{- |\tau|} \over 
 2} \, \delta_{ik} \delta_{jl}, 
 \ee
and for a theory with a quartic interaction 
 \be
 \label{quarticg}
 V_{\rm int}(M)={t \over 4 N} M^4
 \ee
 the Feynman rules are illustrated in \figref{mqmf}. The factor of $N$ in (\ref{quarticg}) 
 is introduced in order to have a standard large $N$ limit, as we will see in more detail later. 
 
 One can use these rules to compute the perturbation series of the ground state 
 energy of MQM, which is obtained by considering connected bubble diagrams, as in conventional QM. Each Feynman diagram leads to a group factor which 
 depends on $N$, i.e. each conventional Feynman 
diagram gives various fatgraphs that can be classified according to their topology. 
A fatgraph with $V$ vertices and $h$ boundaries will have a factor
\be
t^V N^{h-V} =t^V N^{2-2g}, 
\ee
since the number of edges is twice the number of vertices, $E=2V$ and
\be
h+E-V=h-V. 
\ee
Planar diagrams, as usual, are proportional to $N^2$, and the ground state energy has the structure 
    \be
    E(t, N)=\sum_{g=0}^\infty N^{2-2g} \CE_g(t). 
    \ee
The first few terms in the expansion of $\CE_0(t)$ can be easily computed in perturbation theory, 
    \be
    \label{planarquarticg}
    \CE_0(t)={1\over 2} +{1\over 8} t -{17\over 256} t^2 +{75\over 1024}t^3+\cdots
    \ee

 As first found in \cite{bipz}, the planar ground state energy in MQM 
can be obtained {\it exactly} by using 
a free fermion formulation.  This exact result {\it resums} in closed form all the planar 
diagrams of MQM contributing to the ground state energy. This goes as follows. 
After quantization of the system we obtain a Hamiltonian 
operator
\be
H=\tr  \Bigl[ -{1\over 2}{\partial^2 \over \partial M^2} + V(M) \Bigr], 
\ee
where
\be
\label{mdo}
\tr {\partial^2 \over \partial M^2}=\sum_{ab}  
{\partial^2 \over \partial M_{ab} \partial M_{ba}}.
\ee
In order to study the spectrum of this Hamiltonian, it is useful to change variables
\be
\label{Mdiag}
M=U \Lambda U^{\dagger},
\ee
where 
\be
\Lambda ={\rm diag}(\lambda_1, \lambda_2, \cdots, \lambda_N)
\ee
is a diagonal matrix. It can be shown that, when acting on singlet states 
(i.e., states that are invariant under the full $U(N)$ group), the differential 
operator (\ref{mdo}) has the form, 
\be
-{1\over 2} \tr {\partial^2 \over \partial M^2} = -{1\over 2}{1\over \Delta(\lambda)} \sum_{a=1}^N \Bigl( 
{\partial \over \partial \lambda_a}\Bigr)^2 \Delta(\lambda). 
\ee
Due to the residual Weyl symmetry, a singlet state is represented by a symmetric function of the $N$ eigenvalues,
\be
\Psi(\lambda_i).
\ee
We now introduce a completely
{\it antisymmetric} wavefunction
\be
\label{fwf}
\Phi(\lambda) =\Delta(\lambda) \Psi(\lambda),
\ee
where 
\be
\label{vander}
\Delta(\lambda)=\prod_{a<b} (\lambda_a -\lambda_b)
\ee
is the Vandermonde determinant. It is now easy to see that the original problem of calculating the energies for singlet states becomes the problem of calculating the energy of $N$ non-interacting fermions (since the function (\ref{fwf}) is completely antisymmetric) in an external potential $V(\lambda)$. We will now assume that this potential $V(\lambda)$ has good large $N$ scaling properties. More precisely, we will assume 
that the $N$-dependence of the potential $V(\lambda)$ is such that 
\be
V(\lambda) = N v\left( {\lambda \over {\sqrt{N}}} \right), 
\ee
where $v(\lambda)$ does not contain $N$. For example, the quartic potential considered above, 
\be
V(\lambda) ={1\over 2} \lambda^2 + {t \over 4 N} \lambda^4, 
\ee
has good scaling properties. This can be interpreted as saying that 
\be
\label{mqmgs}
t= N g_s
\ee
is the 't Hooft parameter of the model, which is kept fixed as $N\rightarrow \infty$. 
After rescaling $\lambda \rightarrow {\sqrt{N}} \lambda$, we find that the one-body fermion problem reduces to 
\be
\label{hNeq}
\biggl\{ -{\hbar^2\over 2N^2} {\rd^2 \over \rd \lambda^2} +v(\lambda) \biggr\}\phi_n(\lambda)=e_n \phi_n(\lambda)
\ee
where 
\be
e_n={1\over N} E_n
\ee
and $E_n$ are the energy levels in the original one-body problem. The ground state energy is given by
\be
\label{sumfermi}
E(t,N)=\sum_{n=1}^N E_n=N \sum_{n=1}^n e_n=N^2 \CE_0(t) +\cdots 
\ee
To calculate $\CE_0$, we note that the quantum 
effects in (\ref{hNeq}) are controlled by $\hbar/N$. Therefore, large $N$ is equivalent to $\hbar$ small, and in the 
large $N$ limit we can use the semiclassical or WKB approximation. 
In particular, we can use the Bohr--Sommerfeld formula to find the energy spectrum at leading order in $\hbar/N$. 
We will write this semiclassical quantization condition as 
\be
\label{leading}
N J(e_n)=n-{1\over 2}, \qquad n \ge 1,
\ee
where 
\be
J(e)=   {1\over \pi \hbar } \int_{\lambda_1(e)}^{\lambda_2(e)} \rd 
\lambda{\sqrt {2(e  -v(\lambda))}}
\ee
and $\lambda_{1,2}(e)$ are the turning points of the potential. If we denote 
\be
\label{xivar}
\xi={n-{1\over 2} \over N},
\ee
we see that (\ref{leading}) defines implicitly a function $e(\xi)$ through
\be
J\left( e(\xi)\right)=\xi. 
\ee
At large $N$, the spectrum becomes denser and denser, the discrete variable $\xi$ becomes 
a continuous one $\xi \in [0,1]$, and the sum in (\ref{sumfermi}) becomes an integral through the rule
\be
\label{sumrule}
\sum_{n=1}^N \rightarrow N \int_0^1 \rd \xi. 
\ee
One then finds, 
\be
\CE_0 =  \int_0^1 \rd \xi e (\xi).
\ee
We also define the Fermi energy of the system by the condition
\be
\label{fermicond}
J(e_F)=1. 
\ee
After some simple manipulations, one finds the expression
\be
\label{finaleo}
\CE_0= e_F-{1\over 3 \pi \hbar} \int_{\lambda_1(e_F)}^{\lambda_2(e_F)} \rd 
\lambda \Bigl[2(e_F-V(\lambda))\Bigr]^{3/2}. 
\ee

In the case of the quartic potential, the ground state energy can be explicitly computed as a function of the 't Hooft parameter. We first obtain the Fermi energy, which is defined by (\ref{fermicond}) (in this calculation we set $\hbar=1$). 
For the quartic potential, the integral $J(e)$ can be easily computed in terms of elliptic functions. Let us denote, 
\be
\label{abmqm}
a^2=\frac{\sqrt{4 e t+1}-1}{t}, \qquad b^2=\frac{\sqrt{4 e t+1}+1}{t}.
\ee
The turning points of the potential are $\pm a$. We also introduce the elliptic modulus
\be
\label{qmodulus}
k^2={a^2 \over a^2 + b^2}.
\ee
We then find
\be
\label{iintegral}
J(e)={1\over 3 \pi} (2t)^{1\over 2} (a^2+b^2)^{1\over 2}  \Bigl[ b^2 K(k) + (a^2-b^2) E(k)\Bigr], 
\ee
where $E(k)$, $K(k)$ are complete elliptic integrals. The condition (\ref{fermicond}) defines the Fermi energy $e_F(t)$ as an 
implicit function of the 't Hooft parameter. The planar free energy is given by
\be
\CE_0(t)=e_F(t) -{1\over 3 \pi} \left( {t \over 2}\right) ^{3/2} \CI(t, e_F(t)),
\ee
and it involves the integral
\be
\ba
\CI (t,e)&=\int_{-a}^a \rd u \Bigl[(a^2-u^2)(b^2+u^2)\Bigr]^{3/ 2}  \\
& ={2\over 35} {\sqrt{a^2 + b^2}} \Bigl\{ 2(a^2-b^2) (a^4 + 6 a^2 b^2+ b^4) E(k) +b^2 (2 b^4 +9 a^2 b^2 -a^4) K(k)\Big\}. 
\ea
\ee
The final result can be easily expanded in powers of $t$, and one finds
\be
\label{planarser}
\CE_0(t) ={1\over 2}+{t\over 8} - {17 t^2 \over 256}+{75 t^3\over 1024}-{3563 t^4 \over 32678} +\CO(t^5). 
\ee
The first few terms are in perfect agreement with the calculation in planar perturbation theory (\ref{planarquarticg}). 

\begin{figure}[!ht]
\leavevmode
\begin{center}
\includegraphics[height=4.5cm]{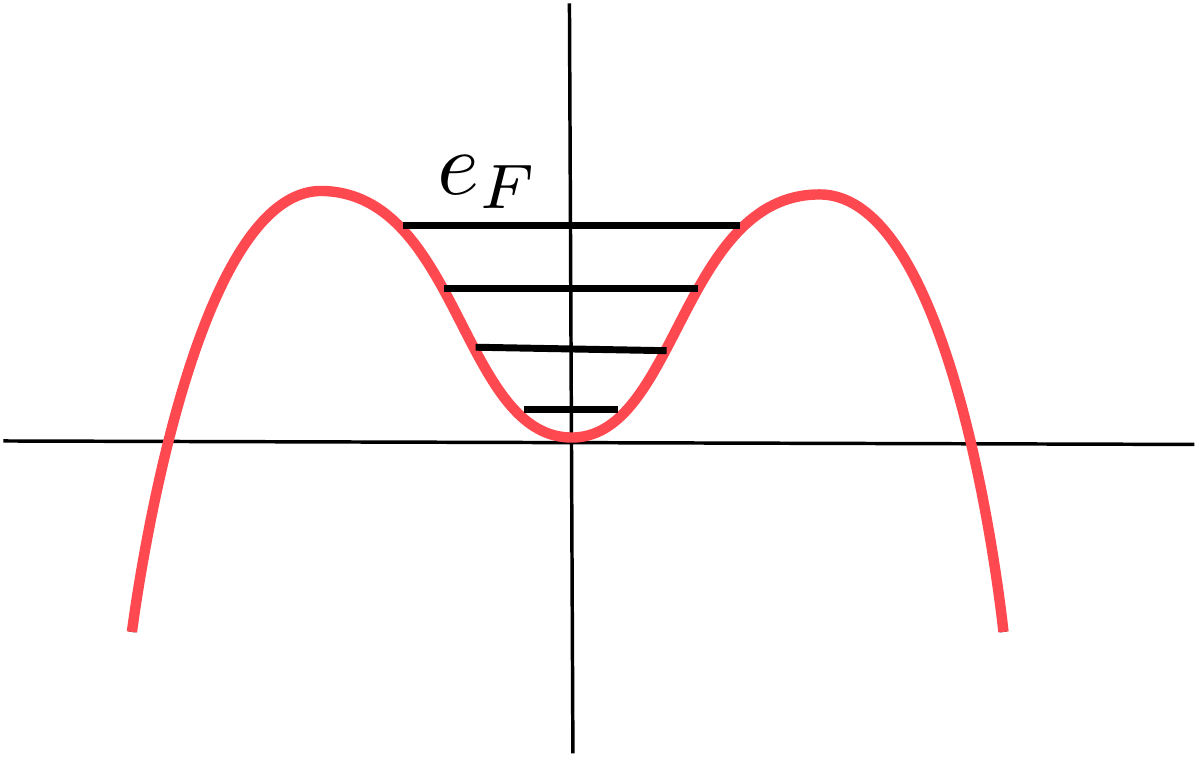}
\end{center}
\caption{The Fermi level $e_F$ in the quartic potential with negative coupling $t<0$. The nearest singularity corresponds to the critical value in which $e_F$ reaches the 
maximum of the potential.}
\label{quarticQM}
\end{figure}

One important remark on this result is that $\CE_0(t)$ is an analytic function of $t$ at $t=0$, in accordance with the general result 
explained above. This follows from the explicit expression for 
$\CE_0$ in terms of elliptic functions. The radius of convergence of the expansion (\ref{planarser}) can be calculated by locating 
the position of the nearest singularity $t_c$ in the $t$ plane. This singularity occurs when the modulus (\ref{qmodulus}) becomes $-\infty$, i.e. when 
\be
e_F (t_c)=-{1\over 4 t_c}. 
\ee
This is also the branch point in (\ref{abmqm}). It can be easily checked that this happens when 
\be
\label{conifold}
t_c=- {2 \sqrt{2} \over 3 \pi}.
\ee
This has a nice interpretation in terms of the 
fermion picture. Since $t_c$ is negative, it corresponds to an {\it inverted} quartic potential and to the precise value of the parameter at which the Fermi level reaches the maximum of the potential, see \figref{quarticQM}. 

We can now try to calculate large $N$ instanton effects in MQM. As in the case of standard QM, we consider an inverted quartic potential, 
\be
V(\lambda) ={1\over 2} \lambda^2 - {\kappa \over 4 N} \lambda^4, 
\ee
In this case, the vacuum at the origin is unstable and we should 
expect an instanton configuration mediating vacuum decay. In principle, one should write down an instanton solution with 
``small" action and calculate the 
path integral around it. This solution can be found by tunneling one single eigenvalue of the matrix $M$, which has an action of order $\CO(1)$, i.e. 
we consider the matrix instanton,  
\be
M_c(t) ={\rm diag}\left(0, \cdots, 0, q_c(t), 0, \cdots, 0\right), 
\ee
where $q_c(t)$ is the bounce (\ref{qcsaddle}) with coupling $\lambda=g_s=\kappa/N$. In principle, one could expand the path integral of MQM 
around this configuration and compute quantum planar fluctuations to determine the large 
$N$ instanton action (this calculation was originally proposed in \cite{neuberger}). 

However, the fermion picture, which gives us a compact way of computing the planar ground state energy, 
should also give us an efficient way to compute the large $N$ instanton action in a single strike. In this picture, the ground state 
is given by a filled Fermi level. As in any Fermi system, tunneling effects will first affect fermions which are near the Fermi surface. An instanton configuration with small action (i.e. of order $\CO(N^0)$) can then be obtained by tunneling a single fermion out of the $N$ particles in the Fermi gas. Since at large $N$ we can use semiclassical methods, the instanton action of such a fermion is 
just given by the standard WKB action, 
\be
\label{instaction}
{A(\kappa) \over g_s} = N \left( 2 \kappa \right)^{1/2} \int_a^b \rd \lambda \, \sqrt{(\lambda^2-a^2)(b^2 - \lambda^2)},
\ee
where $a,\, b$ are the turning points associated to the Fermi energy $e_F$, and they are non-trivial functions 
of the 't Hooft parameter $\kappa$. They are defined by the equations (\ref{abmqm}) with $t=-\kappa$ and an extra minus sign for $b^2$. There is an extra factor of $2$ due 
to the symmetry of the problem, and the factor of $N$ is due to the fact that the effective 
Planck constant in this problem is $1/N$, as we remarked in (\ref{hNeq}) (we are setting $\hbar=1$). 
The integral in (\ref{instaction}) can be explicitly computed by using elliptic functions, and the final result 
is 
\be
\label{mqmins}
A(\kappa)={1\over 3} (2\kappa^3)^{1/2} b  \Bigl[ (a^2+b^2) E(k)-2 a^2 K(k)\Bigr],
\ee
where the elliptic modulus is now given by 
\be
k^2={b^2-a^2 \over b^2}. 
\ee
The above function has the following expansion around $\kappa=0$,
\be
A(\kappa) ={4 \over 3} -\kappa \log\left( {16 e \over \kappa}\right) +{17\kappa^2 \over 16} + {125 \kappa^3 \over 128} +\cdots.
\ee
This is precisely the expected structure (\ref{exlna}) for a large $N$ instanton action: the leading term is the action for the instanton (\ref{bounceaction}) 
in the $N=1$ quantum mechanical problem, the $\log$ term is a one-loop factor in disguise, and the rest of the series is a sum of loop corrections 
in the background of the ``classical" instanton. An interesting 
property of $A(\kappa)$ is that it {\it vanishes} at the critical value 
\be
\kappa_c=-t_c={2 {\sqrt{2} \over 3 \pi}}, 
\ee
which is indeed the singularity (\ref{conifold}) in the complex $t$ plane signaling the convergence radius of the genus 
$g$ ground state energies. It corresponds to the critical point at which the Fermi sea reaches the local maximum; at this point the action for tunneling must indeed vanish 
since the endpoints in (\ref{instaction}) collide: $a(t_c)=b(t_c)$.

Although we have just computed here the leading order term in the $1/N$ expansion of the ground state energy, 
there is a full series of higher genus corrections to this result $\CE_g(t)$, $g\ge 1$, which play the r\^ole of the 
genus corrections to the free energy discussed above. One would expect that these corrections display the asymptotic 
behavior (\ref{largeg}), where $A(t)$ is the action of the large $N$ instanton calculated in (\ref{mqmins}). This was indeed verified in \cite{mpmqm}. 

In \cite{wadia2}, Wadia analyzed a closely related model which shares some properties with the quartic model with negative coupling 
considered above. The model studied in \cite{wadia2} is matrix quantum mechanics on a circle, with a cosine potential, and there are two phases separated by a 
third order phase transition which occurs when the Fermi level reaches the maximum of the potential. As shown in \cite{neuberger}, the action of the instanton 
mediating the tunneling vanishes at the transition point. The model is therefore similar to the one considered above, but due to the compactness of configuration space 
there is no instability: at the critical value of the 't Hooft parameter we rather have a reorganization of the Fermi sea.

\sectiono{Non-perturbative effects in matrix models}

\subsection{Matrix models at large $N$: General aspects}

In this subsection we review some basic facts about matrix models. For more detailed expositions and explanations, the reader is encouraged to look at \cite{dfgzj,mmleshouches,eoreview}. 

We will consider matrix models for an $N \times N$ Hermitian matrix $M$, with a potential $V(M)$. For the moment being 
we will assume that this is a polynomial potential, 
\be
V(\lambda)={1 \over 2}\lambda^2+  \sum_{p\ge 3} {g_p\over p} \lambda^p,
\ee
where the $g_p$ are coupling constants of the model. The partition function is defined by 
\be\label{matrix}
Z(N, g_s)={1 \over {\rm vol}(U(N))}\int \rd M\, \re^{-{1\over g_s}\tr V(M)},
\ee
where $g_s$ is an additional coupling constant, sometimes referred to as the string coupling constant. 
Matrix models have a $U(N)$ ``gauge" symmetry 
\be
M \rightarrow U M U^{\dagger},
\ee
therefore one can go to the ``diagonal gauge" and write this partition function in terms of the 
eigenvalues of $M$, denoted by $\lambda_i$ (this is very similar to the gauge (\ref{Mdiag}) in matrix quantum mechanics). 
The resulting $N$-dimensional integral is given by
\be
\label{zmm}
Z_{\gamma}(N, g_s)={1\over N!} \int_{\gamma} \prod_{i=1}^N {\rd \lambda_i \over 2\pi} \, \Delta^2(\lambda) \, \re^{-{1\over g_s} \sum_{i=1}^N V(\lambda_i)},
\ee
where $\Delta(\lambda)$ is the Vandermonde determinant introduced in (\ref{vander}). 
Here $\gamma$ is a contour in the complex plane, which we take to be the same for the $N$ eigenvalues, 
and which makes the integral convergent.  
Let us denote the critical points of $V(\lambda)$ by 
\be
\lambda^\star_1, \cdots, \lambda^\star_d. 
\ee
It is possible \cite{felder} to choose $d$ integration contours $\gamma_k$ in the complex plane, 
$k=1, \cdots, d$, going to infinity in directions where $\exp(-V(\lambda)/g_s)$ decays exponentially, and in such a way 
that each of them passes through exactly one of the $d$ critical points and is a steepest descent contour. In general, the contours depend on the argument of $g_s$, 
but we can always make a change of variables so as to reabsorb this phase in the coefficients of $V(\lambda)$. The original contour $\gamma$ can be written as a 
linear combination of such contours, 
\be
\gamma =\sum_{k=1}^d C_k \gamma_k. 
\ee
We can then write
\be
\label{sumz}
Z_{\gamma}(N, g_s) =\sum_{N_1 +\cdots +N_d=N} C_1^{N_1} \cdots C_d^{N_d} Z(N_1, \cdots, N_d).
\ee
In this formula, 
\be
\label{genz}
Z(N_1, \cdots, N_d)={1\over N_1! \cdots N_d!} \int_{\lambda^{(1)}_{i_1} \in \gamma_1} \cdots \int_{\lambda^{(d)}_{i_d} \in \gamma_d}  
\prod_{i=1}^N {\rd\lambda_i \over 2 \pi}\, \Delta^2(\lambda) \re^{-{1\over g_s} \sum_{i=1}^N 
V(\lambda_i)},
\ee
and we have split the eigenvalues in $d$ sets, 
\be
\{ \lambda^{(k)}_{i_k}\}_{i_k=1, \cdots, N_k}, \quad k=1, \cdots, d. 
\ee
Each integral (\ref{genz}) defines a possible ``background" or vacuum 
of the original matrix model. There are three important remarks to be made about the above expansion:

\begin{enumerate}

\item The backgrounds (\ref{genz}) of the matrix model are in one-to-one correspondence with 
gauge symmetry breaking patterns, 
\be
U(N) \rightarrow U(N_1) \times \cdots \times U(N_d). 
\ee
This is similar to the vacuum structure (\ref{lens-split}) of CS theory on lens spaces $L(d,1)$. We will denote these vacua by 
\be
(N_1, \cdots, N_d).
\ee

\item These sectors appear as an artifact of the saddle-point approximation. The original integral (\ref{zmm}) is perfectly well-defined, 
and it is only our will to treat the integral in this approximation which led to the appearance of these sectors. 

\item The expression (\ref{sumz}) is the analogue in random matrix theory of a {\it trans-series solution}. 

\end{enumerate}

A choice of background in the matrix model will define a choice of a perturbative sector. As we will see, the remaining sectors can then be regarded as instanton 
sectors w.r.t. the chosen background. A background of the form 
\be
(N, 0, \cdots, 0) 
\ee
is called a {\it one-cut background}, for reasons which will be clear in a moment, while the generic background will be called a {\it multi-cut background}.

Let us now choose a fixed background and let us consider the integral (\ref{genz}) on this background. It can be computed in a 
saddle-point expansion at small $g_s$ but keeping fixed the so-called {\it partial 't Hooft couplings}
\be
\label{partial-thooft}
t_i=g_s N_i.
\ee
This means that we are doing a large $N$ expansion of the matrix integral. The total 't Hooft parameter is
\be
t=g_s N=\sum_{i=1}^d N_i.
\ee
One can use the standard large $N$ counting arguments to see that the structure of the free energy is of the form
\be
\label{largeNas}
\log Z(N_1, \cdots, N_d) =\sum_{g=0}^{\infty} g_s^{2g-2} F_g(t_1, \cdots, t_d) . 
\ee
In the one-cut case $d=1$ this is the standard $1/N$ counting briefly reviewed in the previous section. 
In the multi-cut case one has to do a slightly refined analysis, see the 
Appendix in the published version of \cite{bde} for an explicit derivation. 

\begin{example} {\it The Gaussian matrix model}. The Gaussian matrix model is defined by the matrix integral (\ref{zmm}) with the potential 
\be
V(\lambda)={\lambda^2 \over 2}.
\ee
If $g_s>0$, the integration contour is simply $\gamma=\IR$, the real axis. We will denote the partition function of this model by $Z^{\rm G}(N,g_s)$. 
This is one of the few cases in which the matrix integral can be computed exactly at finite $N$, 
\be
\label{gaussianN}
Z^{\rm G}(N,g_s)= {g_s^{N^2/2} \over (2\pi)^{N/2}}\, G_2(N+1),
\ee
where $G_2(N+1)$ is the Barnes function
\be
G_2(N+1)=\prod_{i=0}^{N-1} i!.
\ee
In this case, the large $N$ expansion (\ref{largeNas}) follows from the asymptotics of this function. One finds, 
\be
\label{gaussiang}
\ba
F^{\rm G}_0(t)&={1\over 2} t^2  \Bigl( \log \, t -{3 \over 2} \Bigr), \\
F^{\rm G}_1(t)&=-{1\over 12} \log \, t +{1\over 12} \log \, g_s +\zeta'(-1),\\
F^{\rm G}_g(t)&= {B_{2g} \over
2g (2g-2)} t^{2-2g}, \quad g>1,
\ea
\ee
where $B_{2g}$ are Bernoulli numbers. Notice that $F^{\rm G}_1(t)$ depends also on $g_s$, but usually this piece (as well as the constant involving the zeta function) is not taken into account. 
\end{example}

The genus $g$ free energies $F_g(t_1, \cdots, t_d)$ appearing in (\ref{largeNas}) can be expanded around $t_i=0$. The first terms in this expansion are just the Gaussian 
free energies for the different $t_i$'s. Once these terms are subtracted, the resulting quantities
\be
\label{g-sub}
F_g(t_1, \cdots, t_d)-\sum_{i=1}^d F^{\rm G}_g(t_i) 
\ee
are {\it analytic} at the origin, in agreement with the general arguments put forward in the previous section. We will regard (\ref{largeNas}) as 
our perturbative expansion. It is interesting to note that this expansion 
is a generalization/deformation of the standard saddle-point expansion of one-dimensional integrals. 
Indeed, let us write (\ref{genz}) as 
\be
{1\over N_1! \cdots N_d!} \int_{\lambda^{(1)}_{i_1} \in \gamma_1} \cdots \int_{\lambda^{(d)}_{i_d} \in \gamma_d}  
\prod_{i=1}^N {\rd\lambda_i \over 2 \pi}\,   \exp\left\{-{1\over g_s} \left( \sum_{i=1}^N V(\lambda_i)-{t\over N} \sum_{i\not=j}\log \left(\lambda_i -\lambda_j\right)^2 \right) \right\}.
\ee
We want to take a limit where $g_s$ is small and $t$ is fixed, so that $N$ is large. The two terms inside the parenthesis are then of the same order (i.e. $\CO(N)$) and it is 
clear that the 't Hooft parameter controls the strength of the Vandermonde interaction. Let us suppose 
that $t$ is very small, so that we can neglect this interaction. In this limit (\ref{genz}) factorizes into a product of 
standard saddle-point integrals, 
\be\label{limitZ}
Z(N_1, \cdots, N_d) \to \prod_{i=1}^d \left( f_{\gamma_i}(g_s)\right)^{N_i}, \qquad t \to 0,  
\ee
where
\be
f_{\gamma_i}(g_s)=\int_{\gamma_i} {\rd \lambda \over 2 \pi} \re^{-{1\over g_s} V(\lambda)} \approx \re^{-{1\over g_s} V(\lambda^\star_i)}
\ee
and we approximated this integral by a saddle-point expansion around $\lambda^\star_i$. When $t$ is no longer small, 
we have to take into account the Vandermonde determinant. Since this induces a repulsion between eigenvalues, 
they will no longer sit at the saddle points of $V(\lambda)$: the $N_k$ eigenvalues in the $k$-th set will sit at an interval or arc $\CC_k$ around the $k$-th 
saddle-point. As we will see in a moment, when $N$ is large but the 't Hooft parameters $t_i$ are fixed, these arcs are compact, and the distribution of the eigenvalues on 
these arcs is given by a {\it density function} $\rho(\lambda)$, whose support is the union of the intervals. The problem of finding this equilibrium distribution can then be regarded as 
a deformation of the saddle-point technique for standard integrals, where the deformation parameters are the 't Hooft parameters $t_i$.

\subsection{The one-cut solution}

The determination of the free energies $F_g(t_i)$, for a given matrix model potential and background, 
has a long story, which starts in \cite{bipz} and culminates in \cite{eo}. We will now review some of the relevant results, 
which we will need to develop the instanton calculus in matrix models. 
To begin with, we will consider as our background a one-cut background. In this case, there is a single 
't Hooft parameter and the free energy has a perturbative genus expansion of the form
\be\label{oneovern}
F = \sum_{g=0}^{\infty} F_g(t)\, g_s^{2g-2}.
\ee
 Another important set of quantities in a matrix model are the connected correlation functions
\be\label{wcor}
W_h (p_1, \ldots, p_h) = \left\langle \tr\, {1\over p_1-M} \cdots \tr\, {1\over p_h-M} \right\rangle_{(\mathrm{c})}, 
\ee
\noindent
where the subscript $(\mathrm{c})$ means connected. These correlation functions are generating functions for multi--trace correlators of the form 
\be\label{scor}
W_h (p_1, \ldots, p_h) = \sum_{n_i \ge 1} \frac{1}{p_1^{n_1+1} \cdots p_h^{n_h+1}}\, \left\langle \tr\, M^{n_1} \cdots \tr\, M^{n_h} \right\rangle_{(\mathrm{c})},
\ee
\noindent
and they have a $g_s$ expansion of the form
\be
\label{wgh} 
W_h (p_1, \ldots, p_h) = \sum_{g=0}^{\infty} g_s^{2g+h-2} W_{g,h} (p_1, \ldots, p_h).
\ee

At large $N$, the one-cut background is characterized by a density of eigenvalues $\rho(\lambda)$ which 
has support on a single, connected interval $\CC=[a,b]$ in the complex plane. This density is completely determined by the condition 
that the so-called {\it effective potential} on an eigenvalue, 
\be\label{veff}
V_{\rm eff}(\lambda) = V(\lambda) - 2t \int {\rm d} \lambda'\, \rho(\lambda') \log |\lambda -\lambda'|,
\ee
\noindent
has to be {\it constant} --at fixed 't~Hooft coupling-- on the interval $\CC$:
\be\label{vconst}
V_{\rm eff}(\lambda) = t \xi(t), \qquad \lambda \in \CC. 
\ee
\noindent
A quantity which is closely related to the density of eigenvalues is the {\it planar resolvent}, which is nothing but the quantity $W_{0,1}(p)/t$, where $W_{0,1}(p)$ has been introduced in (\ref{wgh}). 
The planar resolvent can be computed as 
\be\label{zeroresint}
\omega_0(p) =\int \rd \lambda \,{\rho (\lambda)\over p -\lambda}.
\ee
It also satisfies the asymptotic condition 
\be
\label{asymres}
\omega_0(p) \sim {1 \over p}, \qquad p\rightarrow \infty,
\ee
which follows from the normalization of the density function 
\be
\int_\CC  \rd \lambda \, \rho(\lambda)=1. 
\ee
Once the resolvent is known, the eigenvalue density follows as
\be\label{rhow}
\rho(\lambda) = - {1 \over 2 \pi \ri} \bigl( \omega_0 (\lambda + \ri\epsilon) - \omega_0 (\lambda - \ri \epsilon) \bigr).
\ee
It can be seen that the condition (\ref{vconst}) determines the resolvent as
\be
\label{solwo}
\omega_0(p) ={1 \over 2t} \oint_{\cal C} {\rd z \over 2 \pi \ri} { V'(z) \over p-z} \biggl( { (p-a)(p-b)\over
(z -a) (z -b)}\biggr)^{1 \over 2} .
\ee
Here the integration is around a closed contour which encircles the cut $\CC$. By deforming the integration contour, this solution can also be written as, 
\be
\label{res-sc}
\omega_0(p) ={1\over 2t} \bigl( V'(p) - y(p) \bigr), 
\ee
where $y(p)$ is a function on the complex plane which has a branch cut along $\CC$, called the \textit{spectral curve} of the matrix model. In the one-cut case, the spectral curve has the structure
\be\label{scurve}
y(p) = M(p) {\sqrt{(p-a)(p-b)}}, 
\ee
\noindent
where $M(p)$, known as the \textit{moment function}, is given by 
\be\label{momentf}
M(p) = \oint_{\infty} {\rd z \over 2 \pi \ri}\, {V'(z) \over z-p}\, {1 \over{\sqrt{(z-a)(z-b)}}},
\ee
with the contour of integration being around the point at $\infty$. The endpoints of the cut follow from the asymptotic behavior of the resolvent (\ref{asymres}), 
leading to the equations
\be\label{endpo}
\ba
\oint_{\cal C} {\rd z \over 2\pi \ri}\, {V'(z) \over {\sqrt{(z-a)(z-b)}}}&=0, \\
\oint_{\cal C}{\rd z \over 2\pi \ri}\, {z V'(z) \over {\sqrt{(z-a)(z-b)}}}&=2t.
\ea
\ee
\noindent
%
%
%
%

It turns out that, with the exception of the free energies at genus zero and one, and the planar one-point function $W_{0,1}(p)$, 
the quantities $F_g(t)$ and $W_{g,h} (p_1, \ldots, p_h)$ 
can be computed in terms of the spectral curve alone. More precisely, knowledge of the endpoints of the cut, $a$ and $b$, and of the moment function, is all one needs in order to compute them. This was first made clear in \cite{bipz,ajm,ackm} and 
later culminated in the geometric formalism of \cite{eynard, ce,eo}. For example, the two--point correlator at genus zero is given by \cite{ajm}
\be\label{annmm}
W_{0,2} (p,q) = {1 \over 2 (p-q)^2} \left( {p q - {1\over 2} (p+q) (a+b) + ab \over {\sqrt{(p-a)(p-b)(q-a)(q-b)}}} - 1 \right),
\ee
and only depends on the end-points of the cut. On the other hand, the genus--zero free energy, $F_0(t)$ is given by
\be\label{planarf0}
F_0(t) = -{t\over 2} \int_{\mathcal C} {\rm d}\lambda\, \rho (\lambda) V (\lambda) - {1\over 2} t^2 \xi(t).
\ee
The derivatives of the planar free energy can be computed in terms of the effective potential. One finds (see for example \cite{iy})
\be\label{fzeroders}
\ba
\partial_t F_0 (t) &= - t \xi(t) = - V_{\mathrm{eff}} (b), \\
\partial_t^2 F_0 (t) &= - \partial_t V_{\mathrm{eff}} (b) = 2 \log { b-a \over 4}.
\ea
\ee

\subsection{The multi-cut solution}

We will now consider the more general (and more difficult) background: the multi-cut solution. In this case the support of the eigenvalue
distribution is a disjoint union of $n$ intervals
\be
{\cal C}=\bigcup_{i=1}^n \CC_i, \quad \CC_i= [x_{2i-1}, x_{2i}].
\ee
The density $\rho(\lambda)$ now satisfies the equation,
\be
\label{cpot}
V_{\rm eff}(\lambda)=\Gamma_i,  \qquad \lambda \in \CC_i,
\ee
where $\Gamma_i$ are constants in each interval $\CC_i$. This generalizes (\ref{vconst}) to the multi-cut case. 

The way to implement the multi--cut solution at the level of the planar resolvent is to require $\omega_0(p)$
to have $2n$ branch points. The solution now reads, 
\be
\label{solwmulti}
\omega_0(p) ={1 \over 2t} \oint_{\cal C} {\rd z \over 2 \pi \ri} { V'(z) \over p-z}
\left( \prod_{k=1}^{2n} { p-x_k\over
z-x_k}\right)^{1 /2}.
\ee
This can be also written as in (\ref{res-sc}), in terms of a spectral curve, which is now hyperelliptic, 
\be
y(p)=M(p) \prod_{k=1}^{2n} {\sqrt{p-x_k}}.
\ee
Here, $M(p)$ is the multi-cut moment function, which is given by the obvious generalization of (\ref{momentf}) to the multi-cut case. 
In order to satisfy the asymptotics (\ref{asymres}) the following conditions must hold:
\be
\label{splusone}
\delta_{\ell n}={1\over 2t} \oint_{\cal C} {\rd z \over 2 \pi \ri} {z^{\ell} V'(z)
\over \prod_{k=1}^{2n} (z-x_k)^{1\over 2}}, \qquad \ell=0,1, \cdots, n.
\ee
In contrast to the one-cut case, these are only $n+1$ conditions for the $2n$ variables
$x_k$ representing the endpoints of the cut. The remaining $n-1$ conditions are obtained by fixing the values of the partial 't Hooft parameters:
\be
t_i=g_s \int_{\CC_i} \rd\lambda \, \rho(\lambda). 
\ee
The planar free energy satisfies in addition the equation, 
\be
\label{pdF}
{\partial F_0 \over \partial t_i}-{\partial F_0 \over \partial t_{i+1}} =\Gamma_{i+1}-\Gamma_i, 
\ee
where $\Gamma_i$ are the quantities appearing in (\ref{cpot}).

\FIGURE{
\includegraphics[height=3cm]{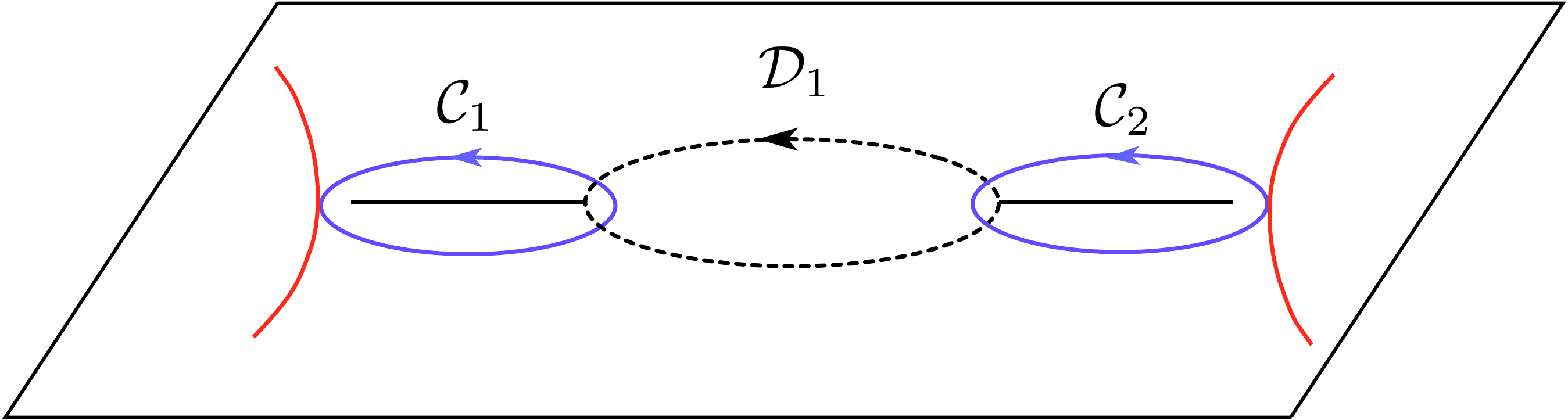}
\caption{A two-cut spectral curve, showing two contours $\CC_{1,2}$ around the cuts where $N_{1,2}$ eigenvalues sit. The ``dual" cycle ${\cal D}_1$ goes from $\CC_2$ to $\CC_1$.}
\label{twocuts}
}

We can write the multi-cut solution in a very elegant way by using contour integrals of the spectral curve. 
First, the partial 't Hooft parameters are given by
\be
\label{tper2}
t_i=-{1\over 4 \pi \ri} \oint_{\CC_i} y(p)\rd p. 
\ee
We now introduce dual cycles ${\cal D}_i$ cycles, $i=1, \cdots, n-1$, going from the $\CC_{i+1}$ cycle to the $\CC_{i}$ cycle counterclockwise, see \figref{twocuts}. In terms of these, we can write (\ref{pdF}) as
\be
\label{dfy}
{\partial F_0 \over \partial t_i}-{\partial F_0 \over \partial t_{i+1}} ={1\over 2} \oint_{{\cal D}_i} y(p) \rd p.
\ee

\subsection{Large $N$ instantons and eigenvalue tunneling}

In the sum (\ref{genz}) over possible ``vacua" of the matrix model, any two vacua are related by a redistribution of the eigenvalues, i.e. by a sequence of operations of the form 
\be
\label{twovac}
N_i \rightarrow N_i-1, \qquad N_j \rightarrow N_j+1.  
\ee
This can be interpreted as a process in which an eigenvalue leaves the cut around the $i$-th critical point and ``tunnels" to the cut around the $j$-th critical point. 
Therefore, given a choice of background or perturbative vacuum, all the other vacua in (\ref{genz}) (which are then regarded as instanton sectors) can be obtained from the reference background by {\it eigenvalue tunneling}. The fact that instantons in the matrix model are due to eigenvalue tunneling was first pointed out in \cite{david, shenker}. 

\begin{figure}[!ht]
\leavevmode
\begin{center}
\includegraphics[height=3cm]{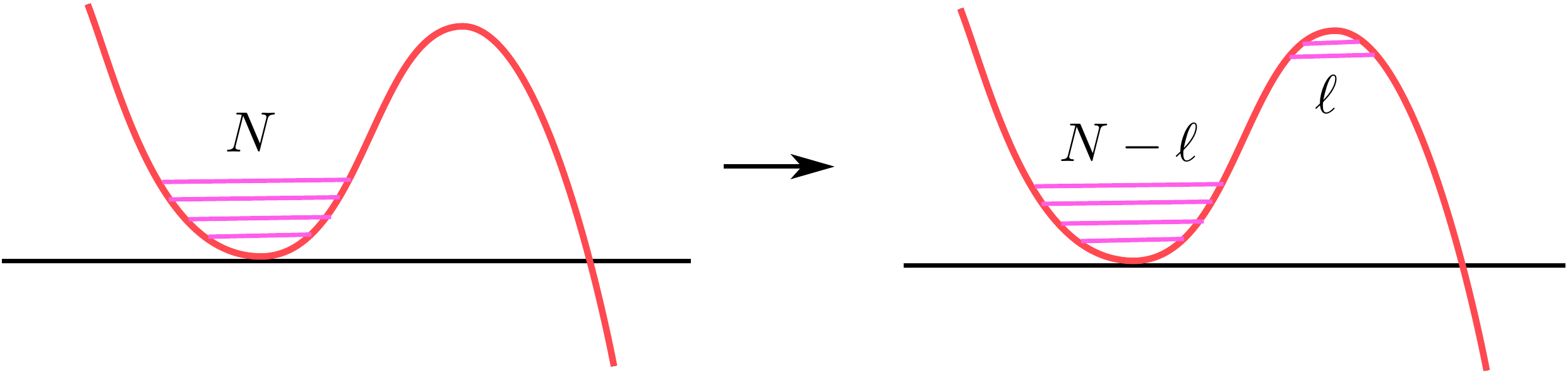} 
\end{center}
\caption{Eigenvalue tunneling in a cubic matrix model.}
\label{tunnel}
\end{figure}

Notice that, when an eigenvalue tunnels as in (\ref{twovac}), the r.h.s. of (\ref{limitZ}) changes, at leading order, by
\be
\label{camm}
 \exp\left( -{1\over g_s} (V(\lambda^\star_j) -V(\lambda^\star_i)) \right).
\ee
We can think about this as the action of the classical instanton connecting the two vacua:
\be
(\cdots, N_i, \cdots, N_j, \cdots) \to (\cdots, N_i-1, \cdots, N_j+1, \cdots).
\ee
As we have explained in the previous section, there is a large $N$ instanton built upon this classical configuration. 
The classical action (\ref{camm}) gets corrected to 
\be
 \exp\left( -{1\over g_s} (V_{\rm eff} (\lambda^\star_j) -V_{\rm eff}(\lambda^\star_i)) \right),
\ee
where the effective potential is defined in (\ref{veff}). This can be regarded as the action of the large $N$ instanton 
connecting the vacua. It follows from (\ref{pdF}) and (\ref{cpot}) that the large $N$ instanton action can be computed in terms of the genus zero free energy as
\be
V_{\rm eff} (\lambda^\star_j) -V_{\rm eff}(\lambda^\star_i)={\partial F_0 \over \partial t_i} -{\partial F_0 \over \partial t_j}. 
\ee
When $t\to 0$, we recover the classical instanton action (\ref{camm}). 
The picture of eigenvalue tunneling is still valid for finite $t$, but now the tunneling is between 
two cuts rather than between two saddle points. A graphical depiction of an $\ell$-eigenvalue tunneling in a cubic potential, 
from the background $(N,0)$ to the background $(N-\ell, \ell)$, is shown in \figref{tunnel}.

\subsection{Large $N$ instantons in the one-cut matrix model}

A particularly simple case of eigenvalue tunneling 
occurs when the reference background is the one-cut configuration. 
The elementary instanton configurations correspond in this case to tunnelings of the form
\be
(N, 0, \cdots, 0) \rightarrow (N-1, 0, \cdots, 0, 1, 0, \cdots, 0).  
\ee
The large $N$ instanton action depends then only on $t=g_s N$ and it has the structure discussed in the 
previous section. Since this is the simplest case, we will now explain 
how to calculate instanton contributions when the reference background is a one-cut solution. It turns out that this computation can be done in 
at least three different ways. The first method is based on a direct calculation of the matrix integral. 
It was introduced by F. David in \cite{david} and further clarified in \cite{lvm,iy}. The calculation of \cite{david,lvm,iy} is done for the so-called double-scaled 
matrix model, in which the 't Hooft parameter is near a critical value. The correct calculation of the instanton contribution at generic values of the 't Hooft parameter 
was presented in \cite{mswone}. There is a second method, which regards the instanton sectors in the one-cut matrix model as limits of the generic multi-cut vacuum. This method has the advantage of giving general expressions for the $k$-th instanton sector, and it was presented in \cite{mswtwo}. Finally, a third method was introduced in \cite{mmnp}, based on the method of orthogonal polynomials, which makes it possible to calculate the instanton sectors as trans-series solutions to difference equations, and therefore it mimicks the structure 
developed for ODEs. We will present the three methods in turn, skipping some of the details which can be found in the original references. 

\subsubsection{A direct calculation}
\label{direct-cal}
\begin{figure}[!ht]
\leavevmode
\begin{center}
\includegraphics[height=5cm]{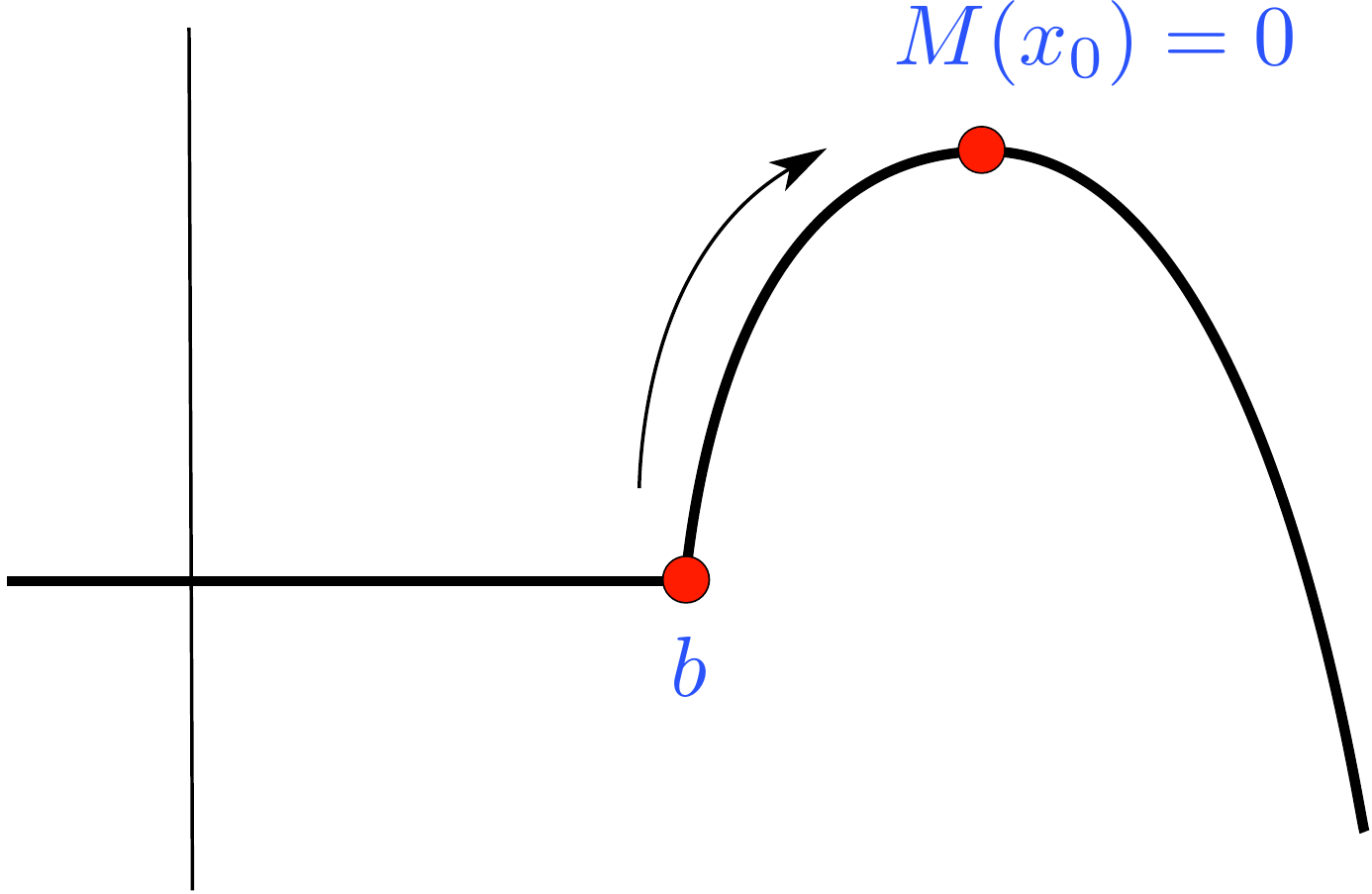}
\end{center}
\caption{The effective potential for the one-cut matrix model. The point $x_0$ is a critical point where eigenvalues can tunnel to, leading to instanton configurations.}
\label{mmeffpot}
\end{figure} 

We thus consider a one--cut matrix model in which the effective potential has the form depicted in \figref{mmeffpot}: it is constant along the location of the cut $\CC=[a,b]$, and there is a critical point $x_0$ which corresponds to another possible vacuum. We will consider the instanton sectors corresponding to eigenvalue tunneling from the cut at $\CC$ to the critical point $x_0$. These instanton sectors are labelled by a positive integer $\ell$, which is the number of tunneling eigenvalues. The total partition function, summing over all these instanton sectors, is given by 
\be
\label{z-trans}
Z(N)= Z^{(0)}(N) \left(1 + \sum_{\ell =1}^\infty C^{\ell} Z^{(\ell)}(N) \right),
\ee
where for convenience we have implicitly normalized the sectors with $\ell\not=0$ by the perturbative, one-cut partition function $Z^{(0)}(N)$, and $C$ is a constant. Notice that this total partition function has the trans-series structure considered in section \ref{ODEs}. The free energy also has a trans-series structure, 
\be
\label{f-trans}
F(N) = F^{(0)}(N) + \log\left(1 + \sum_{\ell =1}^\infty C^{\ell} Z^{(\ell)}(N)\right)=\sum_{\ell =0}^\infty C^{\ell} F^{(\ell)}(N).
\ee

The instanton partition functions $Z^{(\ell)}(N)$ can be computed directly in terms of the defining matrix integral: we consider an integral in which $N-\ell$ eigenvalues are located in the cut $\CC$, while for the $\ell$ eigenvalues which have tunneled, the integration is made around a steepest-descent contour passing through the critical point. In particular, for the one-instanton sector one finds \cite{lvm}
\be\label{onemm}
Z^{(1)}(N) = {N \over N! (2\pi)^N Z^{(0)}(N)} \int_{x\in \CI} \rd x \, \re^{-{1\over g_s}  V (x)}
 \int_{\lambda \in \CI_0}  \prod_{i=1}^{N-1}\rd \lambda_i\, \Delta^2 (x, \lambda_1, \ldots, \lambda_{N-1})\, \re^{-{1\over g_s} \sum_{i=1}^{N-1} V (\lambda_i)},
\ee
\noindent
where the first integral in $x$ is around the critical point at $x_0$, along a saddle-point contour which we have denoted by $\CI$. The rest of the $N-1$ eigenvalues are integrated around the contour $\CI_0$ corresponding to the background one-cut configuration. The overall factor of $N$ in front of the integral is a symmetry factor, counting the $N$ possible distinct ways of choosing one eigenvalue out of a set of $N$. One can easily write similar integrals for the $\ell$--instanton contribution, but these will be easier to calculate with the second method below. The integral (\ref{onemm}) can be written as 
\be\label{oneinstex}
Z^{(1)}(N) = {1\over 2\pi}\, {Z^{(0)}(N-1) \over Z^{(0)}(N)} \int_{x \in \CI} \rd x\, f(x),
\ee
where
\begin{equation}\label{fx}
f(x) = \left\langle \det (x {\bf 1} - M')^2 \right\rangle^{(0)}_{N-1}\, \re^{-{1\over g_s} V(x)}.
\end{equation}
The notation in these equations is as follows. The average is defined as 
\be
\left\langle \CO \right\rangle^{(0)}_N = {\int_{\lambda\in \CI_0} \prod_{i=1}^N \rd \lambda_i\, \Delta^2(\lambda)\, \CO(\lambda)\, \re^{-{1\over g_s} \sum_{i=1}^N V(\lambda_i)} \over \int_{\lambda\in \CI_0} \prod_{i=1}^N \rd \lambda_i\, \Delta^2(\lambda)\, \re^{-{1\over g_s} \sum_{i=1}^N V(\lambda_i)}}
\ee
and it is calculated again in the one-cut bacgrkound. In (\ref{fx}), $M'$ is an $(N-1) \times (N-1)$ hermitian matrix. 
We conclude that the one-instanton sector can be evaluated by calculating correlators in the perturbative sector. 
In fact, by making use of the familiar relation 
\be
\det (x {\bf 1} - M) = \exp\left({\rm tr}\,{\rm ln} (x {\bf 1}-M)\right)
\ee
we obtain 
\be\label{conex}
\left\langle \det (x {\bf 1} - M)^2 \right\rangle = \exp \left[ \sum_{s=1}^{\infty} {2^s\over s!} \left\langle \left( {\rm tr}\, {\rm ln} (x {\bf 1} - M) \right)^s \right\rangle_{(\rm c)} \right],
\end{equation}
\noindent
which is written in terms of connected correlation functions. The correlation functions appearing in (\ref{conex}) are nothing but integrated versions of the $W_{h}$ correlators in (\ref{wcor}), evaluated at coincident points. Let us define
\be\label{theas}
\ba
A_{g,h} (x;t) &= \left. \int^{x_1} \rd p_1 \cdots \int^{x_h} \rd p_h\, W_{g,h} (p_1,\cdots, p_h) \right|_{x_1=\cdots =x_h=x}, \\
\CA_n(x;t) &= \sum_{k=0}^{\left[\frac{n}{2}\right]} \frac{2^{n-2k+1}}{(n-2k+1)!}\, A_{k,n-2k+1} (x;t), \quad n\ge 1.
\ea
\ee
\noindent
In this notation, the general perturbative formula for the determinant reads
\be
\left\langle \det ( x \mathbf{1} - M )^2 \right\rangle = \exp \left( \sum_{n=0}^{\infty} g_s^{n-1} \CA_n (x;t) \right),
\ee
\noindent
where $\CA_n(x;t)$ is the $n$--loop contribution. We have, for example,  
\be
\CA_0(x;t) = 2 A_{0,1} (x;t), \qquad 
\CA_1(x;t) = 2 A_{0,2} (x;t).
\ee
The integration constants involved in the integrations in (\ref{theas}) may be simply fixed by looking at the large $x$ expansion of the correlators. 
Next, we define the \textit{holomorphic} effective potential, which combines the matrix model potential together with $\CA_0(x;t)$, as
\be\label{vheff}
V_{\rm h,eff}(x;t) = V(x) - 2t \int^x \rd p\, \omega_0 (p) = V(x) - 2t \int \rd p\, \rho(p) \log (x -p). 
\ee
It satisfies
\be\label{dery}
V_{\rm h,eff}' (x;t) = y(x)
\ee
\noindent
as well as
\be
{\rm Re}\, V_{\rm h,eff}(x;t) = V_{\rm eff}(x),
\ee
\noindent
where $V_{\rm eff}(x)$ was earlier defined in (\ref{veff}). Altogether, one finally has for the integrand
\be\label{fint}
f(x) = \exp \left( - \frac{1}{g_s} V_{\rm h,eff} (x;t') + \sum_{n=1}^{\infty} g_s^{n-1} \CA_n (x;t') \right),
\ee
\noindent
where
\be
t'=g_s(N-1)=t-g_s.
\ee
\noindent
This shift in the 't~Hooft parameter is due to the fact that the correlation function involved in (\ref{fx}) is computed in a matrix model with $N-1$ 
eigenvalues (recall we removed one eigenvalue from the cut). Since we are computing the one--instanton contribution in the theory with $N$ eigenvalues, we thus have to expand (\ref{fint}) around $t$. This gives further corrections in $g_s$, and one finds 
\be\label{fvphi}
f(x) = \exp \left( - \frac{1}{g_s} V_{\mathrm{h, eff}} (x) + \Phi (x) \right),
\ee
with
\be
\Phi (x) \equiv \sum_{n=1}^{\infty} g_s^{n-1}\, \Phi_{n} (x)= \sum_{n=1}^{\infty} g_s^{n-1} \left[ \frac{(-1)^{n-1}}{n!}\, \partial_t^n V_{\mathrm{h,eff}} (x) + \sum_{k=0}^{n-1} \frac{(-1)^k}{k!}\, \partial_t^k \CA_{n-k} (x) \right].
\ee
One has, for example,
\be
\label{phi-one}
\Phi_1 (x) = \CA_1 (x) + \partial_t V_{\mathrm{h,eff}} (x).
\ee
Further explicit results for these quantities can be found in \cite{mswone}. In the expression (\ref{fvphi}) all quantities now depend on the standard 't~Hooft parameter $t$ for 
the model with $N$ eigenvalues, and we have thus dropped the explicit dependence on $t$. The derivatives with respect to $t$ can be performed by using standard 
results in one-cut matrix models, some of which were listed above. One may now proceed with the integration of $f(x)$, 
\be\label{fxint}
\int_{x \in \CI} \rd x\, \exp \left( - \frac{1}{g_s} V_{\mathrm{h,eff}} (x) + \Phi (x) \right).
\ee
\noindent
If we wish to evaluate this integral as a perturbative expansion when $g_s$ is small, we can do it using a saddle--point evaluation \cite{lvm,iy}. The saddle--point condition
\be
V'_{\mathrm{h,eff}} (x_0) = 0 
\ee
requires $x_0$ to be a critical point of the effective potential, as anticipated above. 
If we use the explicit form of the spectral curve (\ref{scurve}) we find the equivalent condition

\begin{figure}[!ht]
\leavevmode
\begin{center}
\includegraphics[height=4.5cm]{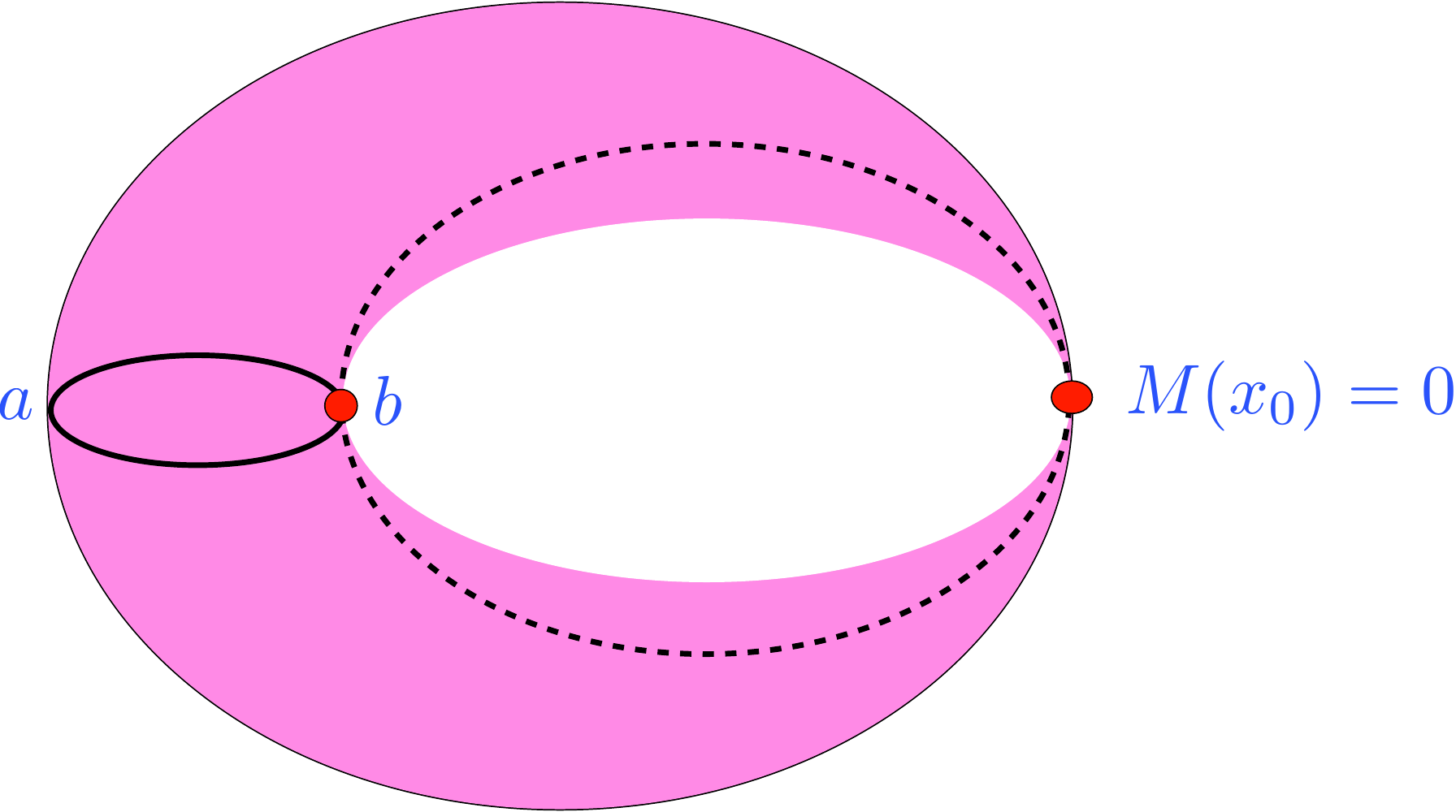}
\end{center}
\caption{The spectral curve $y(x)$ has a singular point at the nontrivial saddle $x_0$.}
\label{riemanninst}
\end{figure}
\be\label{mx}
M(x_0) = 0.
\ee
Geometrically, the spectral curve is a curve of genus zero pinched at $x_0$, as shown in \figref{riemanninst}. This was observed in \cite{seishi} in the context of spectral curves for double--scaled matrix models, and their relation with minimal strings (see also \cite{kk}). Of course, it can happen that there are many different critical points of the effective potential, therefore more than one solution to (\ref{mx}). In this case, there will be various instantons and we will have to add up their contributions (the leading contribution arising from the instanton with the smallest action, in absolute value). 

The calculation of (\ref{fxint}) is now completely standard, and it reduces to Gaussian integrations. The result is
\be\label{fexpansion}
\int_{x\in \CI} \rd x\, f(x) =  \sqrt{\frac{2 \pi g_s}{V''_{\mathrm{h,eff}} (x_0)}}\, \exp \left( - \frac{1}{g_s} V_{\mathrm{h,eff}} (x_0) + \Phi_1 (x_0) \right) \left( 1 + \sum_{n=2}^{\infty} g_s^n\, f_n \right), 
\ee
\noindent
where the $f_n$ can be systematically computed in terms of the functions $\Phi_n(x)$ and their derivatives, evaluated at the saddle--point $x_0$. An explicit expression for $f_2$ can be found in \cite{mswone}. 

With these ingredients, we can already calculate the one-instanton contribution to the free energy. Using (\ref{f-trans}), we find
\be
F^{(1)}(N)=Z^{(1)}(N)={1\over 2\pi} \frac{Z^{(0)}(N-1)}{Z^{(0)}(N)}  \int_{x\in \CI} \rd x\, f(x). 
\ee
The quotient of perturbative partition functions is easy to calculate in terms of the free energies, 
\be
\frac{Z^{(0)}(N-1)}{Z^{(0)}(N)} = \exp \left( F(t')-F(t) \right)= \exp \left( \sum_{n=0}^{\infty} g_s^{n-1} \CG_n \right), 
\ee
where
\be\label{quotex}
\CG_n \equiv \sum_{k=0}^{\left[ \frac{n}{2} \right]} \frac{(-1)^{n-2k+1}}{(n-2k+1)!}\, \partial_t^{n-2k+1} F_k (t).
\ee
\noindent
One has, for example,
\be\label{gex}
\CG_0 =- \partial_t F_0 (t), \qquad 
\CG_1 = \frac{1}{2}\, \partial_t^2 F_0 (t).
\ee
\noindent
Putting together (\ref{fexpansion}) and (\ref{quotex}) above, we finally find that $F^{(1)}$ has the structure
\be\label{muex}
F^{(1)} = \ri\, g_s^{1/2}\,  F_{1,1}\, \exp \left( -\frac{A}{g_s} \right) \left\{ 1 + \sum_{n=1}^{\infty}F_{1,n+1} g_s^n \right\}.
\ee
where, up to one loop, we have:
\be
\ba
A &= V_{\mathrm{h,eff}} (x_0) - \CG_0 (t), \\
F_{1,1} &= -\ri\, \sqrt{\frac{ 1}{2 \pi V''_{\mathrm{h,eff}} (x_0)}}\, \exp \Big( \Phi_1 (x_0)  + \CG_1(t) \Big). 
\ea
\ee

We can give explicit expressions for all the quantities involved in (\ref{muex}) in terms of data which depend only on the spectral curve (\ref{scurve}). First of all, by using (\ref{gex}), (\ref{fzeroders}) and (\ref{dery}) we find
\be\label{instdiff}
A = V_{\rm h,eff}(x_0) - V_{\rm h,eff}(b) = \int_b^{x_0} \rd z \, y(z), 
\ee
\noindent
which is the instanton action (here, we use the fact that $V_{\rm h,eff}(b)=V_{\rm eff}(b)$). As pointed out in \cite{seishi}, this expression also has a geometric interpretation as the contour integral of the one--form $y(z)\, \rd z$, from the endpoint of the cut $\CC$ to the singular point $x_0$ (see \figref{riemanninst} and \figref{curves}, left side). To compute $F^{(1)}$ up to one--loop, we must compute $\Phi_1(x)$, given in (\ref{phi-one}). One can find the result for $A_{0,2}(x;t)$ (which enters in the expression of $\CA_1$) simply by integrating the first formula in (\ref{annmm}) \cite{iy}
\be\label{vann}
A_{0,2}(x;t) = \log \biggl( 1 + \frac{x - (a+b)/2} {\sqrt{(x - a)(x - b)}} \biggr) - \log 2.
\ee
\noindent
Using that (see for example \cite{dfgzj})
\be\label{derom}
{\partial (t \omega_0(p)) \over \partial t}={1\over {\sqrt {(p-a)(p-b)}}}, 
\ee
one further finds, 
\be\label{dervheff}
\partial_t V_{\rm h,eff}(x) = -4 \log \Bigl[ {\sqrt{x-a}} + {\sqrt{x-b}} \Bigr] +4 \log 2,
\ee
and one then obtains
\be
\Phi_1(x) = - \log \Bigl[ (x-a)(x-b) \Bigr].
\ee
\noindent
Adding to $\Phi_1(x)$ the result for $\CG_1(t)$, which follows from (\ref{fzeroders}), it is simple to put all expressions together 
and obtain the contribution, $F_{1,1}$, of the one--loop fluctuations around the one--instanton configuration, 
\be\label{rone}
F_{1,1} = -\ri\, {b-a \over 4} {\sqrt{1 \over  2 \pi M'(x_0) \Bigl[(x_0-a)(x_0-b)\Bigr]^{5\over 2}}}.
\ee
\noindent
This formula is valid for any one--cut matrix model where the potential has an extra critical point $x_0$. Notice that if $x_0$ is a local maximum of $V_{\rm eff}(x)$, one will have that $M'(x_0)<0$, and hence $F_{1,1}$ will be real. This result extends previous calculations in \cite{david,lvm} to one-cut matrix models away from the critical point.

\begin{example} {\it Large $N$ instanton in the quartic matrix model}. The quartic matrix model is defined by the potential 
\be
\label{quarticpot}
V(x) = {1\over 2} x^2 -{\lambda\over 48} x^4,
\ee
where we follow the normalization of \cite{mmnp}. This potential has three critical values, namely 
\be
\label{critical}
x=0, \qquad x=\pm  {2 {\sqrt{3}} \over {\sqrt{\lambda}}}, 
\ee
so the generic matrix model based on this potential has three cuts. We can however consider 
the one-cut model where the eigenvalues sit around the origin $x=0$. The spectral curve for this one-cut background can be easily obtained 
with the standard matrix model techniques reviewed above. It is given by
\be
y=M(x) {\sqrt {x^2-4 \alpha^2}}
\ee
where the moment function is
\be
M(x)=1-{\lambda \over 6} \alpha^2 - {\lambda \over 12} x^2.
\ee
The cut is located at $\CC=[-2\alpha, 2\alpha]$, and the endpoints are determined by
\be
\label{asq}
\alpha^2={2\over \lambda} \biggl( 1- {\sqrt { 1 - t \lambda}}\biggr).
\ee
The moment function has two zeros which give two non--trivial saddle--points of the effective potential, namely $\pm x_0$ with 
\be\label{qsaddle}
x_0^2 ={12 \over \lambda} -2 \alpha^2.
\ee
As $t \rightarrow 0$, $\alpha \rightarrow 0$ and they become the two non-trivial critical points of the potential in (\ref{critical}). 
Since the potential is symmetric, there are {\it two} one-instanton solutions, corresponding to one eigenvalue tunneling from $\CC$ to the two saddles $\pm x_0$. Both instantons have the same action, which is computed by integrating the spectral curve as in (\ref{instdiff}), 
\be
\label{quartic-inst}
\ba
A(t)&= {3\over \lambda }  \sqrt{\left(1-\frac{\alpha ^2 \lambda }{2}\right)
   \left(1-\frac{\alpha ^2 \lambda }{6}\right)}\\
   &-\frac{1}{4} \alpha ^2
     \left(\alpha ^2 \lambda -4\right) \left(\log
   \left(\frac{\alpha ^2 \lambda }{3}\right)-2 \log
   \left(\sqrt{1-\frac{\alpha ^2 \lambda }{2}}+\sqrt{1-\frac{\alpha ^2
   \lambda }{6}}\right)\right).
 \ea
\ee
This function has the small $t$ expansion
\be
A(t)={3\over  \lambda}+t \left( -1+ \log \left( {\lambda t \over 12} \right) \right) +\CO(t^2).
\ee
The first term (i.e. the instanton action as $t\rightarrow 0$) is given by 
\be
V\left( {2 {\sqrt{3}}  \over {\sqrt{\lambda}}} \right) -V(0), 
\ee
in agreement with (\ref{camm}). The structure of the next-to-leading correction is in agreement with the general expectation (\ref{exlna}). 
\end{example}

\begin{example} {\it Large $N$ instanton in the cubic matrix model}. Let us consider the matrix model analogue of the Airy integral (\ref{lamint}). The potential is 
\be
\label{stokespot}
 V(x)= -\re^{\ri \kappa} x + {x^3 \over 3}.
\ee
This has two critical points 
\be
x^{{\rm L}, {\rm R}}=\mp \zeta^{1/2}
\ee
where we have introduced the variable
\be
\zeta=\re^{\ri \kappa}.
\ee
The most general background of a matrix model based on this potential 
is a two-cut configuration, labelled by $(N_1, N_2)$, where $N_1$, $N_2$ are the number of eigenvalues near the critical points $x^{{\rm L}, {\rm R}}$, respectively.
Let us consider the one-cut background where all the eigenvalues sit near $x^{\rm L}$, i.e. 
\be
(N_1, N_2)=(N,0). 
\ee
The endpoints of the cut $(a,b)$ are determined by the equations
\be
x_0(x_0^2-\zeta)=t, \qquad \delta^2=2(\zeta-x_0^2), 
\ee
where
\be
a=-x_0+\delta, \qquad b=-x_0-\delta.
\ee
The spectral curve is given by
\be
y(x) =(x-x_0) {\sqrt{x^2+2xx_0 + 3x_0^2-2\zeta}}.
\ee
The effective potential is (up to a constant) the integral of $y(x)$, which gives
\be
\ba
V_{\rm eff}(x)&=\frac{1}{3}\left( x (x -x_0) -2 \zeta\right) {\sqrt{x^2+2xx_0 + 3x_0^2-2\zeta}} \\
&-2 x_0 \left(x_0^2-\zeta \right) \log \left(x_0+x+{\sqrt{x^2+2xx_0 + 3x_0^2-2\zeta}}\right).
\ea
  \ee
The instanton action is given by 
   \be
   A=\int_b^{x_0} y(x) \rd x =V_{\rm eff}(x_0)-V_{\rm eff}(b).
   \ee
It has the small $t$ expansion
\be
A=-{4\over 3} \zeta^{3/2}+t \left( -1+ \log {t\over 8 \zeta^{3/2}} \right) +\CO(t^2), 
\ee
and as expected the leading term as $t\rightarrow 0$ is given by (\ref{camm}), 
\be
-{4\over 3}\zeta^{3/2} =V(x^{\rm R})-V(x^{\rm L}).
 \ee
\end{example}

\subsubsection{Large $N$ instantons from multi-cuts}
\label{multi-multi}

It is clear that the instanton configurations in the one-cut matrix model that we have just analyzed, in which $\ell$ eigenvalues tunnel to other 
critical point, can be regarded as particular cases of a {\it two-cut} configuration with
\be
N_1=N-\ell, \qquad N_2=\ell, \qquad \ell\ll N.
\ee
Therefore, we should be able to calculate the general instanton partition function $Z_N^{(\ell)}$ as a particular case of a two-cut partition function, i.e. we expect
\be
\label{onefromtwo}
Z^{(\ell)}(N)= Z(N-\ell, \ell).  
\ee
This was made explicit in \cite{mswtwo} and gives compact formulae for the general $\ell$-instanton sector of the one-cut matrix model. 
We can evaluate the r.h.s. of (\ref{onefromtwo}) by expanding the free energy of the two-cut matrix model 
\be
F\left(t_1=t-g_s\ell, t_2=g_s \ell \right)
\ee
in powers of $g_s$. There is however one subtlety: the free energies $F_g(t_1, t_2)$ are \textit{not} analytic at $t_2=0$. 
Geometrically, this corresponds to the fact that we are expanding around a configuration in which the second cut of the 
spectral curve is completely pinched, as shown in \figref{riemanninst}. However, as we mentioned in (\ref{g-sub}), 
this non-analyticity is due only to the Gaussian contribution to the two-cut integral. 
Therefore, the functions $\widehat F_g (t_1, t_2)$ defined by
\be
F_g(t_1, t_2) = F^{\rm G}_g(t_2) + \widehat F_g(t_1, t_2), 
\ee
\noindent
where $F^{\rm G}_g(t)$ are the genus $g$ Gaussian free energies written down in (\ref{gaussiang}), are analytic at $t_2=0$. 

Physically, the reason for the appearance of this singularity is that in this problem $t_2=\ell g_s$, and $\ell$ is small as compared to $N$. 
Therefore, it is not appropriate to treat the integration over the $\ell$ tunneling eigenvalues from the point of view of the large $N$ 
expansion. Instead, they should be integrated exactly. This argument also suggests that, in order to regularize the computation, 
we should subtract $F^{\rm G}(t_2)$ from the total free energy and at the same time multiply $Z^{(\ell)}$ by the 
{\it exact} partition function $Z^{\rm G}_\ell$, which is given in (\ref{gaussianN}). The appropriate expression 
for the partition function around the $\ell$--instanton configuration is then
\be\label{interzn}
Z^{(\ell)} = Z^{\rm G}_\ell\, \exp \biggl[ \sum_{g\ge 0} g_s^{2g-2} \left( \widehat F_g (t-\ell g_s, \ell g_s) - F_g (t) \right)\biggr].
\ee
By expanding in $g_s$, this expression leads to the general formula
\be
\label{npboundary}
\ba
Z^{(\ell)}_N=& {g_s^{\ell^2/2} \over (2\pi)^{\ell/2}}\, G_2(\ell+1)\, \, \hat q^{\frac{\ell^2}{2}}\, \exp \left( - \frac{\ell  A}{g_s} \right) \\
& \times \sum_{k} \sum_{m_i>0}\sum_{g_i>1-{m_i\over 2}} {g_s^{\sum_i (2g_i+m_i-2)}\over k! m_1!\,\dots\, m_k!}\,\,\, 
\widehat F_{g_1}^{(m_1)}\dots \widehat F_{g_k}^{(m_k)}  (-\ell)^{\sum_i l_i}.
\ea
\ee
In this equation we have introduced the following notations: the $m$-th derivative 
of $\widehat F_g$ is taken w.r.t. the variable $s$, which is defined as
\be
\label{average}
s={1\over 2}(t_1 -t_2). 
\ee
All derivatives are evaluated at $t_1=t$ and $t_2=0$. The exponential factor involves the instanton action, which is given by 
\be\label{hata}
A(t) = \partial_{s}  \widehat F_0 
\ee
and
\be
\hat q = \exp \Bigl(  \partial_s^2 \widehat F_0 \Bigr).
\ee
At leading order in $g_s$, we have
\be
\ba
&\sum_{k} \sum_{m_i>0}\sum_{g_i>1-{m_i\over 2}} {g_s^{\sum_i (2g_i+m_i-2)}\over k! m_1!\,\dots\, m_k!}\,\,\, \widehat F_{g_1}^{(m_1)}\dots \widehat F_{g_k}^{(m_k)}  (-\ell)^{\sum_i l_i}
\\
&=1 - g_s \Bigl( \ell\, \partial_{s} \widehat F_1 (t) + {\ell^3 \over 6}\, \partial_s^3 \widehat F_0 (t) \Bigr) + \CO(g_s^2).
\ea
\ee
By using the general formulae for the free energies in the two-cut model, one can check \cite{mswtwo} 
that (\ref{npboundary}) reproduces the one-instanton contribution to the free energy 
computed in the previous subsection and generalizes it to arbitrary $\ell$.

\subsubsection{Large $N$ instantons from orthogonal polynomials}
\label{diff-matrix}

So far we have analyzed the ``perturbative" sector of the one-cut matrix model from the point of view of the spectral curve: the procedure explained in \cite{ce,eo} gives 
a systematic way to compute the ``perturbative" $1/N$ corrections in terms of data of this curve, and the results reviewed above 
show that the formal trans-series giving the instanton corrections can be also computed 
using data of the spectral curve. For example, the instanton action is given by a period of the 
differential $y(x) \rd x$ along a cycle, and the loop corrections can be computed from 
the knowledge of the curve only. 

However, in the one-cut case there is a completely different way to obtain the $1/N$ corrections in the perturbative sector, 
by using orthogonal polynomials \cite{bessis,biz} (see \cite{alvarezmedina} for some recent developments). It was shown in \cite{mmnp} that this procedure can be generalized to 
compute systematically multi-instanton effects in the one-cut sector (partial results in this direction can be found in \cite{lattice,akk,sy}). 
This technique is very powerful in order to go to higher loops in simple matrix models. In addition, it is formally very similar to the trans-series 
method that we developed in section \ref{ODEs}. Therefore, it makes it possible to apply many of the techniques typical of the theory of resurgence (the connection 
to resurgence was further developed in \cite{asv}). 
Another advantage of this method is that it can be applied to unitary matrix models. 

The basic idea of the technique of orthogonal polynomials is the following (see \cite{dfgzj,mmleshouches} for reviews): in the matrix integral (\ref{zmm}), if we regard 
\be
\rd\mu = \re^{-{1\over g_s} V(\lambda)} {\rd\lambda \over 2 \pi}
\ee
as a measure in $\IR$, we can introduce orthogonal polynomials $p_n(\lambda)$ defined by
\be
\int \rd\mu \, p_n(\lambda) p_m(\lambda)  = h_n \delta_{nm},\quad n\ge 0,
\label{ortho}
\ee
where $p_n(\lambda)$ are normalized by requiring the behavior $p_n(\lambda)=\lambda^n +\cdots$. 
One then easily finds, 
\be
\label{parth}
Z =\prod_{i=0}^{N-1} h_i = h_0^N \prod_{i=1}^N r_i^{N-i},
\ee
where we have introduced the coefficients
\be
\label{rcoeff}
r_k= {h_k \over h_{k-1}}, \qquad k\ge 1,
\ee
which appear in the recursion relations for the $p_n(\lambda)$, 
\be
\label{recurs}
(\lambda + s_n ) p_n(\lambda) = p_{n+1}(\lambda) + r_n p_{n-1}(\lambda).
\ee
In this subsection $F$ will denote the normalized free energy, which is obtained by subtracting the Gaussian free energy,
\be
F=\log Z -\log Z^{\rm G}.  
\ee
At finite $N$, $F$ is given by the following formula:
\be
g_s^2 F= {t^2 \over N} \log {h_0\over h_0^G} + {t^2 
\over N}  \sum_{k=1}^N \biggl( 1-{k\over N} \biggr)
\log {r_k \over k g_s},
\label{allf}
\ee
where $h_0^G$ is the coefficient $h_0$ for the Gaussian model. 
In order to proceed, we introduce a continuous variable as $N \rightarrow \infty$, 
\be
\label{continuumvar}
 g_s k \rightarrow z, \qquad 0\le z \le t, 
\ee
and we assume that in this continuum, $N \rightarrow \infty$ limit, $r_k$ becomes a function of $z$ and $g_s$, 
\be
r_k  \rightarrow R (z,g_s).
\ee
It will be useful to consider the function 
\be
\Xi(z,g_s) = {R(z,g_s) \over z}
\ee
which can be regarded as the continuum limit of $r_k/(k g_s)$. It is easy to see that, for polynomial potentials of the form
\be
V(M) = {1\over 2}M^2 +\cdots,
\ee
one has $r_k \sim k g_s +\cdots$, therefore the function $\log (r_k/(k g_s))$ is regular at $k=0$ and we 
can use the standard Euler--Maclaurin summation formula to evaluate (\ref{allf}). One then obtains \cite{bessis,biz}:
\be
\label{ofgex}
\ba
g_s^2 F &= \int_0^t \rd z\,  (t-z) \log \Xi (z)+  \sum_{p=1}^{\infty} g_s^{2p}\, \, {B_{2p} \over (2p)!}\, \frac{\rd ^{2p-1}}{\rd z^{2p-1}} \biggl[ \left( t-z \right) \log \Xi(z,g_s) \biggr] \bigg|_{z=0}^{z=t} \\
&+ {t g_s \over 2 } \biggl[ 2 \log {h_0 \over h_0^{\rm G}} - \log \Xi (0,g_s)\biggr].
\ea
\ee
From this expression one can deduce the difference equation, 
\be
\label{diffeq}
F(t+g_s) + F(t-g_s)-2 F(t) =\log \Xi, 
\ee
which can be also obtained by starting from the identity
\be
{Z_{N+1} Z_{N-1} \over Z_N^2} =r_N. 
\ee

In order to compute the $g_s$ expansion of the free energy, one has to first find an expansion for the function $R(z,g_s)$ of the form 
\be
\label{zeroex}
R^{(0)}(z,g_s) = \sum_{s=0}^{\infty} g_s^{2s} R_{0,2s}(z).
\ee
Once this expansion is plugged in $\Xi(z,g_s)$ and then in (\ref{ofgex}), the genus expansion  follows. 
In order to obtain (\ref{zeroex}) one has to use the so-called 
{\it pre-string equation}. This is a difference 
equation for $R(z,g_s)$ which can be derived as the continuum limit of the recursion relations obeyed by the coefficients (\ref{rcoeff}). 
The pre-string equation can be explicitly written for any polynomial potential \cite{biz,dfgzj}. For example, in the 
case of the quartic matrix model with potential (\ref{quarticpot}), the difference equation for $R(z,g_s)$ reads as
\be
\label{diff}
R(z,g_s)\Bigl\{ 1 -{\lambda\over 12} (R(z,g_s) + R(z+g_s,g_s) + R(z-g_s,g_s) \Bigr\}=z. 
\ee
These types of equations have a solution of the form (\ref{zeroex}), and they determine $R_{0,s}(z)$ in terms of the $R_{0,s'}(z)$, $s'<s$. 
When this solution is plugged in (\ref{diffeq}), one obtains the perturbative expansion of the total free energy in powers of $g_s$, 
which is the standard $1/N$ expansion of the matrix model \cite{bessis,biz}. 

Difference equations, just like differential equations, admit 
trans-series solutions, and one could guess that the trans-series solution to the difference equation governing 
$R(z,g_s)$ encodes the multi-instanton amplitudes of the full matrix model. This was first proposed in \cite{mmnp}. It was then verified that the results obtained 
with this technique agree with the previous techniques in the quartic matrix model \cite{mmnp,asv} and in the cubic matrix model \cite{mswtwo}. 
To obtain the trans-series solutions to the pre-string equation, we consider a more general ansatz than (\ref{zeroex}), 
\be
\label{transr}
R(z,g_s) = \sum_{\ell=0}^{\infty} C^{\ell} R^{(\ell)}(z,g_s), 
\ee
where $R^{(0)}(z,g_s)$ is given by (\ref{zeroex}), and for $\ell\ge1$ we have
\be
\label{kpert}
 R^{(\ell)}(z,g_s)=  \re^{-\ell A(z)/g_s} R_{\ell,1}(z) \Bigl( 1+\sum_{n=1}^{\infty} g_s^{n} R_{\ell,n+1}(z) \Bigr),\qquad \ell \ge 1.
 \ee
 Once this ansatz is plugged in the difference equation for $R(z,g_s)$, one obtains a recursive system of equations for the different quantities involved. The 
 quantity $A(z)$, which is a parameter-dependent instanton action, is determined by an equation of the form 
 \be
 A'(z) =f(R_{0,0}(z)), 
 \ee
 where $f$ is a function fixed by the difference equation. For $\ell=1, n>0$, one 
 obtains an equation which determines 
\be
{\rd R_{1,n} (z) \over \rd z}
\ee
in terms of $R_{1,n'}(z)$ with $n'<n$. For $n=1$, we have a differential equation for the logarithmic derivative, i.e. for
\be
{1\over R_{1,1} (z)} {\rd R_{1,1} (z) \over \rd z}. 
\ee
The integration constant for $R_{1,1}(z)$ can be reabsorbed in the parameter $C$, and for $A(z)$ and the $R_{1,n} (z)$, $n>1$ the integration constants are fixed by using 
appropriate boundary conditions. For $\ell>1$, the difference equation determines $R_{\ell,n}$ in terms of $R_{\ell',n'}$ with $\ell<\ell'$. 

The trans-series structure of $R(z,g_s)$ leads to a trans-series structure for the full free energy, as in (\ref{f-trans}). We will write it as
\be
\label{fullf}
F(t,g_s) = \sum_{\ell=0}^{\infty}  C^{\ell} F^{(l)}(t,g_s), 
\ee
where
\be
\label{fkinst}
 F^{(\ell)}(z,g_s)=  \re^{-\ell  A(t)/g_s} F_{\ell,1}(z) \Bigl( 1 +\sum_{n=1}^{\infty} g_s^{n} F_{\ell,n+1}(z)\Bigr), \qquad \ell\ge 1.
 \ee
Once (\ref{transr}) is known, one can plug it in (\ref{diffeq}) to deduce the $F^{(\ell)}(t,g_s)$, see \cite{mmnp} for more details. 
The resulting amplitude is nothing but the $\ell$-instanton 
amplitude of the full matrix model. 

\begin{example} {\it The quartic matrix model}. As an example, let us present some results for multi-instanton corrections 
in the quartic matrix model with 
the potential (\ref{quarticpot}). The perturbative solution (\ref{zeroex}) has been much studied since 
it was first worked out in the pioneering papers \cite{bessis,biz}. The planar part is given by
\be
R_{0,0}(z)={2\over \lambda} \Bigl(1-{\sqrt{1-\lambda z}}\Bigr). 
\ee
As already noticed in \cite{biz}, it turns out to be useful to express all results in terms of 
\be
r=R_{0,0}(z). 
\ee
For the higher $g_s$ corrections one finds, 
\be
\ba
R_{0,2}(z)&={2 \lambda^2 \over 3} {r\over (2-\lambda r)^4},\\
R_{0,4}(z)&={28 \lambda^4\over 9} {r (5 + \lambda r) \over (2- \lambda r)^9},\\
R_{0,6}(z)&= {4 \lambda^6 \over 27} \frac{r \left(111 \lambda^2 r^2+5728 \lambda r+7700\right)}{(2-\lambda r)^{14}},
\ea
\ee
and so on. If we now plug the trans-series ansatz (\ref{transr}) in the difference equation (\ref{diff}), we find a system of recursive difference equations  
for the $R^{(\ell)}(z,g_s)$ which can be solved by using the ansatz (\ref{kpert}). Let us focus on 
$\ell=1$, the one-instanton solution. The first thing to compute is $A(z)$, which 
corresponds physically to the instanton action. One finds, at leading order in $g_s$, 
\be
\cosh (A'(z)) = 2 {3-\lambda r \over \lambda r}. 
\ee
This can be integrated to find $A(z)$ up to an additive constant and an overall sign (since $\cosh\, z$ is even). Since $z$ stands here for the 
't Hooft parameter, both 
ambiguities can be fixed by requiring that, as $z\rightarrow 0$, the instanton action becomes (\ref{camm}). The result is
 \be
 \ba
A(z)&= - \int  \rd r \cosh^{-1} \Bigl( 2 {3-\lambda r \over \lambda r}\Bigr) \Bigl(1-{\lambda r\over 2}\Bigr)\\
= &\frac{1}{4} r  (\lambda r-4) \cosh ^{-1}\left(\frac{6}{\lambda r}-2\right)+{1\over 2 \lambda}
   \sqrt{3 (2- \lambda r)(6-\lambda r)}.
   \ea
   \label{quarticinstanton}
 \ee
It can be checked that (\ref{quarticinstanton}) coincides with the instanton action 
of the quartic matrix model computed in terms of its spectral curve in (\ref{quartic-inst}). 
Once the instanton action is known, we can proceed to compute $R_{1,1}(z)$. The equation one obtains at the next order in $g_s$ is
\be
{R'_{1,1}(z) \over R_{1,1}(z)}=-{1\over 2} \coth (A'(z)) A''(z), 
\ee
which can be immediately integrated as
\be
R_{1,1}(z)= \Bigl( \sinh (A'(z))\Bigr)^{-1/ 2}. 
\ee
The rest of the coefficients can be found by integrating the resulting equations for $R_{1,n}(z)$, and from (\ref{diffeq}) one finds 
the loop expansion of the one-instanton free energy, see \cite{mmnp,asv} for explicit formulae. The results agree with those obtained with the method 
of direct calculation in subsection \ref{direct-cal} and with the result obtained from the multicut matrix model in subsection \ref{multi-multi}. 
\end{example}

\subsection{Large $N$ instantons, large order behavior and the spectral curve}

\label{inst-spectral}
In the previous section we have discussed various techniques to compute large $N$ instanton effects in one-cut matrix models. The total free energy, including multi-instanton sectors, 
is the analogue of the trans-series solution in this situation. We should then expect a connection between the large order behavior of the perturbative genus expansion and the instanton trans-series, 
generalizing the connections that we have seen in ODEs and in QM. Let us consider the total free energy
\be
\CF (g_s) = g_s^2\, F(g_s), 
\ee
which is defined in such a way that the perturbative sector has no negative powers of $g_s$. The one-instanton contribution, or first trans-series, yields an expansion of the form 
\be\label{curlfa}
\CF^{(1)} (z) = \ri z^{\beta/2} \re^{-\frac{A}{\sqrt{z}}} F_{1,1}\left( 1+ \sum_{n=1}^{\infty} F_{1,n+1} z^{n/2}\right),
\ee
\noindent
where $z=g_s^2$. This is an important feature distinguishing matrix models and string theory from field theory and QM: the action of an instanton goes like $1/\sqrt{z}$, and not as $1/z$. Similarly, the perturbation series around the instanton sector is a series in powers of $\sqrt{z}$, and not a series in powers of $z$. We may now write
\be
\CF^{(0)} (z)=\sum_{g=0}^{\infty} F_g(t)\, z^g.
\ee
Let us assume that there is a ``resurgent" relation between the perturbative sector and the first trans-series. In analogy with the calculation in (\ref{per-np-odes}), we expect
\be\label{lostringl}
\ba
F_g & = {1\over 2\pi} \int_0^{\infty} {\rd z \over z^{g+1}} z^{\beta/2} \re^{-\frac{A}{\sqrt{z}}}  F_{1,1}\left( 1+ \sum_{n=1}^{\infty} F_{1,n+1} z^{n/2}\right)\\
& \sim {F_{1,1} \over \pi} \left( A^{-2g+\beta}  \Gamma(2g-\beta)+ \sum_{n=1}^{\infty} F_{1,n+1} A^{-2g+\beta+n} \Gamma(2g-\beta-n)\right).
\ea
\ee
This leads to an asymptotic expansion of $F_g$ in powers of $1/g$. Up to two loops we can write it as 
\be
\label{large-g-beh}
F_g \sim {A^{-2g+\beta} \over \pi}\, \Gamma(2g-\beta)\, F_{1,1} \left[1 + {F_{1,2} A \over 2g} + \cdots \right].
\ee
This gives a prediction for the large-order behavior of genus $g$ free energies in the one-cut matrix model. 

In writing (\ref{large-g-beh}) we have implicitly assumed that there is a single instanton solution that contributes to the asymptotic behavior. 
In general there might be various instanton configurations in the system, with the same action in absolute value, and in this case $F^{(1)}$ 
will denote the sum of all these contributions. It is also common to have complex instanton solutions which give 
complex conjugate contributions to $F^{(1)}$, as we saw in QM. In this case the asymptotic behavior of $F_g$ is again 
obtained by adding their contributions. If we write
\be\label{thetas}
A =|A| \re^{-\ri \theta_A}, \qquad F_{1,1}=|\mu|\re^{\ri \theta}, 
\ee
the leading asymptotics will read in this case
\be\label{acos}
F_g \sim {|A|^{-2g+\beta}  \over \pi}\, \Gamma(2g-\beta)\, |\mu|\,  \cos \bigl( (2g-\beta) \theta_A + \theta \bigr).
\ee

We have also assumed so far that the large order behavior is dominated by the instanton associated to eigenvalue tunneling. 
However, there is a subtlety concerning the large order behavior and related to the Gaussian part of the free energy. For small $t$, the behavior of the free energy is dominated by 
its Gaussian part, 
\be
F_g(t) \approx {B_{2g} \over 2g (2g-2) t^{2g}}, \qquad t \rightarrow 0.
\ee
Using the asymptotics (\ref{bernoulli-as}), we can see that the Gaussian part indeed displays the large $g$ behavior (\ref{large-g-beh}), with 
\be
\label{triv-inst}
A=-2 \pi t, \qquad \beta=1. 
\ee
However, this behavior is due to the universal Gaussian part of the free energy. For other values of $t$, the large order 
behavior is indeed controlled by the non-trivial instantons due to eigenvalue tunneling and discussed above. 
In the quartic matrix model, for example, there is a non-trivial instanton with action (\ref{quartic-inst}), 
and one sees from the explicit calculation in (\ref{muex}) that this instanton has
\be
\beta={5\over 2}. 
\ee
One can test the conjectural asymptotics (\ref{large-g-beh}) governed by this non-trivial instanton by considering the normalized free energy obtained, i.e. 
by subtracting the Gaussian free energy. This was done numerically in \cite{mswone,asv} by looking at the sequence of the first $F_g$, and one can see that (\ref{large-g-beh}) is indeed satisfied. 

\FIGURE[ht]{\label{curves}
\includegraphics[height=4cm]{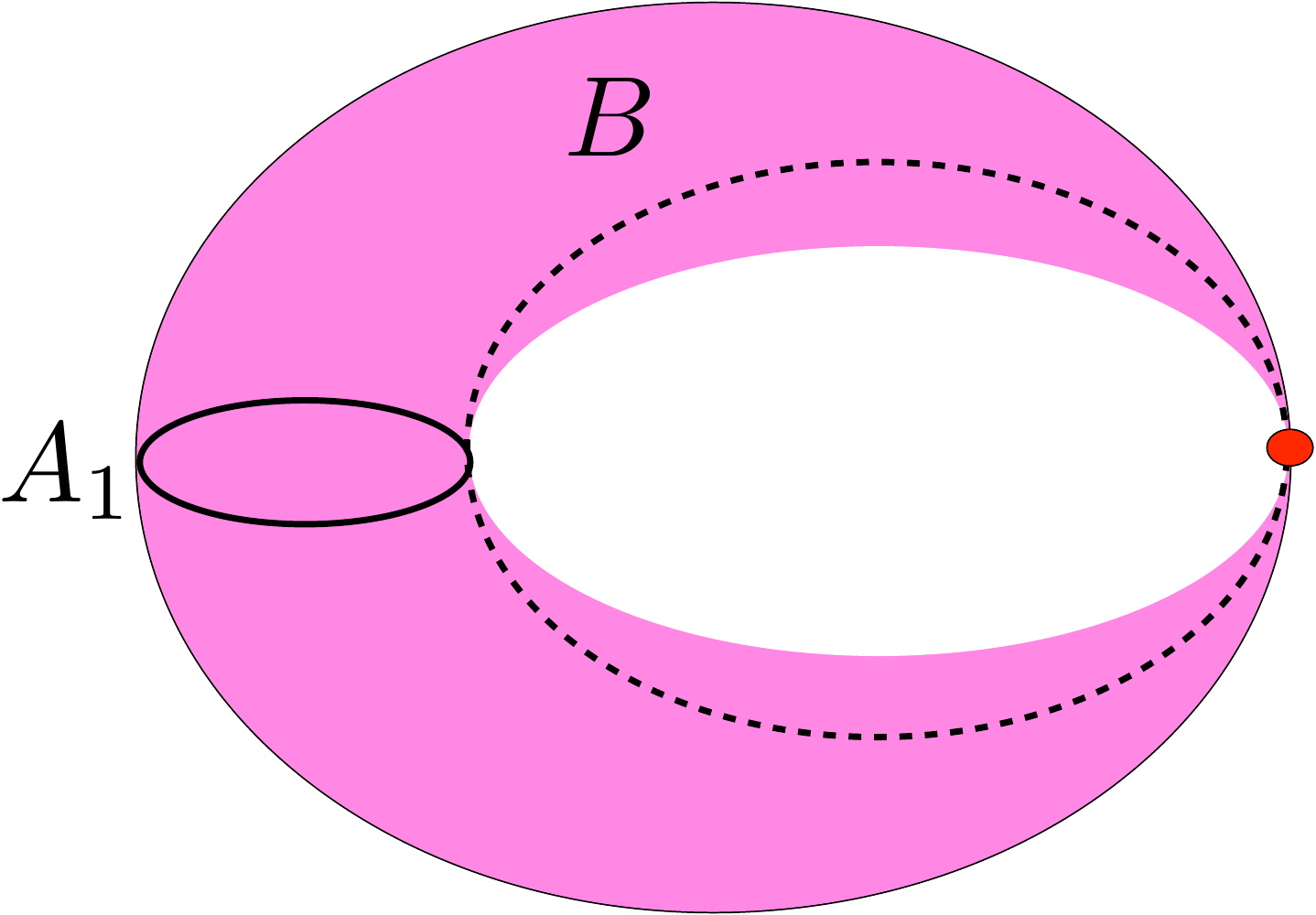} \qquad \qquad
 \includegraphics[height=4cm]{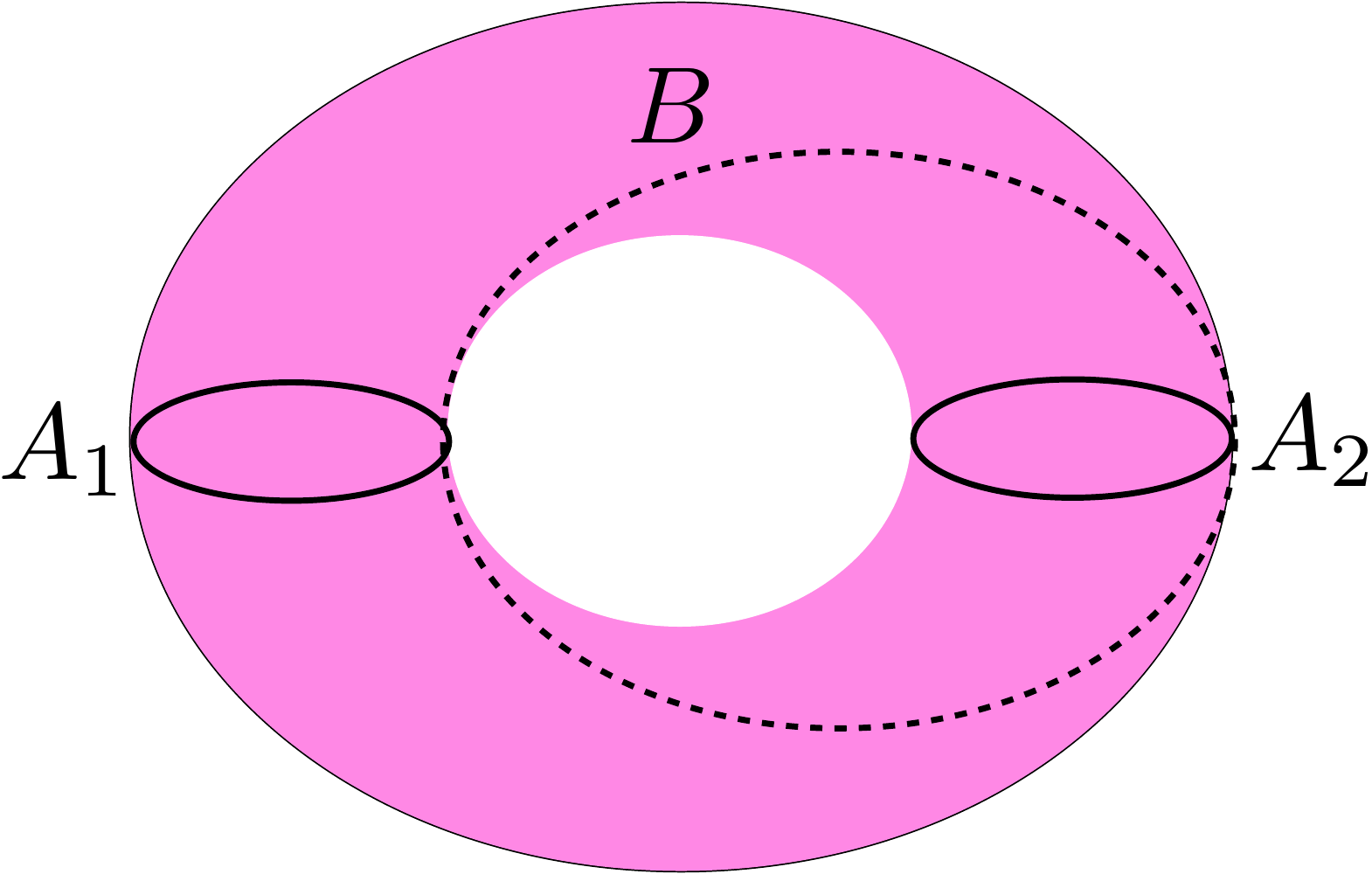} 
\caption{The left-hand side shows the spectral curve in the one-cut phase of the cubic matrix model. The instanton action relevant in the double-scaling limit is obtained by 
calculating the $B$-period of the one-form $y(x)\rd x$, which goes from the filled cut $A_1$ to the pinched point. The two-cut phase, in which the pinched point becomes a filled interval, is shown on the right hand side. The instanton action is still given by the $B$-period integral.}}

The appearance of two different instanton actions, corresponding to the ``trivial" action (\ref{triv-inst}) and the non-trivial instanton action associated to eigenvalue tunneling, has a natural geometric interpretation \cite{dmpnp}. The instanton action describing eigenvalue tunneling in the one-cut phase can be written in terms of the spectral curve $y(x)$ as in (\ref{instdiff}). This can be in turn written as a period integral of the natural meromorphic form $y(x) \rd x$ along a 
$B$ cycle which goes from the filled cut to the critical point:
\be\label{onecutia}
A_B={1\over 2} \oint_B y(x) \rd x.
\ee
In \figref{curves} (left) we show the pinched curve describing the one-cut curve. The $A_1$ cycle corresponds to the filled cut, and the $B$ cycle goes from $A_1$ to the pinched cycle. The $A_1$-period going around the filled cut is just proportional to the 't Hooft parameter: 
\be
\label{gaussper}
A_{A_1} (t) =2\pi \ri t ={1\over 2} \oint_{A_1} y(x) \rd x. 
\ee
As we have just seen, both the $A_1$ and the $B$ periods give rise to instantons in the matrix model. The instanton corresponding to the $A_1$ period is the ``trivial" one and governs the asymptotics near $t=0$, through the Gaussian part of the potential. In other regions of the $t$-plane, the large genus behavior will be controlled by $B$-periods $A_B(t)$ of the form (\ref{onecutia}). In general, the action controlling the large order behavior at a given point $t$ will be proportional to the smallest period of the meromorphic form $y(x)\rd x$ (in absolute value). The B-type periods $A_B(t)$ vanish
at critical values of the 't Hooft parameter, and the $A_1$ period vanishes at $t=0$, 
so in both cases the instanton action is given by a ``vanishing cycle." 

This result can be generalized to two-cut phases, where the pinched point is now resolved into a second cut $A_2$: 
the instanton action is still given by the $B$-cycle integral, now going from the first cycle $A_1$ to the second cycle $A_2$, see \figref{curves} (right). 
This instanton action controls the large order behavior of the free energies in the appropriate regions of moduli space. This was verified in \cite{kmr} for the two-cut, cubic matrix model.  

This picture of instanton actions as periods of the spectral curve, pointed out in \cite{dmpnp}, seems to be the most general framework explaining the large order structure of a 
very general class of matrix models. 
For example, in the Chern--Simons matrix model for $\IS^3$ introduced in \cite{mm} and briefly reviewed in the Example \ref{cs-example}, 
the instanton actions are given by \cite{ps}
\be
\label{cs-periods}
2 \pi \ri  \left( t+ 2\pi \ri n \right), \qquad n \in \IZ. 
\ee
For $n=0$ one recovers the action governing the Gaussian behavior. The instantons with $n=\pm 1$ can be detected through the large order behavior of the genus $g$ free energies, once the Gaussian part is subtracted \cite{ps}. Since the spectral curve describing this model admits a constant period, one can regard (\ref{cs-periods}) as a linear combination of periods. A more detailed 
discussion of this general point of view on instanton actions can be found in \cite{dmpnp}.

\subsection{Classical asymptotics and the Stokes phenomenon in matrix models}

In the previous subsections we have introduced the basic ingredients to study the large $N$ asymptotics of matrix models. Once a background is chosen, 
it leads to a formal perturbative series in $g_s$, as in (\ref{largeNas}), and the rest of the sectors in (\ref{sumz}) lead to a formal trans-series. 

We would now like to proceed to step two in the program 
sketched in the introduction, i.e. we want solve the problem of classical asymptotics in the 
case of matrix models at large $N$, or, equivalently, we want to find the formal expansions which provide the asymptotics in each 
region of the complex space of parameters, paying attention to the Stokes phenomenon. 

This turns out to be a non-trivial and rich problem. We will discuss some of its aspects by looking at a concrete example: we will consider the matrix model 
version of the Airy function. This model is defined by the partition function (\ref{zmm}) where $\gamma=\gamma_1$ 
is the contour shown in \figref{airycontours} and the potential $V(x)$ is given by (\ref{stokespot}). Without loss of generality, 
we will assume that $g_s$ is real and positive, and we will study how this partition function changes as we change $t >0$ (the 't Hooft parameter) and the 
parameter $\kappa$ in the potential. 

For small $t$, as we discussed in (\ref{limitZ}), the saddle point structure should be the same 
as for the Airy function. Therefore, at small $t$ we expect the following phase structure. 
For $|\kappa|<2 \pi/3$, the dominant saddle is the one-cut configuration
where all the 
eigenvalues sit near $x^{\rm L}=-\zeta^{1/2}$, i.e. 
\be
(N, 0),
\ee
and 
\be
Z_{\gamma_1} (N, g_s)\sim Z(N,0). 
\ee
The other saddles, where eigenvalues will tunnel to the critical point at $x^{\rm R}$, are not relevant since the integration path does not pass through $x^{\rm R}$. 

 For $2\pi/3\le |\kappa|<\pi$ the integration path gets deformed into the sum of the two steepest-descent paths, and the other instanton configurations will start contributing by eigenvalue tunneling. However, at least at small $t$ they should be exponentially suppressed. In other words, the 
 dominant configuration to the matrix integral in this region should still be the boundary saddle $(N,0)$, but there will be exponentially small corrections due to tunneling to the other critical point $x^{\rm R}$, and we will have
 \be
 Z_{\gamma_1}(N, g_s)= Z(N,0) \left( 1+ \sum_{\ell=1}^\infty C^\ell Z^{(\ell)}(N)\right),
 \ee
where $Z^{(\ell)}(N)$ is given by (\ref{npboundary}) and $C$ is an appropriate constant. In contrast to the case of the Airy function, in the large $N$ limit there is an infinite number of corrections 
involving the subleading saddle points. The above formula is the analogue of (\ref{stokesints}) for the ``matrix Airy integral." Of course, from the point of view of classical asymptotics, 
this infinite series of exponentially small corrections 
is not taken into account, but it will be needed if we want to perform Borel resummations. Notice that the exponentially small corrections in (\ref{npboundary}) 
correspond to the eigenvalue tunneling depicted in \figref{tunnel}, i.e. 
\be
(N, 0) \rightarrow (N-\ell, \ell). 
\ee

Now, a very interesting thing happens when we reach the anti-Stokes line $\kappa=\pi$. The real part of the 
action $A(t)$ vanishes at $t=0$, and it is actually negative for small $t$, therefore the subleading corrections are no longer negligeable. This is 
the familiar ``return of the small exponential" along the anti-Stokes line. There is however a crucial difference with the 
case of the Airy function. The reason is that, in the large $N$ limit, there is an {\it infinite} number of corrections that come to life on that line, namely all the sub-leading terms in 
(\ref{npboundary}). In fact, the same situation arises in non-linear ODEs, where this leads to the formation of 
singularities along anti--Stokes lines, see for example \cite{costincostin}.

It is clear that in order to obtain a reasonable asymptotics we must somehow sum all the corrections. We proceed as follows. It is clear that the saddle $(N,0)$ is no longer a good starting point for the expansion. We have to find a new dominant saddle, which must necessarily have $N^*_2 \not=0$. In order for this saddle to be stable, the effective potentials on both cuts have to be the same. This requires
\be
\label{realvanish}
{\rm Re} \,  \left({F_0'(N_i^*)\over g_s} \right)=0,
\ee
where the derivative is w.r.t. the variable $s$ defined in (\ref{average}). This condition was first proposed in \cite{jur}, and it is an equipotential condition which 
blocks the ``flow" of eigenvalues. A way to derive this condition is to write the sum (\ref{sumz}), in the case $d=2$, as
\be
C_1^N \sum_{n=0}^N \left( {C_2 \over C_1} \right)^n \re^{g_s^2 F_0(n) + \cdots}
\ee
where we denoted $N_1=N-n$, $N_2=n$. Requiring this sum to have a saddle-point for $n^*$ leads to (\ref{realvanish}). In the large $N$ limit, it has been proposed in \cite{bde} to 
replace the sum over $n$ by a {\it theta function}. One then gets the following asymptotic expansion in $g_s$, 
\be
\label{npinterior}
\ba
& Z_{\gamma_1}(N, g_s)
= Z(N_1^*, N_2^*) \sum_{k} \sum_{m_i>0}\sum_{g_i>1-{m_i\over 2}} {g_s^{\sum_i (2g_i+m_i-2)}\over k! m_1!\,\dots\, m_k!}\,\,\, F_{g_1}^{(m_1)}\dots F_{g_k}^{(m_k)}  \,\, \Theta_{\mu,\nu}^{(\sum_i l_i)}(F'_0/g_s,\tau) \\
&\quad =Z(N_1^*, N_2^*)  \biggl\{ \Theta_{\mu,\nu} +g_s \Bigl(\Theta'_{\mu,\nu} F_1' + {1\over 6} \Theta_{\mu,\nu}'''\,F_0'''\Bigr) + \CO(g_s^2) \biggr\}. 
\ea
\ee
The derivatives of the free energies $F_g$ are again w.r.t. $s$. 
The theta function $\Theta_{\mu,\nu}$ with characteristics $(\mu,\nu)$ is defined by
\be
\label{biget}
\Theta_{\mu,\nu}(u,\tau) = \sum_{n\in {\mathbb Z}} {\rm e}^{(n+\mu-N \epsilon)u}\,\,\re^{\pi \ri  (n +\mu-N \epsilon)\tau (n+\mu-N\epsilon)}\,\,{\rm e}^{2 \ri\pi (n+\mu) \nu},
\ee
and it is evaluated at 
\be
u={F_0'(N_i^*)\over g_s},
\qquad
\tau =  {1\over 2\pi \ri} F_0''.
\ee
In the above equation, we have denoted
\be
\epsilon={N_2^* \over N}, \qquad C={C_2 \over C_1}={\rm e}^{2 \ri\pi\nu}.
\ee
In standard matrix models we have $\mu=0$, although the term $N\epsilon$ might give an effective characteristic depending on the parity of $N$ \cite{bde}. 

The asymptotics (\ref{npinterior}), discovered in \cite{bde}, was interpreted in in \cite{mpp} as a matrix model generalization of the oscillatory asymptotics along an anti-Stokes line. The singularities along anti--Stokes lines in nonlinear ODEs correspond here to zeros of the partition function, which are made possible due to the presence in (\ref{npinterior}) of the 
theta function at leading order: when the theta function vanishes, the partition function vanishes at leading order in $g_s$.  These zeros are then nothing but Lee--Yang zeros for the partition function 
of the matrix model, which are known to occur along anti--Stokes lines \cite{ipz,ps}. 

Notice that the term involving the theta function is 
{\it not} analytic in $N$, in the same way that the asymptotics (\ref{oscila}) is not analytic at $x=\infty$ (even when expressed in terms of $z=x^{3/2}$). We have, for the free energy, the expansion
\be
F=g_s^{-2} F_0(N_i^*)+ F_1(N_i^*) + \log  \Theta_{\mu,\nu} +\cdots,
\ee
so the oscillatory behavior is already present at next-to-leading order. Notice that this asymptotics has contributions which are not present in the standard large $N$ expansion (\ref{largeNas}). 

Let us now come back to the example of the 
cubic matrix model along $\kappa=\pi$. We have to find a new saddle satisfying (\ref{realvanish}). For ${\rm arg}(\kappa)=\pi$, $\zeta^{1/2}$ is purely imaginary, and it is easy to see from the structure of the genus zero free energy that its {\it real} part is {\it symmetric} in $t_1$, $t_2$. Therefore, 
\be\label{cubicts}
t_1=t_2
\ee
solves the saddle-point equation 
\be
{\rm Re}\, {\partial F_0 \over \partial s}=0,
\ee
at least for small $t$. We conclude that, on the anti--Stokes line and for 
$t$ small enough, the partition function of the cubic matrix model is given by an expansion of the form (\ref{npinterior}). Since $C_1=C_2=1$ (this follows from 
the fact that the path $\gamma_1$ is deformed into a sum of two steepest-descent paths), the characteristics of the theta function are $\mu=\nu=0$. 

\begin{figure}[!ht]
\leavevmode
\begin{center}
\includegraphics[height=7cm]{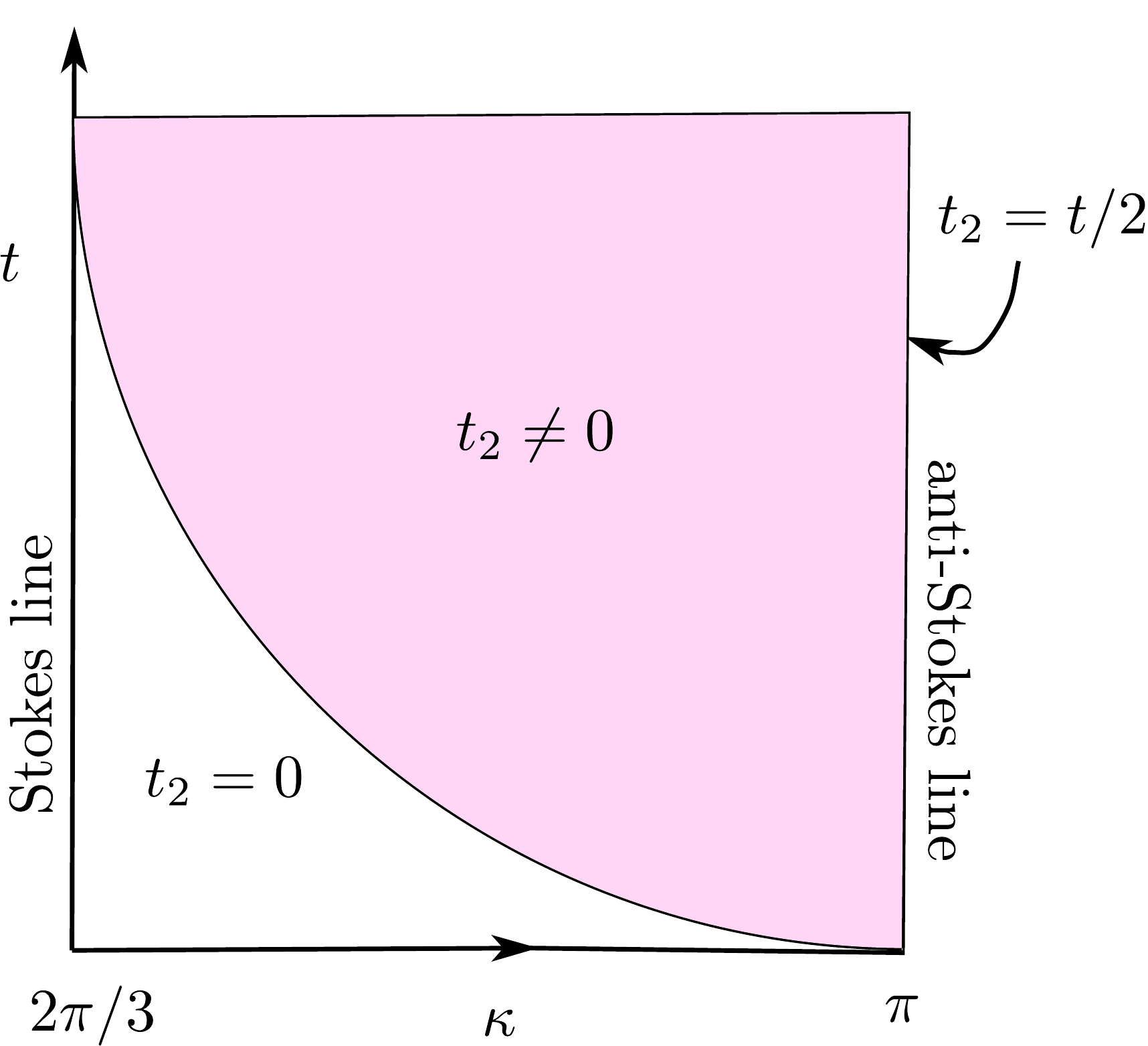}
\end{center}
\caption{The phase diagram of the cubic matrix model with potential (\ref{stokespot}), as a function of $2\pi/3\le \kappa\le \pi$ and real $t=g_s N$. The parameter $t$ is also supposed to 
be small. The Stokes and anti--Stokes lines of the $N=1$ case, which occur for $\kappa=2\pi/3$ and $\kappa=\pi$, respectively, determine to a large extent the 
phase diagram for small $t$. On the anti--Stokes line, the saddle value of $t_2$ is $t/2$, and the classical $1/N$ asymptotics requires corrections beyond the genus expansion. }
\label{cubicphase}
\end{figure} 

The above analysis is only valid, strictly speaking, at $t\sim 0$. When $t$ is small but nonzero, the real part of the instanton action vanishes at a finite $t_c(\kappa)$ for each value of $2\pi/3<\kappa<\pi$. Therefore, the transition to a new saddle will occur along a line in the $t-\kappa$ plane defined by 
\be
{\rm Re}\left(A(t_c(\kappa), \kappa)\right) =0. 
\ee
We display the phase diagram of the model in \figref{cubicphase}. For a generic point along the 
curve $t_c(\kappa)$ separating the two phases, there is a phase transition in which $N_2^*$ increases smoothly away from zero (this has been verified very explicitly in another, related model in \cite{mpp}). This transition is then of the type ``birth-of-a-cut" studied in for example \cite{eynardbirth}. 
Notice that the smooth transition becomes discontinuous as $t\to 0$ i.e. it becomes the jump in asymptotics along the anti-Stokes line. For $t>0$, the discontinuity is smoothed out. 
We then have a {\it deformed Stokes phenomenon} as we turn on the 't Hooft parameter. 

The phase diagram that we have just sketched is 
valid for small $t$. For arbitrary $t$ there is a very rich phase structure first described in \cite{david} and 
which can be formalized in terms of so-called Boutroux curves, see \cite{bertola}. A recent analysis of the phase structure of the cubic matrix model can be found in \cite{alvarez}. 

Based on this detailed example, we can extract some general conclusions for the study of asymptotics and non-perturbative effects in matrix models:

\begin{enumerate}

\item As we change the parameters of the model, the partition function (\ref{sumz}) is dominated by different large $N$ saddle points in different ``phases". There are phase transitions 
among these different phases. The transitions occur as in asymptotic analysis: as we move in parameter space, large $N$ saddles which were exponentially suppressed are no longer 
so and start contributing to the asymptotics. These transitions, triggered by large $N$ instantons, are called {\it large $N$ phase transitions}, and occur along generalized anti-Stokes lines. 

\item The Stokes phenomenon is ``smoothed out" at finite 't Hooft parameter, even in the sense of classical asymptotics. 

\item The $1/N$ expansion breaks down along anti-Stokes lines. In the Hermitian case, the classical asymptotics involves 
theta functions with a non-analytic dependence on $N$.  

\end{enumerate}
A more detailed study of these phenomena, together with an analysis of the so-called lens space matrix model, can be found in \cite{mpp}. 

\sectiono{Two applications in String Theory}
 
It is impossible to provide here a detailed overview of non-perturbative effects in string theory. We will content ourselves with a discussion of 
some string theory models where one can use large $N$ matrix models to describe non-perturbative effects. 

From the point of view discussed in these lectures, it is useful to focus on string free energies and to regard string theory as a quantum mechanical system with two different Planck constants. 
The {\it worldsheet} Planck constant is given by the square of the string length,
\be
\hbar_{\rm ws} =\ell_s^2.  
\ee
The free energy at genus $g$, $F_g(t)$, depends on a set of target moduli $t$ (this could be a compactification radius or a K\"ahler parameter), 
and the perturbative regime of the worldsheet 
theory corresponds to the regime in which these moduli are very large as compared to the string length. Non-perturbative effects with respect to 
this Planck constant are of the form 
\be
\exp\left( -A_{\rm ws}(t)/\ell_s^2\right), 
\ee
where $A_{\rm ws}(t)$ is a worldsheet instanton action. These effects are usually 
realized in string theory as instantons of the non-linear sigma model underlying perturbative string theory. 

There 
is however a second, {\it spacetime} Planck constant called the string coupling constant,
\be
\hbar_{\rm st}=g_s. 
\ee
The contribution of a genus $g$ worldsheets to the free energy is weigthed by the factor $g_s^{2g-2}$, and the perturbative free energy is given by a genus expansion 
of the form
\be
F^{(0)}(t,g_s) =\sum_{g=0}^{\infty} F_g(t) g_s^{2g-2}.
\ee
This has the same structure that the large $N$ expansion of a gauge theory or a matrix model, compare to (\ref{largeNas}), and it is believed that, 
generically, the genus $g$ free energies grow as $(2g)!$, just as in large $N$ theories \cite{grossperiwal, shenker}.
We should then expect non-perturbative effects w.r.t. this second Planck constant, of the form
\be
\label{twop}
\exp\left( -A_{\rm st} (t)/g_s\right)
\ee
where $A_{\rm st} (t)$ is the action of a {\it spacetime} instanton. These effects 
were first found experimentally in non-critical string theory \cite{david,shenker}, and later on it 
was proposed by Polchinski that they are realized by D-branes \cite{polchinski}. 

In general, the calculation of these non-perturbative effects directly in string theory is not easy. 
However, when the string theory model has a large $N$ gauge theory dual, one can in principle compute them in the gauge theory. 
Worldsheet instantons appear as exponentially small corrections in the 't Hooft expansion at strong 't Hooft coupling, 
while spacetime instantons should correspond to large $N$ instantons. 
In some cases, the large $N$ dual further reduces to a matrix model, and the techniques described in these 
lectures can then be applied to calculate non-perturbative effects in string theory. In particular, the free energy of non-critical strings is described by non-linear ODEs, and non-perturbative effects in these theories can be analyzed with the trans-series formalism presented in section \ref{ODEs} 
(see also \cite{taiwan} for recent developments on non-critical strings).  
We will now describe two examples where the ideas developed in these lectures lead to non-perturbative results in string theory, 
and which go beyond non-critical string theory.

 \subsection{A toy model: Trans-series and Hurwitz theory}
 
 Hurwitz theory can be regarded as a toy model for string theory (or, more precisely, for topological string theory). In Hurwitz theory one considers maps from a Riemann surface or string, to another Riemann surface, and ``counts" these maps in an appropriate form, by using Hurwitz numbers. Here, for simplicity, we will assume that the target is $\IP^1$, i.e. a two-sphere. The only parameters in Hurwitz theory are the string coupling constant $g_H$ and a target parameter $t_H$ which can be regarded as measuring the ``size" of the target. The partition function of this model is of the form 
\be
Z(t_H,g_H)=\sum_{g\ge 0} g_H^{2g-2} \sum_{d\ge 0} {H_{g,d}^{\IP^1} (1^d)  \over (2g-2+2d)!}Q^d,
\ee
where $Q=\re^{-t_H}$. The dependence on $g_H$ is the standard one for a sum over genera in string theory. The quantity 
$H_{g,d}^{\IP^1} (1^d)$ is a simple Hurwitz number counting degree $d$ covering maps of $\IP^1$, with simple branch points only, and by Riemann surfaces 
of genus $g$ (see for example \cite{mswone} for an explicit expression for this number in terms of representation theory data). Here, in the partition function, we consider maps from generally disconnected worldsheets. The free energy $F=\log\, Z$ describes connected, simple 
Hurwitz numbers $H_{g,d}^{\IP^1} (1^d)^{\bullet}$,
\be\label{freehurwitz}
F(g_H,t_H)   = \sum_{g\ge 0} g_H^{2g-2} F_g(Q), 
\ee
where
\be
F_g(Q)=\sum_{d\ge 0}  {H_{g,d}^{\IP^1} (1^d)^{\bullet} \over (2g-2+2d)!}Q^d.
\ee
This theory is in fact a string theory in disguise. It can be realized as a special limit of topological string theory on certain toric Calabi--Yau manifolds, 
see for example \cite{bp,italy} for detailed derivations. It was also conjectured in \cite{mm} and proved in \cite{eynardproof} that Hurwitz theory can be described in terms of 
matrix integrals, and this in turn was used in \cite{mswone} to compute instanton effects, by using the techniques reviewed in subsection \ref{direct-cal}. As shown in \cite{mmnp} there is however another way to understand non-perturbative effects in this theory, by using difference equations as in subsection \ref{diff-matrix}.

It was proved in \cite{p} that the free energy of Hurwitz theory satisfies a difference equation of the Toda type, 
\be
\label{toda}
\exp \Bigl( F(t_H+g_H) +F(t_H-g_H)-2 F(t_H)\Bigr)= g_H^2 \re^{t_H} \partial_{t_H}^2 F(t_H,g_H). 
\ee
As we did in subsection \ref{diff-matrix}, we can try to solve this equation with a trans-series ansatz for the free energy of the form (\ref{fullf}). Doing this one immediately obtains the following equation for the one-instanton amplitude, 
\be
\label{hurwitzone}
\exp \Bigl( \Delta_{g_H} F^{(0)}(t_H)\Bigr) \Delta_{g_H} F^{(1)}(t_H) = g_H^2 \re^{t_H} \partial_{t_H}^2 F^{(1)},
\ee
where we have written
\be
\Delta_{h} f(t)=
f(t+h) +f (t-h)-2 f(t)
\ee
to denote the discrete Laplace operator with step $h$. The first term in the expansion of (\ref{hurwitzone}) in 
powers of $g_H$ gives an equation for $A'(t_H)$, 
\be
\label{hurins}
2\Bigl[ \cosh (A'(t_H)) -1\Bigr]=\exp\left\{ t_H-\partial_{t_H}^2 F^{(0)}_0(t_H) \right\} (A'(t_H))^2. 
\ee
It is straightforward to use (\ref{toda}) 
to derive a one-parameter, trans-series solution for the free energy of the Hurwitz model. In addition, one can check that the function 
$A(t_H)$ defined implicitly by (\ref{hurins}) coincides with 
the instanton action computed in \cite{mswone} from the matrix model realization of \cite{mm,eynardproof}. This instanton action controls the large genus behavior 
of the free energies $F_g(Q)$ of Hurwitz theory. We conclude that, in this toy model of string theory, the perturbative sector obtained by considering Hurwitz coverings can be generalized to a general trans-series. However, the geometric or physical meaning of this trans-series is not yet known. 

The Hurwitz model that we have discussed can be generalized to a family of topological string theories on non-compact Calabi--Yau manifolds 
(the so-called ``local curves") where non-perturbative effects can be also computed by using matrix model techniques, see \cite{mmopen,mswone,mmlocal}.

 \subsection{Stringy instantons in ABJM theory}

We will now describe an application in physical superstring theory, namely ABJM theory \cite{abjm} and its type IIA dual. 

ABJM theory \cite{abjm,abjmreview} is a Chern--Simons--matter theory in 
three dimensions with gauge group $U(N)_k \times U(N)_{-k}$ and $\CN=6$ supersymmetry. It is a conformally invariant theory. 
The Chern--Simons actions for the gauge groups are of the form (\ref{csact}) and they 
have couplings $k$ and $-k$, respectively. The theory contains as well four hypermultiplets $C_I$, $I=1, \cdots, 4$, 
in the bifundamental representation of the gauge group. The 't Hooft parameter of this theory is defined as
\be
\lambda={N\over k}.
\ee
In \cite{kwy} it was shown, through a beautiful application of localization 
techniques, that the partition function of ABJM theory on the three-sphere can be computed by a matrix model (see \cite{mmcslectures} for a pedagogical review). This matrix model is given by 
\be
\label{kapmm}
Z_{\rm ABJM}(N, g_s)={1\over N!^2} \int \prod_{i=1}^{N}{ \rd \mu_i \rd \nu_j  \over (2 \pi)^2} {\prod_{i<j} \sinh^2 \left( {\mu_i -\mu_j \over 2}\right)  \sinh^2 \left( {\nu_i -\nu_j \over 2}\right) \over 
\prod_{i,j}  \cosh^2 \left( {\mu_i -\nu_j \over 2}\right)} \re^{-{1\over 2g_s}\left(  \sum_i \mu_i^2 -\sum_j \nu_j^2\right)}, 
\ee
where the coupling $g_s$ is related to the Chern--Simons coupling $k$ of ABJM theory as
\be
g_s={2 \pi \ri \over k}.
\ee
The reduction of the partititon function of the theory to a matrix integral makes it possible to address some of the questions considered in these lectures in a very concrete way. For example, one can wonder, as we did in subsection \ref{nonpert-cs} for Chern--Simons theory, what is the large order behavior of perturbation theory for the partition function or the free energy. For fixed $N=2$, it has been shown in \cite{russo} that the perturbative expansion in powers of $g_s$ grows factorially but it is Borel summable. A different, more complicated question concerns the behavior of the free energy in the large $N$ expansion, i.e.: what is the behavior of the genus $g$ free energies? What are the non-perturbative effects at large $N$? These issues were addressed in \cite{dmp,dmpnp}, which we now review.

The above matrix integral turns out to be a close cousin of the matrix integral for CS theory mentioned in (\ref{betain}). Although in principle it is not of the form (\ref{zmm}), it can be put in such a form by an appropriate change of variables. Moreover, all the notions introduced in the previous section to analyze the large $N$ limit 
of conventional matrix models --such as resolvent, spectral curve, and the like-- can be generalized to the ABJM matrix model.
 It turns out that, at large $N$, the ABJM matrix model is a two-cut matrix model. In terms of the variables 
\be
Y=\re^y,\;X=\re^x,
\ee
the spectral curve is given by the equation \cite{akmv,hy,mp,dmp}
\begin{equation}
\label{abjmcurve}
 Y+\frac{X^2}{Y}-X^2+\ri\kappa\,X-1=0\,. 
\end{equation}
The Riemann surface of (\ref{abjmcurve}) can be represented by two $X$-planes glued along the cuts $[1/a,a]$ and $[-b,-1/b]$. The position of the endpoints can be determined from
\begin{equation}
 a+\frac{1}{a}+b+\frac{1}{b}=4,\qquad 
 a+\frac{1}{a}-b-\frac{1}{b}=2\ri\kappa\,.
\end{equation}
The variable $\kappa$ can be regarded as the complex modulus of the spectral curve. 
As in (\ref{tper2}), the 't Hooft parameter can be obtained as a period of the spectral curve and 
it is related to $\kappa$ by \cite{mp}
 \be
 \label{lamkap}
 \lambda(\kappa)={\kappa \over 8 \pi}   {~}_3F_2\left(\frac{1}{2},\frac{1}{2},\frac{1}{2};1,\frac{3}{2};-\frac{\kappa^2
   }{16}\right).
\ee
The free energy of the ABJM matrix model has a $1/N$ expansion 
of the form 
\be
\label{fgs}
F(\lambda,g_s)=\sum_{g=0}^{\infty} g_s^{2g-2} F_g(\lambda).
\ee
The genus zero free energy can be obtained from (\ref{dfy}) as \cite{dmp,mmcslectures}
\be
\label{comf}
 \partial_\lambda F_0={\kappa \over 4} G^{2,3}_{3,3} \left( \begin{array}{ccc} {1\over 2}, & {1\over 2},& {1\over 2} \\ 0, & 0,&-{1\over 2} \end{array} \biggl| -{\kappa^2\over 16}\right)+{ \pi^2 \ri \kappa \over 2} 
  {~}_3F_2\left(\frac{1}{2},\frac{1}{2},\frac{1}{2};1,\frac{3}{2};-\frac{\kappa^2 
   }{16}\right),
\ee
wheere $G^{2,3}_{3,3}$ is a Meijer function. Near $\lambda=0$ this planar free energy has a logarithmic singularity, but this is due to the Gaussian 
part of the free energy, and once this part is subtracted (as in (\ref{g-sub}), with $d=2$ since there are two cuts) we obtain an analytic function at $\lambda=0$, as expected. There are singularities in the complex plane of the 't Hooft parameter, signaling the finite radius of convergence of the series. The location of the singularities can be found explicitly from the above expressions. Indeed, the singularities correspond to the branch point of the hypergeometric functions involved, at 
\be
\kappa =\pm 4 \ri, 
\ee
which correspond to 
\be
\label{critpoint}
\lambda=\mp{2 \ri K \over \pi^2},
\ee
where $K$ is Catalan's constant. For reasons explained in \cite{dmp}, these points are called the {\it conifold} points. 

In \cite{dmp} it was also shown that, for $g\ge 1$, the free energies can be written in terms of quasi-modular forms of the 
modular parameter of the elliptic curve (\ref{abjmcurve}). 
\be
\label{tauex}
\tau=\ri  {K'\left({\ri \kappa \over 4}\right)\over K \left({\ri \kappa \over 4}\right)}.
\ee
For $g=1$, one simply has 
\be
F_1=-\log \, \eta (\tau), 
\ee
where $\eta$ is the usual Dedekind eta function. For $g\ge 2$, the $F_g$ can be written in terms of the quasi-modular forms $E_2$ (the standard Eisenstein series), $b$ and $d$, where 
\be
b=\vartheta_2^4(\tau), \qquad d=\vartheta_4^4(\tau),
\end{equation} 
are standard Jacobi theta functions. More precisely, we have the general structure  
\be
F_g(\lambda)={1\over \left( b d^2 \right)^{g-1}} \sum_{k=0}^{3g-3}
E_2^{k}(\tau) p^{(g)}_k(b,d), \qquad g\ge 2, 
\ee
where $p^{(g)}_k(b,d)$ are polynomials in $b, d$ of modular weight $6g-6-2k$. In \cite{dmp} a recursive procedure was described which gives 
all the genus $g$ free energies unambiguously. The genus $g$ free energies $F_g(\lambda)$ obtained in this way are exact interpolating functions of the 't Hooft parameter, and they can be studied in various regimes. When $\lambda\rightarrow 0$ they reproduce the perturbation theory of the matrix model (\ref{kapmm}) around the Gaussian point $\lambda=0$, where they behave, as expected, as two copies of the Gaussian matrix model, 
\be
F_g(\lambda)=- {B_{2g} \over g (2g-2)} (2 \pi \ri \lambda)^{2-2g}+\CO(\lambda). 
\ee
The genus $g$ free energies can be also studied in the strong coupling regime $\lambda \rightarrow \infty$, or, equivalently, at $\kappa \rightarrow \infty$. In this regime it is more convenient to use the shifted variable 
\be
\label{hatl}
\hat \lambda =\lambda -{1\over 24}={\log ^2\kappa\over 2 \pi^2}+\CO( \kappa^{-2}), \qquad \kappa \gg 1. 
\ee
One finds the following structure. For $F_0$ and $F_1$ one has, at strong coupling, 
\be
\ba
F_0&={4 \pi^3 {\sqrt{2}} \over 3} \hat \lambda^{3/2} + {\zeta(3) \over 2} + \CO\left(\re^{-2 \pi {\sqrt{2 \hat \lambda}}}\right), \\
F_1&={1\over 6} \log \kappa -{1\over 2} \log\left[ {2 \log \kappa \over \pi} \right]+ \CO\left( {1\over  \kappa^2} \right).
\ea
\ee
For $g\ge 2$ one has
\be
\label{fgl}
F_g=c_g+  f_g\left( {1\over \log \, \kappa} \right)+ \CO\left( {1\over  \kappa^2} \right),
\ee
where 
\be
c_g=- {4^{g-1} |B_{2g} B_{2g-2}| \over g (2g-2) (2g-2)!}
\ee
can be interpreted as the contribution from constant maps to the free energy, and 
\be
f_g(x)=\sum_{j=0}^g c_j^{(g)}x^{2g-3+j}
\ee
is a polynomial\footnote{The interpolating functions $F_g$ computed in \cite{dmp} did not include the constant map contribution. This was corrected in \cite{hanada}.}. The leading, strong coupling behavior is then given by 
\be
\label{leadingfg}
 F_g(\lambda)-c_g\sim \lambda^{{3\over 2}-g}, \qquad \lambda \to \infty, \quad g\ge2.  
\ee

We can now ask what is the large order behavior of $F_g(\lambda)$ and what are the possible instanton configurations underlying this behavior. 
First of all, we notice that the $\lambda$-independent part $c_g$ of $F_g$ has the large genus behavior, 
\be
\label{loconstant}
c_g \sim -(2\pi^2)^{-2g}  \Gamma(2g-1), 
\ee
and corresponds to a constant instanton action governing this constant map contribution \cite{ps}, 
\be
A_m=2 \pi^2.
\ee
In order to identify other instanton actions contributing to the asymptotics, one can look at vanishing periods of the meromorphic form $y(x) \rd x$, 
as suggested by the discussion in subsection \ref{inst-spectral}. There are two obvious vanishing periods, namely the ``Gaussian" period vanishing at $\lambda=0$, 
\be
\label{wca}
A_{w}(\kappa)=-4 \pi^2 \,  \lambda (\kappa).  
\ee
There is another period which vanishes at the conifold point $\kappa=\pm 4 \ri$, given by 
\be
\label{wiac}
A_{c}(\kappa)= {\ri \over \pi} {\partial F^{(w)}_0 \over \partial \lambda} +4 \pi^2 \lambda \pm  \pi^2  \\
=\frac{\ri \kappa}{4\pi} G^{2,3}_{3,3} \left(\left. 
  \begin{array}{ccc} 
    \frac{1}{2}, & \frac{1}{2},& \frac{1}{2} \\ 
    0, & 0,&-\frac{1}{2} 
  \end{array} \right| -\frac{\kappa^2}{16}\right)\pm  \pi^2.
\end{equation} 
Finally, there is a linear combination of both instanton actions, 
\be
\label{sca}
A_{s}(\kappa)=A_{w}(\kappa)+A_{c}(\kappa). 
\ee
It was shown in \cite{dmpnp} that these three instanton actions control the large order behavior of $F_g(\lambda)-c_g$ in different regions of the complex plane of the 
't Hooft parameter. Near the Gaussian or weakly coupled point, it is (\ref{wca}) which controls the large order behavior. Near the conifold point (\ref{critpoint}), it is (\ref{wiac}) which dominates. Finally, in the strong coupling region $\lambda \gg 1$, the large order behavior is controlled by (\ref{sca}). Notice that we can regard the behavior of (\ref{loconstant}) as controlled by a constant period, which always exists in these models. 

We can now discuss Borel summability. In the physical ABJM theory, $\lambda$ is real and $g_s$ is purely imaginary. The expansion (\ref{fgs}) should be written in terms of the real coupling constant 
$2\pi/k$, i.e. as
\be
F(\lambda, k)=\sum_{g=0}^{\infty} \left( {2\pi \over k}\right)^{g-2} (-1)^{g-1} F_g(\lambda). 
\ee
We get an extra $(-1)^{g-1}$ sign at each genus. 
Equivalently, this leads to an extra $-\ri$ factor in the instanton actions computed above. We can now ask whether this factorially divergent 
 series is Borel summable or not. At strong coupling $\lambda\gg 1$, the behavior of the genus $g$ free energy $(-1)^{g-1} F_g(\lambda)$ 
 is dominated by the constant period $-\ri A_m=-2 \ri \pi^2$, which is purely imaginary. We then obtain a Borel summable series. Even after subtracting the constant map contribution, we obtain a Borel-summable series at strong coupling. Indeed, the strong coupling action (\ref{sca}), which controls the asymptotics in this regime, is complex: 
 \be
 {\rm Im}\left(  -\ri A_s(\lambda)\right) =\pi^2, 
 \ee
and for large $\lambda$ we have, 
\be
\label{iaction}
-\ri A_s(\lambda)=2 \pi^2 {\sqrt{2 \lambda}} + \pi^2 \ri + \CO\left(\re^{-2 \pi {\sqrt{2\lambda}}}\right), \qquad \lambda \gg 1. 
\ee
Interestingly, it was shown in \cite{dmpnp} that the first, leading term of this action at strong coupling coincides with the action of a D2-brane wrapping $\IR\IP^3$ inside $\IC \IP^3$. This seems to indicate that this instanton of the matrix model corresponds indeed to a D2-brane in the dual type IIA string theory, as expected from general arguments in superstring theory \cite{polchinski}. 

\sectiono{Concluding remarks}

In these lectures I have tried to provide a pedagogical introduction to some aspects of non-perturbative effects in quantum theory. I have 
focused on topics which are not covered in detail in the classic reviews on the subject, like large $N$ gauge theories and matrix models, with a view towards 
applications to string theory. It is clear that, in many respects, this topic is still in its infancy. Even in the context of large $N$ matrix models, where many explicit results are available, there are still some important open problems. For example, there are no detailed results on the large order asymptotics for the genus $g$ free energies in multi-cut phases, i.e. 
there is no analogue of equation (\ref{lostringl}) for these phases; only the instanton action $A$ has been identified in some cases, as discussed in subsection \ref{inst-spectral}. 
The ideas of resurgence, which have been very powerful in the context 
of ODEs and Quantum Mechanics, have scarcely been used in QFT (see however \cite{au} for recent work in this direction). In the case of string theory, 
these ideas have been explored in \cite{mmnp,asv}, but mostly in the case of non-critical strings and toy models with large $N$ duals, and much remains to be done. I only hope that these lectures 
will be useful for future research along these directions.

\section*{Acknowledgments}
A first version of these lectures was presented in the research program ``Topological String Theory, Modularity and Non-Perturbative Physics"
at the Erwin Schr\"odinger Institute in Vienna, in the summer of 2010. I would like to thank the organizers, and particularly Albrecht Klemm, for the invitation to 
 give these lectures. I would also like to thank all my collaborators in the work described here:  Nadav Drukker, 
Bertrand Eynard, Stavros Garoufalidis, Alexander Its, 
Andrei Kapaev, Albrecht Klemm, Sara Pasquetti, Pavel Putrov, 
Marco Rauch, Ricardo Schiappa, and Marlene Weiss. Special thanks to 
Ricardo Schiappa, who was kind enough to read the whole set of lectures and made 
many invaluable comments. This work is supported in part by the 
Fonds National Suisse, subsidies 200020-126817 and 
200020-137523. 


\end{document}